\documentclass[twocolumn,fleqn]{svjour3}         % twocolumn
\smartqed  % flush right qed marks, e.g. at end of proof
\usepackage{graphicx}
\usepackage{mathptm,times}
\usepackage{tikz}
\usepackage{xspace}
\usepackage{pgfkeys}
\usepackage{pgfplots}
\usepackage{pgfplotstable}
\usetikzlibrary{shapes}
\usetikzlibrary{arrows,automata}
\usepackage{color}
\usepackage{xcolor}
\usepackage{graphicx}
\usepackage{subfigure}
\usepackage{algorithm}
\usepackage{algorithmic}
\usepackage[framed]{ntheorem}
\usepackage{xspace}
\usepackage{amsmath}
\usepackage{amssymb}
\usepackage{cite}

\newtheorem{observation}{Observation}

\newcommand{\univ}{\mathcal{U}}
\newcommand{\W}{W}
\newcommand{\WSET}{\mathcal{W}}

\newcommand{\Dom}{\mathcal{D}}

\newcommand{\attrset}{{\bf \mathcal{A}}}
\newcommand{\prefset}{\mathcal{H}}
\newcommand{\formset}{\mathcal{F}}
\newcommand{\formvars}{Var}

\newcommand{\U}{\mathcal{U}}
\newcommand{\parcomp}{ \ \otimes \ }
\newcommand{\parcompsymbol}{\otimes}
\newcommand{\pricomp}{ \ \& \ }
\newcommand{\pricompsymbol}{\&}

\newcommand{\skyline}{\emph{sky}}
\newcommand{\crhs}[1]{\mathcal{R}_{#1}}
\newcommand{\clhs}[1]{\mathcal{L}_{#1}}
\newcommand{\noedges} {\sim}

\newcommand{\nodes}{N}

\newcommand{\negsystem}{\mathcal{N}}
\newcommand{\possystem}{\mathcal{P}}
\newcommand{\oset}{\mathcal{O}}

\newcommand{\nodesof}[1]{N(#1)}
\newcommand{\edgesof}[1]{E(#1)}
\newcommand{\edgeof}[3]{(#1, #2) \in #3}
\newcommand{\edgeofx}[3]{(#1, #2) $\in$ #3}
\newcommand{\notedgeof}[3]{(#1, #2) \not \in #3}
\newcommand{\notedgesymof}[3]{#3 \models #1 \sim #2}
\newcommand{\edgepartition} {$\rightarrow$-partition\xspace}
\newcommand{\noedgepartition} {$\sim$-partition\xspace}
\newcommand{\pair}[2]{\langle #1,#2\rangle}

\newcommand{\pskylinephi}{q}

\newcommand{\attrseq}[2]{\Psi_{{#1}, {#2}}}
\newcommand{\tupleseq}[2]{\Sigma_{{#1}, {#2}}}

\newcommand{\envelope}{{\tt Envelope}\xspace}
\newcommand{\genenvelope}{{\tt Ge\-ne\-ral\-En\-ve\-lo\-pe}\xspace}
\newcommand{\spoenvelope}{{\tt SPO+Enve\-lope}\xspace}
\newcommand{\spoenvelopex}{{\tt SPO+ Envelope}\xspace}
\newcommand{\spo}{{\tt SPO}\xspace}
\newcommand{\thmname}[1]{{\tt{\bf (#1)}}}

\newcommand{\equivset}[1]{\ensuremath{\approx_{#1}}}
\newcommand{\atomic}{\succ}
\newcommand{\attr}{>}

\newcommand{\sat}{\texttt{SAT}\xspace}
\newcommand{\fsat}{\texttt{FSAT}\xspace}

\newcommand{\fnp}{\texttt{FNP}\xspace}
\newcommand{\fnpcomplete}{\texttt{FNP}-complete}
\newcommand{\fnphard}{{\tt FNP}-hard}

\newcommand{\betterin}{BetIn}
\newcommand{\winnow}{\omega}

\newcommand{\set}[1]{\ensuremath{{\bf #1}}}

\newcommand{\prooflabel}[1]{\noindent \begin{tikzpicture} 
    \node[draw, inner sep=3pt, rounded corners=2pt] at (0,0) 
         {{\scriptsize #1}};;\end{tikzpicture}\ \ }

\newcommand{\leftsideproof}{\prooflabel{$\Leftarrow$}}
\newcommand{\rightsideproof}{\prooflabel{$\Rightarrow$}}
\newcommand{\equivproof}[2]{\prooflabel{$#1 \Leftrightarrow #2$}}

\newcommand{\immparents}[1]{Pa^*_{#1}}

\newcommand{\derivingseq}{derivation sequence\xspace}

\tikzstyle{p-graph-node} = [draw, rounded corners=2pt, font=\fontsize{8}{8}, outer sep=1pt, inner sep=2pt]
\tikzstyle{p-graph-edge-left} =  [->, bend left=60]
\tikzstyle{p-graph-edge-right} = [->, bend right=60]
\tikzstyle{p-graph-edge-left-80} =  [->, bend left=80]
\tikzstyle{p-graph-edge-right-80} = [->, bend right=80]
\tikzstyle{p-graph-edge} = [->]

\newcommand{\parents}[1]{Pa_{#1}}
\newcommand{\children}[1]{Ch_{#1}}
\newcommand{\sibl}[1]{Sibl_{#1}}
\newcommand{\anc}[1]{Anc_{#1}}
\newcommand{\ancself}[1]{Anc\mbox{-}self_{#1}}
\newcommand{\desc}[1]{Desc_{#1}}
\newcommand{\descself}[1]{Desc\mbox{-}self_{#1}}

\begin{document}

\title{Preference Elicitation in Prioritized Skyline Queries}
\author{Denis Mindolin         \and
        Jan Chomicki}

\institute{D. Mindolin \at
           201 Bell Hall\\
           University at Buffalo, Buffalo, NY 14260-2000, USA\\
           Tel.: +1-718-408-0833\\
           \email{mindolin@buffalo.edu}           %  \\
           \and
           J. Chomicki \at
           201 Bell Hall\\
           University at Buffalo, Buffalo, NY 14260-2000, USA\\
           Tel: +1-716-645-4735\\
           Fax: +1-716-645-3464\\
           \email{chomicki@buffalo.edu}
}

\date{Received: date / Accepted: date}

\maketitle

\begin{abstract}
\emph{Preference queries}  incorporate the notion of binary \emph{preference relation}
into relational database querying. Instead of returning \emph{all} the answers,
such queries return only the \emph{best} answers, according to a given preference
relation.

Preference queries are a fast growing  area of database research.
\emph{Skyline} queries constitute one of the most thoroughly studied
classes of preference queries.  A well known limitation of skyline
queries is that skyline preference relations assign the same
importance to all attributes. In this work, we study
\emph{p-skyline} queries that generalize skyline queries
by allowing varying attribute importance in preference relations.

We perform an in-depth study of the properties of p-skyline preference
relations.  In particular, we study the problems of containment and
minimal extension.  We apply the obtained results to the central
problem of the paper:
\emph{eliciting relative importance  of attributes}.
Relative importance is implicit in the constructed p-skyline
preference relation. The elicitation is based on user-selected sets of
\emph{superior} (positive) and \emph{inferior} (negative) examples. We
show that the computational complexity of elicitation depends on
whether inferior examples are involved. If they are not, 
elicitation can be achieved in polynomial time. Otherwise, it is
NP-complete.  Our experiments show that the proposed elicitation
algorithm has high accuracy and good scalability.

\end{abstract}

\section{Introduction}\label{sec:intro}
Effective and efficient \emph{user preference management} is a crucial part of any
successful sales-oriented business. Knowing \emph{what} customers like
and more importantly \emph{why} they like that and what they
\emph{will} like in the future is an essential part of the modern risk
management process.  The essential components of preference management
include preference specification, preference elicitation, and querying
using preferences.  Many preference handling frameworks
have been developed
\cite{Borz01theskyline,kiessling02sql,brafman2002,Chomicki2003,Pu2003,Hansson1995,Fishburn1970}.

Our starting point here is the 
\emph{skyline framework} \cite{Borz01theskyline}. 
The skyline preference relation is defined on top of a set of
preferences over individual attributes.  It represents the
\emph{Pareto improvement} principle: \emph{a tuple $o_1$ is preferred
to a tuple $o_2$ iff $o_1$ is as good as $o_2$ according to all the
attribute preferences, and $o_1$ is strictly better than $o_2$
according to at least one attribute preference}. Now given a set of
tuples, the set of the \emph{best} tuples according to this principle
is called a \emph{skyline}.

\begin{example} \label{ex:skyline}

Assume  the following cars are available for sale. 

\begin{center}
\begin{tabular}{c||c|c|c} 
        & make & price & year \\
 \hline
  $t_1$ & ford & 30k & 2007\\
  $t_2$ & bmw & 45k & 2008 \\
  $t_3$ & kia & 20k & 2007 \\
  $t_4$ & ford & 40k & 2008 \\
  $t_5$ & bmw & 50k & 2006 \\
\end{tabular}
\end{center}

Also, assume that  Mary wants to buy a car and her attribute preferences 
are as follows:

\begin{center}
\begin{tabular}{l||l}
  $>_{make}$ & BMW is better than Ford,  Ford is better than Kia\\
  $>_{year}$ & the car should be as new as possible \\
  $>_{price}$ & the car should be as cheap as possible.\\
\end{tabular}
\end{center}

Then the skyline is 

\begin{center}
\begin{tabular}{c||c|c|c} 
        & make & price & year \\
 \hline
  $t_1$ & ford & 30k & 2007 \\
  $t_2$ & bmw & 45k & 2008 \\
  $t_3$ & kia & 20k & 2007 \\
  $t_4$ & ford & 40k & 2008 \\
\end{tabular}
\end{center}
\end{example}

A large number of algorithms for computing skyline que\-ries have been
developed
\cite{Borz01theskyline,Godfrey2003,godfrey2005,Lin2005}.
Elicitation of skyline
preference relations based on user-provided feedback has also been studied \cite{Pei2008}.

\medspace

One of the reasons of the popularity of the skyline framework is the
simplicity and intuitiveness of skyline semantics.  Indeed, in
order to define a skyline preference relation, one needs to provide
only two parameters: the set $\attrset$ of relevant attributes
and the set  $\prefset$ of corresponding preferences  over each individual
attribute in $\attrset$.  
(In Example \ref{ex:skyline}, $\attrset=\{make,price,year\}$
and $\prefset=\{>_{make},>_{price},>_{year}\}$.)

At the same time, the simplicity of skyline semantics comes with a
number of well known limitations.  One of them is the inability of skyline
preference relations to capture the important notion of \emph{difference in attribute
importance}.
The Pareto improvement principle implies that all relevant
attributes have the same importance.
However, in real life, it is often the case that benefits in
one attribute may outweigh losses in one or more attributes.  For
instance, given two cars that differ in age and price, for some people the
age is crucial while the price is secondary.  Hence, in that case,
\emph{the price has to be considered only when the benefits in age cannot
  be obtained}, i.e., when the age of the two cars is the same. 

\begin{example}\label{ex:motiv}
 Assume that Mary decides that \emph{year} is more important for her
 than \emph{make} and \emph{price}, which in turn are equally
 important.  Thus, regardless of the values of \emph{make} and
 \emph{price}, a newer car is always better than an old one. At the
 same time, given two cars of the same age, one needs to compare their
 \emph{make} and \emph{price} to determine the better one.  The set of
 the best tuples according to this preference relation is

\begin{center}
\begin{tabular}{c||c|c|c} 
        & make & price & year \\
 \hline
  $t_2$ & bmw & 45k & 2008 \\
  $t_4$ & ford & 40k & 2008 \\
\end{tabular}
\end{center}
\smallskip

Namely, $t_2$ and $t_4$ are better than all other tuples in {\tt
  year}, but $t_2$ is better than $t_4$ in {\tt make}, and $t_4$ is
better than $t_2$ in {\tt price}.
\end{example}

Another drawback of the skyline framework is that the size of a
skyline may be exponential in the number of attribute preferences \cite{DBLP:conf/foiks/Godfrey04}. A query result of that size is likely
to overwhelm the user. In \emph{interactive preference elicitation scenarios}
\cite{Balke2007}, user preferences are elicited in a stepwise
manner. A user is assumed to analyze the set of the
best tuples according to the \emph{intermediate} preference relation
and criticize it in some way. Clearly, if such a tuple set is too
large, it is hard to a expect high quality feedback from the
user.
The large size of a skyline is
caused by the looseness of the Pareto improvement
principle. \emph{Pareto improvement} implies that if a tuple $o$ is
better than $o'$ in one attribute, then the existence of some attribute
in which $o'$ is better than $o$ makes the tuples
\emph{incomparable}. Thus, every additional attribute increases the
number of incomparable tuples.

\medskip

Here we develop the \emph{p-skyline} framework which generalizes the
skyline framework and addresses its limitations listed above:
the inability to capture differences in attribute importance and large
query results.
The skyline semantics is enriched with the notion of
\emph{attribute importance} in a natural way.  
Assuming two relevant attributes $A$ and $B$ such that
$A$ is more important than  $B$, a tuple
with a better value of $A$ is \emph{unconditionally} preferred to all
tuples with worse values of $A$, regardless of their values of
$B$. However, given a tuple with the same value of $A$, the one with a
better value of $B$ is preferred (assuming no other attributes are involved). 
For equally important
attributes, the Pareto improvement principle applies. Therefore,
skyline queries are also representable in our framework.

Relative attribute importance implicit in a p-skyline preference
relation is represented explicitly as a \emph{p-graph}: a graph
whose nodes are attributes, and edges go from more to less important
attributes.  Such graphs satisfy the properties quite natural for
importance relationships: transitivity and irref\-le\-xi\-vi\-ty.
We show that, in addition to representing attribute importance,
p-graphs play another important role in the p-skyline framework: they
can be used to determine \emph{equivalence} and \emph{containment} of
p-skyline relations, and tuple \emph{dominance}.

We notice that two p-skyline relations may differ in the following
aspects: 
\begin{itemize}
\item the set $\attrset$ of relevant attributes,
\item the set $\prefset$ of  preferences over
those attributes, and 
\item the relative importance of the corresponding
attributes, represented by a p-graph.
\end{itemize}
In this work, we are
particularly interested in the class $\formset_\prefset$ of \emph{full p-skyline relations}
for which the set of relevant attributes $\attrset$ consists of all the attributes
and the set of corresponding attribute preferences is  $\prefset$.
Hence, two different p-skyline relations from $\formset_\prefset$ are different
only in the corresponding p-graphs. 
We show the following properties of  such relations:
\begin{itemize}
\item the containment and equivalence of p-skyline relations are equivalent to the containment and equivalence
of their p-graphs;
\item four transformation rules are enough to generate all minimal extensions 
of a p-skyline  relation;
\item the number of all minimal extensions of a p-skyline relation
is \emph{polynomial} in $|\attrset|$;
\item every $\subset$-chain in  $\formset_\prefset$ is of \emph{polynomial} length, 
although $\formset_\prefset$ contains at least $|\attrset|!$
relations.
\end{itemize}

The properties listed above are used to develop the elicitation algorithm
and prove its correctness.
Incorporating attribute
importance into skyline relations allows not only to model user
preferences more accurately but also to make the size of the
corresponding query results more manageable.

\medskip

At the same time, enriching the skyline framework with attribute
importance comes at a  cost. To construct a p-sky\-line
preference relation from a skyline relation, one needs to provide a p-graph describing
relative attribute importance.  However, requiring
users to describe attribute importance explicitly seems impractical
for several reasons.  First, the number of pairwise attribute comparisons required may be 
large. Second, users themselves may be not fully aware of their
own preferences.

To address this problem, we develop a method of \emph{elicitation} of
p-skyline relations based on simple \emph{user-provided feedback}.  The type
of feedback used in the method consists of two sets of tuples
belonging to a given set: \emph{superior examples} \cite{Pei2008},
i.e., the \emph{desirable} tuples, and \emph{inferior examples}
\cite{Pei2008} i.e., the \emph{undesirable} tuples.  This type
of feedback is quite natural in real life: given a set of tuples, a
user needs to examine them and identify some tuples she likes and
dislikes most. Moreover, it is advantageous from the point of view of
user interface design -- a user is required to perform a number of simple
``check off'' actions to identify such tuples. Finally, such feedback can be elicited
automatically \cite{Kiessling2003}.

We consider the problems related to the construction of p-skyline relations covering
the given superior and inferior examples. Specifically, we need to guarantee that
the superior examples are among the best tuples and that the inferior examples
are dominated by at least one other tuple.
Also, to guarantee an optimal fit we postulate that the constructed relation be
maximal. We show that 
determining the existence of a p-skyline relation
covering the given examples is {\tt NP}-complete and constructing a maximal such relation
{\tt FNP}-complete.

In real-life scenarios of preference elicitation using superior and
inferior examples, users may only be indirectly involved in the
process of identifying such examples. For instance, the click-through
rate may be used to measure the popularity of products. Using this
metric, it is easy to find the superior examples -- the tuples with
the highest click-through rate.  However, the problem of identifying
inferior examples -- those which the user confidently dislikes -- is
harder. Namely, low click-through rate may mean that a tuple is
inferior, the user does not know about it, or it
simply does not satisfy the search criteria. Thus, there is a need for
eliciting p-skyline relations based on superior examples only.  We
address that problem here.  We show a polynomial-time algorithm
for checking the existence of a p-skyline relation covering a given
set of superior examples, and a poly-nomial-time algorithm for
constructing a maximal p-skyline relation of that kind.  The latter
algorithm is based on checking the satisfaction of a
\emph{system of negative constraints}, each of which captures
the fact that one tuple does not dominate another according to the
p-skyline relation being constructed.

We provide two effective methods for \emph{reducing} the size of
systems of negative constraints and hence improving the performance of
the elicitation algorithm. At the same time, we show that the problem
of \emph{minimizing} the size of such a system is unlikely to be
efficiently solvable.  The experimental evaluation of the algorithms
on real life and synthetic data sets demonstrates high accuracy and
scalability of the elicitation algorithm, as well as the efficacy of the proposed
optimization methods.

The paper is organized as follows. In section
\ref{sec:basic-notation}, we introduce the concepts used throughout the
paper.  In section \ref{sec:pskylines}, we describe p-skylines --
skylines enriched with relative attribute importance information. We also
discuss the fundamental  properties of such relations. In section
\ref{sec:p-skyline-elicitation}, we study the problem of eliciting
p-skyline relations based on superior and inferior examples. In Section
\ref{sec:experiments}, we show the results of the experimental
evaluation of the proposed algorithms. Section \ref{sec:relwork}
concludes the paper with a discussion of related and future work.  The proofs of all
the results presented in the paper are provided in the Appendix.

\section{Basic notations}\label{sec:basic-notation}

\subsection{Binary relations}
  A \emph{binary relation} $R$ over a (finite of infinite) set $S$ is
  a subset of $S \times S$. Binary relations may be \emph{finite} or
  \emph{infinite}. 
To denote $(x,y) \in R$, we may write $R(x,y)$ or $x
  \ R\ y$. Here we list some typical properties of binary
  relations. 
  A binary relation $R$ is
  \begin{itemize}
    \item \emph{irreflexive} iff $\forall x\ .\ \neg R(x,x)$,
    \item \emph{transitive} iff $\forall x,y,z\ .\ R(x,y) \wedge R(y,z)
      \rightarrow R(x,z)$,
    \item \emph{connected} iff $\forall x,y,z\ .\ R(x,y) \vee R(y,x) \vee x = y$,
    \item \emph{a strict partial order (SPO)} if it is 
      irreflexive and transitive,
    \item \emph{a weak order} iff it is an SPO such that
      \[\forall x,y,z\ .\ R(x,y) \rightarrow R(x,z) \vee R(z,y),\]
    \item \emph{a total order} if it is a connected SPO.
  \end{itemize}
  The \emph{transitive closure} $TC(R)$ of a binary relation $R$
  is defined as
  \[(x,y) \in TC(R)\ \mbox{iff}\ R^m(x,y)\ \mbox{for some}\ m > 0,\]
  where
  \begin{align*}
    R^1(x,y) 	& \equiv R(x,y)\\
    R^{m+1}(x,y) 	& \equiv \exists z\ .\ R(x, z) \wedge R^m(z,y)
  \end{align*}

  A binary relation $R \subseteq S \times S$ may be
  viewed as a directed graph.  The
  set $S$ is called \emph{the set of nodes of $R$} and denoted as
  $\nodes(R)$.  We say that the tuple $xy$ is an \emph{$R$-edge from
    $x$ to $y$} if $(x, y) \in R$.  A \emph{path in $R$} (or an
  \emph{$R$-path}) from $x$ to $y$ for an $R$-edge $xy$ is a sequence
  of $R$-edges such that the start node of the first edge is $x$, the
  end node of the last edge is $y$, and the end node of every edge
  (except the last one) is the start node of the next edge in the
  sequence. The \emph{length of an $R$-path} is the number of
  $R$-edges in the path.  An \emph{$R$-sequence} is the sequence of
  nodes participating in an $R$-path.  The \emph{length of an
    $R$-sequence} is the number of nodes in it.

  Given a directed graph $R$ and its node $x$,
  \begin{itemize}
  \item $\children{R}(x) = \{y\ |\ (x, y) \in R\}$ is the set of
    \emph{children of $x$ in $R$},
  \item $\parents{R}(x) = \{y\ |\ (y, x) \in R\}$ is the set of
    \emph{parents of $x$ in $R$},
  \item $\immparents{R}(x) = \parents{R}(x) -
\parents{R}(\parents{R}(x))$ is the set of
    \emph{immediate parents of $x$ in $R$},
  \item $\desc{R}(x) = \{y\ |\ (x,y) \in TC(R)\}$ is the set of
    \emph{descendents of $x$ in $R$},
  \item $\anc{R}(x) = \{y\ |\ (y,x) \in TC(R)\}$ is the set of
    \emph{ancestors of $x$ in $R$},
  \item $\sibl{R}(x) = \nodesof{R} - (\desc{R}(x) \cup \anc{R}(x) \cup
    \{x\})$ is the set of \emph{siblings of $x$ in $R$}
  \end{itemize}

  We also write $\descself{R}(x)$ and $\ancself{R}(x)$ as shorthands
  of $(\desc{R}(x) \cup \{x\})$ and $(\anc{R}(x) \cup \{x\})$,
  respectively.
  Similarly, we define set versions of the above definitions,
  e.g., $\children{R}(X) = \{y\ |\ \exists x\in X. (x, y) \in R\}$.

  Given two nodes $x$ and $y$ of $R$ and two sets of nodes $X$ and $Y$ of $R$, we write
  \begin{itemize}
  \item $\notedgesymof{x}{y}{R}$ \ iff \ $\notedgeof{x}{y}{R}$ and
    $\notedgeof{y}{x}{R}$;
  \item $\notedgesymof{X}{Y}{R}$ \ iff \ $\forall x \in X, y \in
    Y\ .\ \notedgesymof{x}{y}{R}$;
  \item $\edgeof{X}{Y}{R}$ \ iff \ $\forall x \in X, y \in
    Y\ .\ \edgeof{x}{y}{R}$.
  \end{itemize}

\subsection{Preference relations}
  
  Below we describe some concepts of a variant of the 
  preference framework \cite{Chomicki2003}, which we adopt here.
  
  Let $\attrset = \{A_1,...,A_n\}$ be a finite set of attributes (a
  relation schema). Every attribute $A_i \in \attrset$ is
  associated with an \emph{infinite domain} $\Dom_{A_i}$.  The domains
  considered here are rationals and uninterpreted constants (numerical
  or categorical).  We work with the \emph{universe of tuples} $\U =
  \prod_{A_i \in \attrset} \Dom_{A_i}$.  Given a tuple $o \in \U$, we
  denote the value of its attribute $A_i$ as $o.A_i$.
  
  Preference relations we consider in this paper are of two types:
  \emph{attribute} and \emph{tuple}.
  \begin{definition}
    {\bf(Attribute preference relation)}
    An \emph{attribute preference relation} $\attr_{A_i}$ for an attribute $A_i \in
    \attrset$ is a subset of $\Dom_{A_i}\times\Dom_{A_i}$, which is a \emph{total order} over $\Dom_{A_i}$.
  \end{definition}

  An attribute preference relation describes a preference over the values of
  a single attribute e.g., the \emph{red} color is preferred to
  the \emph{blue} color, or the make \emph{BMW} is preferred to the
  make \emph{Kia}.
  
  \begin{definition}
    {\bf (Tuple preference relation)}
    A \emph{tuple preference relation} $\succ$ is a subset of $\U \times \U$, which is a strict partial
    order over $\U$.
  \end{definition}

  In contrast to an  attribute preference relation, a tuple preference
  relation describes a preference over \emph{tuples}, e.g., a
  \emph{red} \emph{BMW} is preferred to a \emph{blue} \emph{Kia}.  We
  say that 
\begin{itemize}
\item a tuple $o_1$ \emph{dominates} (\emph{is preferred to, is better than}) a tuple $o_2$, and
\item $o_2$ is  \emph{dominated by} (\emph{is worse than})  $o_1$, 
\end{itemize}
according to a preference relation $\succ$, 
  iff $t_1 \succ t_2$.  In the remaining part of the paper, tuple
  preference relations are simply referred to as preference relations.

We assume that both attribute and tuple preferences are defined as quantifier-free
formulas over some appropriate signature.
In this way both finite and infinite preference relations can be captured.
For instance, the following formula
defines an \emph{infinite} tuple preference relation over the domains of the attributes
\emph{make}, \emph{year}, and \emph{price}.
\begin{align*}
 o_1 \succ_1 o_2 \ = & \ o_1.\mbox{{\small \texttt{year}}} \geq o_2.\mbox{{\small\tt year}} 
			\wedge o_1.\mbox{{\small\tt price}} \leq o_2.\mbox{\tt {\small price}} \wedge \\
			& (o_1.\mbox{\tt {\small make}} = BMW \wedge o_2.\mbox{\tt{\small make}} = Ford \ \vee \\
			& o_1.\mbox{\tt {\small make}} = Ford \ \wedge o_2.\mbox{\tt {\small make}} = Kia \vee \\
			& o_1.\mbox{\tt {\small make}} = BMW \wedge o_2.\mbox{\tt {\small make}} = Kia \vee \\
			& o_1.\mbox{\tt {\small make}} = o_2.\mbox{\tt {\small make}}) \wedge 
			(o_1.\mbox{\tt {\small year}} \neq o_2.\mbox{\tt {\small year}} \ \vee \\
			& o_1.\mbox{\tt {\small price}} \neq o_2.\mbox{\tt {\small price}} \vee 
			o_1.\mbox{\tt {\small make}} \neq o_2.\mbox{\tt {\small make}})
\end{align*}

Given a tuple preference relation, the two most common tasks are:
\begin{enumerate}
\item \emph{dominance testing}: checking if a tuple is preferred to another
one, and
\item  \emph{computing the best
  (most preferred) tuples} in a given finite set of tuples.
\end{enumerate}

The first problem is easily solved  by
checking if the formula representing the preference relation evaluates
to true for the given pair of tuples.  (Nevertheless, we will revisit this problem  in
section  \ref{sec:pskylines}.)
To deal with the second
problem, a new \emph{winnow} relational algebra operator was
proposed \cite{Chomicki2003,kiesling2002}.
	
\begin{definition}\label{def:winnow} 
{\bf (Winnow)}
If $\succ$ is a tuple preference relation 
over \hspace{0.05mm} $\U$, then the \emph{winnow} operator 
$\winnow_\succ(\attrset)$ is defined as  
\[\winnow_{\succ}(r) = \{t \in r\ | \ \neg \exists t' \in r\ .\  t' \succ t\}.\]
for every finite  subset $r$ of $\U$.
\end{definition}

\section{p-skylines}\label{sec:pskylines}

Let $\attrset = \{A_1,...,A_n\}$ be a finite set of attributes and $\prefset = \{\attr_{A_1}, \ldots, \attr_{A_n}\}$ be a set of
the corresponding attribute preference relations. 
Below we define the syntax and the semantics of p-skyline
relations.

\noindent{\bf Notation:} We use ``$=$'' for syntactic identity of expressions
and ``$\equiv$'' for equality of relations viewed as sets of tuples.

\begin{definition}\label{def:p-expression}{\bf (p-expression)}
An expression $\pi$ is a \emph{p-ex\-pres\-sion} if
  \begin{itemize} 
        \item $\pi\ {\rm is}\ \attr_{A_i}$ for $A_i \in \attrset$, or
        \item $\pi\ =\ \pi_1 \parcomp \pi_2$ for two p-expressions
              $\pi_1$ and $\pi_2$, or
        \item $\pi\ =\ \pi_1 \pricomp \pi_2$, for two p-expressions
              $\pi_1$ and $\pi_2$.
  \end{itemize}
\end{definition}

\begin{definition}\label{def:rel-attributes}{\bf (Relevant attributes)}
  Given a p-expression $\pi$, the corresponding \emph{set 
  of relevant attributes} $\formvars(\pi)$ is:
  \begin{itemize} 
        \item $\{A_i\}$, if $\pi\ {\rm is}\ \attr_{A_i}$;
        \item $\formvars(\pi_1) \cup \formvars(\pi_2)$ 
         for $\pi \ =\ \pi_1 \pricomp \pi_2$ or
        $\pi =\ \pi_1 \parcomp \pi_2$, where 
        $\pi_1$ and $\pi_2$ are p-expressions.
  \end{itemize}
\end{definition}

\noindent
Given a set of attributes $X$
\[o_1\equivset{X} o_2 \ {\rm iff}\ \forall A\in X. o_1.A=o_2.A.\]

\begin{definition}\label{def:p-expr-induced}{\bf (Preference relation induced by p-expression)}
%    For a set of attribute preference relations $\prefset$ and
%    attributes $\attrset$, 
The preference relation $\succ_\pi$ \emph{induced by a  p-expression $\pi$} is defined as
\begin{enumerate}
   \item if $\pi\ {\rm is}\ \attr_{A_i}$ and $A_i \in \attrset$, 
        \[\succ_{\pi}\ \equiv \ \{ (o,o')\ |\ o,o' \in \univ\ .\ o.A \attr_{A_i} o'.A \},\]
and $\succ_\pi$ is also written as $\succ_{A_i}$, and called an \emph{atomic}
preference relation,
   \item for $\pi = \pi_1 \pricomp \pi_2$,
\[\succ_\pi \ \equiv \ \succ_{\pi_1} \cup
      \ (\equivset{\formvars(\pi_1)} \ \cap \ \succ_{\pi_2}),\]
   \item for $\pi = \pi_1 \parcomp \pi_2$,
\begin{align*}
      \succ_\pi \ \equiv \  (\succ_{\pi_1} \cap \ \equivset{\formvars(\pi_2)}) \ \cup\ \notag
      		        (\succ_{\pi_2} \cap \ \equivset{\formvars(\pi_1)}) \ \cup\ \\
			(\succ_{\pi_1} \cap \succ_{\pi_2}),
\end{align*}
\end{enumerate}
where $\succ_{\pi_1}$ and $\succ_{\pi_2}$ are preference relations induced by
the p-expres\-sions $\pi_1$ and $\pi_2$.
\end{definition}

In the second case, we say that $\succ_\pi\ \equiv\ \succ_{\pi_1} 
\pricomp \succ_{\pi_2}$ and in the third case, that $\succ_\pi\ \equiv\ \succ_{\pi_1} \parcomp \succ_{\pi_2}$.
We also refer to the set of relevant attributes $\formvars(\pi)$ of
$\pi$ as $\formvars(\succ_\pi)$. When the context in clear, we may
omit the subscript $\pi$ and refer to p-skyline relations as
$\succ,\succ_1,\succ_2,\ldots$. 
Note the difference between the \emph{attribute} preference relation $\attr_A$
and the \emph{tuple} preference relation $\atomic_A$. However, the correspondence
between those two relations is straightforward.

The intuition behind Definition \ref{def:p-expr-induced} is as
follows.  In the first case, $\succ_{A_i}$ is the tuple preference
relation corresponding to the attribute preference relation
$\attr_{A_i}$. In the second case, $\succ_\pi$ is composed of
$\succ_{\pi_1}$ and $\succ_{\pi_2}$ in such a way that $\succ_{\pi_1}$
has
\emph{higher importance} than $\succ_{\pi_2}$: a tuple $o$ is preferred to $o'$ according to
$\succ_\pi$ iff $o$ is preferred to $o'$ according to $\succ_{\pi_1}$
(regardless of $\succ_{\pi_2}$), or $o$ and $o'$ are \emph{equal} in
all the relevant attributes of $\succ_{\pi_1}$ and $o$ is preferred to
$o'$ according to $\succ_{\pi_2}$. The operator $\pricomp$ is called
\emph{prioritized accumulation}\cite{kiesling2002}.  Similarly, if
$\pi\ =\ \pi_1 \parcomp\
\pi_2$, then $\succ_{\pi_1}$ and $\succ_{\pi_2}$ are considered to be \emph{equally
important} in $\succ_\pi$. The operator $\parcomp$ is called
\emph{Pareto accumulation}\cite{kiesling2002}.  Some known properties
of these operators are summarized below.

\begin{proposition}\label{prop:op-pros}\emph{\cite{kiesling2002}}
  The operators $\parcomp$ and $\pricomp$ are associative.  The
  operator $\parcomp$ is commutative.
\end{proposition}

Since accumulation operators are associative, we extend them from
binary to n-ary operators.

\begin{proposition}\label{prop:op-spo-kiessling}\emph{\cite{kiesling2002}}
  A relation induced by a p-expression is an SPO, i.e., a tuple
  preference relation.
\end{proposition}

\begin{definition}\label{def:p-skyline} {\bf (p-skyline relation)}
   A \emph{p-skyline relation} $\succ_\pi$ is the relation induced by
   a p-expression $\pi$ such that for all subexpressions of $\pi$ of the form
   $\pi_1 \pricomp \pi_2$ or
   $\pi_1 \parcomp \pi_2$:
  \begin{itemize}
        \item $\formvars(\pi_1) \cap \formvars(\pi_2) = \emptyset$;
        \item the relations induced by $\pi_1$ and $\pi_2$ are p-skyline
relations.
  \end{itemize}
A p-skyline relation $\succ_\pi$ induced by $\pi$ is \emph{full} iff $\formvars(\pi) = \attrset$.

\end{definition}

Essentially, p-skyline relations are induced by those p-ex\-pres\-sions
in which every member of $\prefset$ is used at most once
(exactly once in the case of full p-skyline relations).
The set of \emph{all full p-skyline relations} for $\prefset$ is denoted by
$\formset_\prefset$.
Further we consider only full p-skyline relations. 

A key property of p-skyline relations is that the
\emph{skyline preference relation $\skyline_\prefset$} is the p-skyline 
relation induced by the p-expression 
$\attr_{A_1} \parcomp \ldots \parcomp \attr_{A_n}.$
That is, the p-skyline framework is an \emph{extension}
 of the skyline framework.

\subsection{Syntax trees}\label{sec:syntax-tree}

Dealing with p-skyline relations, it is natural to represent the
corresponding p-expressions as \emph{syntax trees}. This
representation is used in Section \ref{sec:p-skyline-min-ext}
for constructing minimal extensions of a p-skyline relation.

\begin{definition}\label{def:p-skyline-syntax-tree}
{\bf (Syntax tree)} A \emph{syntax tree $T_{\succ_\pi}$} of a
p-skyline relation $\succ_\pi$ is an ordered rooted tree representing
the p-expression $\pi$.
\end{definition}
Every \emph{non-leaf node} of the syntax tree is labeled with an
accumulation operator and corresponds to the result of applying the
operator to the p-skyline relations represented by its children, from
left to right. Every \emph{leaf} node of the syntax tree is labeled with an
attribute $A \in \attrset$ and corresponds to the attribute preference
relation $\attr_A \ \in
\ \prefset$ (and the atomic preference relation $\atomic_A$).

%Given a node $C$ of the syntax tree, we denote the $i$-th
%child node of $C$ as $\treenodechild{C}{i}$.  

\begin{definition}\label{def:normalized}
{\bf (Normalized syntax tree)} A syntax tree is 
\emph{normalized} iff each of its non-leaf nodes is labeled differently
from its parent.
\end{definition}

Clearly, for every p-skyline relation, there is a normalized syntax
tree which may be constructed in polynomial time in the size of the
original tree. To do that, one needs to find all occurrences of syntax
tree nodes $C_1$ and their children $C_2$ such that $C_1$ and $C_2$
have the same label. After that, $C_2$ has to be removed from the list
of children of $C_1$, and the list of children of $C_2$ has to be
added to the list of children of $C_1$ in place of $C_2$.  The
correctness of this procedure follows from Proposition
\ref{prop:op-pros}.

We note that a normalized syntax tree is not unique for a p-skyline
relation. That is due to the commutativity of $\parcomp$ (Proposition
\ref{prop:op-pros}).

\begin{figure}[ht]
	\centering
	\subfigure[Unnormalized]{\label{pic:parse-tree-unnorm}
	\begin{tikzpicture}[>=stealth, xscale=0.5, yscale=0.4]
		\tikzstyle{nonleaf} = [draw=black,circle,inner sep=1pt]
		\tikzstyle{leaf} = [draw=black,inner sep=2pt]
		\node[nonleaf, fill = gray!20] (PAR1) at (1,4) {{\tiny $\parcomp$}};
		\node[leaf] (A)    at (0,3) {{\tiny $A$}};
		\node[nonleaf] (PRI1) at (2,3) {{\tiny $\pricomp$}};
		\node[leaf] (B)    at (1,2) {{\tiny $B$}};
		\node[leaf] (C)    at (3,2) {{\tiny $C$}};

		\node[nonleaf, fill = gray!20] (PAR0) at (2.5,5) {{\tiny $\parcomp$}};

		\node[nonleaf] (PRI2)    at (4,4) {{\tiny $\pricomp$}};
		\node[leaf] (D)       at (3.0,3) {{\tiny $D$}};
		\node[nonleaf] (PAR2)    at (5.5, 3) {{\tiny $\parcomp$}};
		\node[leaf] (E)       at (4,2) {{\tiny $E$}};
		\node[leaf] (F)       at (7,2) {{\tiny $F$}};

		\draw[-, thick] (PAR1) -- (A);
 		\draw[-, thick] (PAR0) -- (PAR1);
 		\draw[-, thick] (PAR0) -- (PRI2);

		\draw[-, thick] (PRI1) -- (B);
		\draw[-, thick] (PRI1) -- (C);
		\draw[-, thick] (PAR1) -- (PRI1);
		\draw[-, thick] (PRI2) -- (D);
		\draw[-, thick] (PRI2) -- (PAR2);
		\draw[-, thick] (PAR2) -- (E);
		\draw[-, thick] (PAR2) -- (F);
	\end{tikzpicture}
	}
	\subfigure[Normalized]{\label{pic:parse-tree-norm1}
	\begin{tikzpicture}[>=stealth, xscale=0.5, yscale=0.4]
		\tikzstyle{nonleaf} = [draw=black,circle,inner sep=1pt]
		\tikzstyle{leaf} = [draw=black,inner sep=2pt]
		\node[nonleaf] (PAR1) at (2,4) {{\tiny $\parcomp$}};
		\node[leaf] (A)    at (0,3) {{\tiny $A$}};
		\node[nonleaf] (PRI1) at (2,2.5) {{\tiny $\pricomp$}};
		\node[leaf] (B)    at (1,1.5) {{\tiny $B$}};
		\node[leaf] (C)    at (3,1.5) {{\tiny $C$}};

		\node[nonleaf] (PRI2)    at (4,3) {{\tiny $\pricomp$}};
		\node[leaf] (D)       at (4.0,2) {{\tiny $D$}};
		\node[nonleaf] (PAR2)    at (5.5, 2) {{\tiny $\parcomp$}};
		\node[leaf] (E)       at (4,1) {{\tiny $E$}};
		\node[leaf] (F)       at (7,1) {{\tiny $F$}};

		\draw[-, thick] (PAR1) -- (A);
		\draw[-, thick] (PAR1) -- (PRI1);
		
		\draw[-, thick] (PRI1) -- (B);
		\draw[-, thick] (PRI1) -- (C);
		\draw[-, thick] (PAR1) -- (PRI2);
		\draw[-, thick] (PRI2) -- (D);
		\draw[-, thick] (PRI2) -- (PAR2);
		\draw[-, thick] (PAR2) -- (E);
		\draw[-, thick] (PAR2) -- (F);
	\end{tikzpicture}
	}
	\subfigure[Equivalent normalized]{\label{pic:parse-tree-norm2}
	\begin{tikzpicture}[>=stealth, xscale=0.5, yscale=0.4]
		\tikzstyle{nonleaf} = [draw=black,circle,inner sep=1pt]
		\tikzstyle{leaf} = [draw=black,inner sep=2pt]
		\node[nonleaf] (PAR1) at (2,4) {{\tiny $\parcomp$}};
		
		\node[leaf] (A)    at (2,2.5) {{\tiny $A$}};
		\node[nonleaf] (PRI1) at (0,2.5) {{\tiny $\pricomp$}};
		\node[leaf] (B)    at (-1,1.5) {{\tiny $B$}};
		\node[leaf] (C)    at (1,1.5) {{\tiny $C$}};

		\node[nonleaf] (PRI2)    at (4,3) {{\tiny $\pricomp$}};
		\node[leaf] (D)       at (4.0,2) {{\tiny $D$}};
		\node[nonleaf] (PAR2)    at (5.5, 2) {{\tiny $\parcomp$}};
		\node[leaf] (E)       at (7,1) {{\tiny $E$}};
		\node[leaf] (F)       at (4,1) {{\tiny $F$}};

		\draw[-, thick] (PAR1) -- (A);
		\draw[-, thick] (PAR1) -- (PRI1);
		
		\draw[-, thick] (PRI1) -- (B);
		\draw[-, thick] (PRI1) -- (C);
		\draw[-, thick] (PAR1) -- (PRI2);
		\draw[-, thick] (PRI2) -- (D);
		\draw[-, thick] (PRI2) -- (PAR2);
		\draw[-, thick] (PAR2) -- (E);
		\draw[-, thick] (PAR2) -- (F);
	\end{tikzpicture}
	}
	\vspace{-3mm}
	\caption{Syntax trees of $\succ$}
	\label{pic:parse-tree}
	\vspace{-2mm}
  \end{figure}
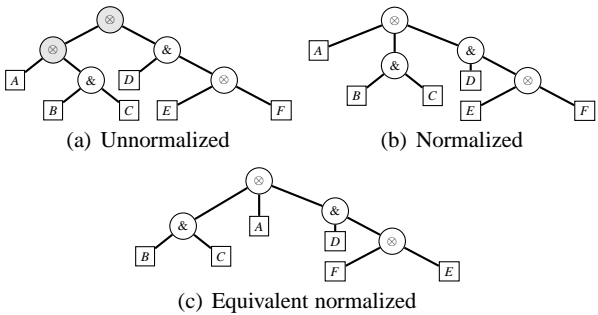

\begin{example}\label{ex:parse-tree}
 Let a p-skyline relation $\succ$ \footnote{Strictly speaking, we should
use attribute preference relations from $\prefset$, instead of
atomic preference relations. However, due to the close correspondence
of the two kinds of relations, we abuse the notation a bit.} be defined as
\[\succ \ = \ (\atomic_{A} \parcomp (\atomic_{B} \pricomp \atomic_{C}))
 \parcomp ( \atomic_{D} \pricomp ( \atomic_{E} \parcomp \atomic_{F}))\] 
An unnormalized syntax tree of $\succ$ is shown in Figure
 \ref{pic:parse-tree-unnorm}. Two normalized syntax trees of
 $\succ$ are shown in Figures \ref{pic:parse-tree-norm1} and
 \ref{pic:parse-tree-norm2}.
\end{example}

Every node of a syntax tree is itself a root of another
syntax tree. Let us associate with every node $C$ of a
syntax tree the set $\formvars(C)$ of attributes which are
descendants of $C$ in the syntax tree or $C$ itself (if it is a leaf). Essentially, $\formvars(C)$
corresponds to $\formvars(\pi_C)$ where $\pi_C$ is the p-expression
represented by the subtree with the root node $C$.

\newcommand{\wha}[1][]{{(\WSET^{#1}, \attrset^{#1})}}
\newcommand{\whax}[1][]{{$(\WSET^{#1},$ $\attrset^{#1})$}}
\newcommand{\wham}[1][]{{(\WSET^{#1}, \attrset)}}
\newcommand{\whamx}[1][]{{$(\WSET^{#1},$ $\attrset)$}}

\subsection{Attribute importance in p-skyline relations}\label{sec:up-struct}

Recall that the p-skyline relations  composed using
$\pricompsymbol$ (resp. $\parcompsymbol$) have different (resp. equal)
importance in the resulting relation. However, the composed p-skyline
relations do not have to be
\emph{atomic}  and may themselves be composed using
$\pricompsymbol$ or $\parcompsymbol$. The problem we discuss in this
section is
\emph{how to represent relative importance of attributes in different subtrees.}
For this purpose, we define another graphical representation of a
p-skyline relation -- the \emph{p-graph}.

\begin{definition}\label{def:p-graph}
{(\bf p-graph)} 
The \emph{p-graph} $\Gamma_\succ$ of a p-skyline relation $\succ$ has the set of nodes
$\nodes(\Gamma_\succ) = \formvars(\succ)$ and the set of edges
$\edgesof{\Gamma_\succ}$:
\begin{itemize}
        \item $\edgesof{\Gamma_\succ}=\emptyset$, if $\succ$ is an atomic preference
        relation;
        \item $\edgesof{\Gamma_\succ}=\edgesof{\Gamma_{\succ_1}} \ \cup \ \edgesof{\Gamma_{\succ_2}}$,
        if $\succ\ =\ \succ_1 \parcomp \succ_2$;
        \item $\edgesof{\Gamma_\succ}=\edgesof{\Gamma_{\succ_1}} \ \cup \ \edgesof{\Gamma_{\succ_2}} 
\ \cup\  (\formvars(\succ_1) \times \formvars(\succ_2))$, 
if $\succ\ =\ \succ_1 \pricomp \succ_2$,
\end{itemize}
for two p-skyline relations $\succ_1$ and $\succ_2$.
\end{definition}

A p-graph represents the attribute importance relationships implicit
in a p-skyline relation $\succ$ in the following way: an edge in
$\edgesof{\Gamma_{\succ}}$ goes from a
\emph{more important} attribute to a \emph{less important} attribute. This
follows from Definition \ref{def:p-graph}: if $\succ\ =\
\succ_1
\parcomp
\succ_2$ (i.e., $\succ_1$ and $\succ_2$ are equally important in $\succ$),
then no new attribute importance relationships are added to
$\edgesof{\Gamma_{\succ}}$, and those which exist in
$\edgesof{\Gamma_{\succ_1}}$ and $\edgesof{\Gamma_{\succ_2}}$ are
preserved in $\edgesof{\Gamma_{\succ}}$. Similarly, if $\succ\ =\
\succ_1 \pricomp \succ_2$, then the attribute importance relationships
in $\edgesof{\Gamma_{\succ_1}}$ and $\edgesof{\Gamma_{\succ_2}}$ are
preserved in $\edgesof{\Gamma_{\succ}}$, but new importance
relationships are added: every attribute relevant to $\succ_1$ is more
important than every attribute relevant to $\succ_2$.

\begin{figure}
	\centering 	
	\subfigure[p-graph $\Gamma_{\succ_1}$]{
	\begin{tikzpicture}[yscale=0.5]
		\node[p-graph-node] (A) at (0, 1.2) {{\small $A$}};
		\node[p-graph-node] (B) at (1, 1.2) {{\small $B$}};
		\node[p-graph-node] (C) at (0.5, 0) {{\small $C$}};
	
		\draw[->] (A)--(C);
		\draw[->] (B)--(C);

		\node at (-1, 1) {};
		\node at (2, 1) {};
	\end{tikzpicture}
		\label{pic:p-graph-1}
	}
%	\hspace{1cm}
	\subfigure[p-graph $\Gamma_{\succ_2}$]{
	\begin{tikzpicture}
		\tikzstyle{cir} = [draw=black,rounded corners,inner sep=2pt]
		\node[p-graph-node] (A) at (0, 0.5) {{\small $A$}};
		\node[p-graph-node] (B) at (1, 0.5) {{\small $B$}};
		\node[p-graph-node] (C) at (2, 0.5) {{\small $C$}};

		\node at (-1, 1) {};
		\node at (3, 0) {};	
	\end{tikzpicture}
		\label{pic:p-graph-2}
	}
	\caption{P-graphs from Example \ref{ex:p-graphs}}
	\label{pic:p-graph}
\end{figure}
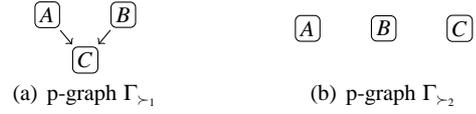

\begin{example}\label{ex:p-graphs}
Take the p-skyline relations $\succ_1$ and $\succ_2$ as below. 
Their p-graphs are shown in Figure \ref{pic:p-graph}.
\begin{align*}
\succ_1 \ \equiv &\ (\atomic_A \parcomp \atomic_B) \pricomp \atomic_C\\
\succ_2 \ \equiv &\ \atomic_A \parcomp \atomic_B \parcomp \atomic_C
\end{align*}
\end{example}

In the previous section, we showed that the skyline relation
$\skyline_\prefset$ is constructed as the Pareto accumulation of all the
members of $\prefset$. Hence, the following holds.

\begin{proposition}\label{cor:skyline-p-graph}
 The p-graph $\Gamma_{\skyline_\prefset}$ of the skyline relation 
 $\skyline_\prefset$ has the set of  nodes
 $\nodes(\Gamma_{\skyline_\prefset}) = \attrset$ and the set of edges
 $\edgesof{\Gamma_{\skyline_\prefset}} = \emptyset$.
\end{proposition}

Theorem \ref{thm:up-skyline} shows that p-graphs indeed represent
attribute importance. According to the theorem, a p-skyline relation
can be decomposed into ``dimensions'' which are attribute preference
relations. This decomposition shows which attribute preferences (resp.
the corresponding attributes) are \emph{less important} than a given
attribute preference (resp. the corresponding attribute) in a preference
relation.

\begin{theorem}\label{thm:up-skyline}
  Every p-skyline relation ${\succ} \in \formset_\prefset$
  is equal to
   \[\succ \ \equiv\ TC \left(\bigcup_{A \in \attrset}
  \pskylinephi_{A}\right),\]
  where 
  \[\pskylinephi_{A} \equiv \{(o_1, o_2)\ |\ o_1.A \attr_A o_2.A \} 
  \cap \equivset{\attrset - (\children{\Gamma_\succ}(A) \cup \{A\})}.\]
\end{theorem}

The relation $\pskylinephi_{A}$ may be viewed as a ``projection''
of the p-skyline relation $\succ$ to a ``dimension'' which is a
preference relation over $A$. Comparing tuples on the attribute $A$
one needs to consider only the attributes $\attrset -
(\children{\Gamma_\succ}(A) \cup \{A\})$ The values of the remaining
attributes $\children{\Gamma_\succ}(A)$ do not matter: those
attributes are
\emph{less important} than $A$. The relation $\succ'$ above
can also be viewed as a relaxed \emph{ceteris paribus preference
relation}
\cite{boutilier03cpnets}, for which attribute preferences are
unconditioned on each other, and \emph{``everything else being equal''} is
replaced with \emph{``$\attrset -$  $(\children{\Gamma_\succ}(A)$ $\cup
\{A\})$ being equal''}.

\medskip

Now let us take a closer look at the properties of p-graphs.  Since
p-graphs represent attribute importance implicit in p-skyline
relations, there are some properties of importance relationships that
p-graphs are expected to have, for example \emph{SPO}. In particular:
\begin{itemize}
\item  no attribute should be more important than itself (irreflexivity),
and 
\item if an attribute $A$ is more important than an attribute $B$
which is more important than an attribute $C$, $A$ is expected to be
more important than $C$ too (transitivity).
\end{itemize}

 As Theorem
\ref{thm:graph-props} shows, a p-graph is indeed an SPO\footnote{The
SPO properties of p-graphs should not be confused with the SPO
properties of the p-skyline relations. In the former case, we are
talking about ordering \emph{attributes}; in the latter, about
ordering {\emph{tuples}.}}.

However, a graph needs to satisfy some additional properties in order
to be a p-graph of some p-skyline relation. In particular, there is a
requirement that the p-expression inducing the p-skyline relation
contain exactly one occurrence of each member of $\prefset$. This
requirement is captured by the \envelope property visualized in Figure
\ref{pic:envelope}: if a graph $\Gamma$ has the three bold edges, then
it must have at least one dashed edge.

\begin{theorem}\label{thm:graph-props}\thmname{\spoenvelope}\\
 A directed graph $\Gamma$ with the set of nodes $\attrset$ is a p-graph 
 of some p-skyline relation iff 
 \begin{enumerate}
  \item $\Gamma$ is an SPO, and
  \item $\Gamma$ satisfies the \envelope property:
	\begin{align*}
	\forall A, B, C, D & \in \attrset, 
        \mbox{all different}\\ \edgeof{A}{B}{\Gamma} & \wedge
        \ \edgeof{C}{D}{\Gamma} \wedge \edgeof{C}{B}{\Gamma}
        \Rightarrow \\
        & \edgeof{C}{A}{\Gamma} \vee \edgeof{A}{D}{\Gamma}
        \vee \edgeof{D}{B}{\Gamma}
	\end{align*}
\end{enumerate}
\end{theorem}
  \begin{figure}[ht]
%        \vspace{-5mm}
	\centering
	\begin{tikzpicture}[>=stealth, scale=0.8]
		\node[p-graph-node] (B) at (0,0) {{\tiny $B$}};
		\node[p-graph-node] (D) at (2,0) {{\tiny $D$}};
		\node[p-graph-node] (A) at (0,1) {{\tiny $A$}};
		\node[p-graph-node] (C) at (2,1) {{\tiny $C$}};

		\draw[->, thick] (A) -- (B);
		\draw[->, thick] (C) -- (D);
		\draw[->, thick] (C) -- (B);
		
		\draw[->, dashed] (C) -- (A);
		\draw[->, dashed] (A) -- (D);
		\draw[->, dashed] (D) -- (B);
	\end{tikzpicture}
	\caption{The \envelope property}
	\label{pic:envelope}
  \end{figure}

%We showed above that a p-graph represents the attribute importance
%induced by a p-skyline relation.  Hence, the SPO
%properties of a p-graph are quite intuitive -- they capture the
%rationality of the importance relationship. The
%\envelope property of a p-graph is due to the fact that each attribute
%preference relation can have only one occurrence in a p-skyline
%p-expression. According to that property, if a graph $\Gamma$
%has the three edges shown bold in Figure \ref{pic:envelope}, then it 
%must have at least one dashed edge. 

We note that so far we have introduced two graph notations for
p-skyline relations: syntax trees and p-graphs. Although
these notations represent different concepts, there is a
correspondence between them shown in the next proposition.

\begin{proposition}\thmname{Syntax tree and p-graph correspondence}
\label{prop:common-anc}
 Let $A$ and $B$ be leaf nodes in a normalized syntax tree $T_{\succ}$
 of a p-skyline relation $\succ \ \in \ \formset_\prefset$. Then
 $\edgeof{A}{B}{\Gamma_{\succ}}$ iff the least common ancestor $C$ of
 $A$ and $B$ in $T_\succ$ is labeled by $\pricomp$, and $A$ precedes
 $B$ in the left-to-right tree traversal.
\end{proposition}

\subsection{Properties of p-skyline relations}

In this section, we show several fundamental properties of p-skyline
relations. These properties are used later to efficiently perform
essential operations on p-skyline relations: checking equivalence and
containment of relations and (tuple) dominance testing.  Before going
further, we note that p-skyline relations are representable as
formulas constructed from the corresponding p-expressions. So one can
use such formulas to perform the operations mentioned above. For
example, relation containment corresponds to formula implication.
However, we show below more direct ways of performing the operations
on p-skyline relations.  The results presented in this section are
used in sections
\ref{sec:p-skyline-min-ext} and \ref{sec:p-skyline-elicitation}.

%\medskip

%We note that so far we have introduced two graph notations for
%p-skyline relations: syntax trees and p-graphs. Although
%these notations represent different concepts, there is a
%correspondence between them shown in the next proposition.

%\begin{proposition}\thmname{Syntax tree and p-graph correspondence}
%\label{prop:common-anc}
% Let $A$ and $B$ be leaf nodes in a syntax tree $T_{\succ}$ of $\succ
% \ \in \ \formset_\prefset$. Then $\edgeof{A}{B}{\Gamma_{\Gamma}}$ if
% and only if the least common ancestor $C$ of $A$ and $B$ in $T_\succ$
% is of type $\pricomp$, and $A$ precedes $B$ in left-to-right tree
% traversal.
%\end{proposition}

Recall Example \ref{ex:parse-tree}, where we showed that a 
p-skyline relation may have more than one
syntax tree (and hence p-expression) defining it. 
In contrast, as shown in the next theorem, the p-graph corresponding to a 
p-skyline relation is \emph{unique}.

\begin{theorem}\label{thm:p-graph-uniqueness}\thmname{p-graph uniqueness}
 Two p-skyline relations \mbox{$\succ_1,$} $\succ_2 \in \formset_\prefset$
 are equal iff their p-graphs are identical.
\end{theorem}

According to Theorem \ref{thm:p-graph-uniqueness}, to check
equality of p-skyline relations, one only needs to compare their
p-graphs.  As the next theorem shows, containment of p-skyline
relations may be also checked using p-graphs.

\begin{theorem}\thmname{p-skyline relation containment}\label{thm:cnf-subset}
  For p-skyline relations $\succ_1, \succ_2 \ \in
  \formset_\prefset$,\ 
  $\succ_1 \ \subset\ \succ_2 \ \ \Leftrightarrow
  \ \ \edgesof{\Gamma_{\succ_1}} \subset \edgesof{\Gamma_{\succ_2}}.$
\end{theorem}

Theorem \ref{thm:cnf-subset} implies an important result. Recall that
in Corollary \ref{cor:skyline-p-graph} we showed that the edge set of
the p-graph $\Gamma_{\skyline_\prefset}$ of the skyline preference
relation $\skyline_\prefset$ is empty. Hence, the following facts are
implied by Theorem \ref{thm:cnf-subset}.

\begin{corollary}\label{cor:skyline-superset}
 For every relation instance $r$ and p-skyline relations
$\succ_1, \ \succ_2 \ \in \ \formset_\prefset$, s.t. $\Gamma_{\succ_2}
 \ \subset \ \Gamma_{\succ_1}$, we have $\winnow_{\succ_1}(r) \subseteq
 \winnow_{\succ_2}(r) \subseteq \winnow_{\skyline_\prefset}(r)$
\end{corollary}

The importance of Corollary \ref{cor:skyline-superset} is that  for
every p-skyline relation, the winnow query result will always be
contained in the corresponding skyline. In real life, that means that
if user preferences are modeled as a p-skyline relation instead of
a skyline relation, the size of the query result will not be larger than
the size of the skyline, and may be smaller.

\begin{figure}
	\centering
	\subfigure[$\Gamma_{\skyline_\prefset}$]{
		\begin{tikzpicture}[yscale=0.75, xscale=0.6]
			\node[p-graph-node] (A1) at (0,3) {$A_1$};
			\node[p-graph-node] (A2) at (1,3) {$A_2$};
			\node[p-graph-node] (A3) at (2,3) {$A_3$};

			\node at (0, 1) {};		
		\end{tikzpicture}
		\label{pic:p-skyline-containment-skyline}
	}
        \hspace{1mm}
	\subfigure[$\Gamma_{\succ_1}$]{
		\begin{tikzpicture}[yscale=0.75, xscale=0.75]
			\node[p-graph-node] (A1) at (0,3) {$A_1$};
			\node[p-graph-node] (A2) at (1,3) {$A_2$};
			\node[p-graph-node] (A3) at (0,2) {$A_3$};		

			\draw[p-graph-edge] (A1) to (A3);

			\node at (0, 1) {};		
		\end{tikzpicture}
		\label{pic:p-skyline-containment-g1}
	}
        \hspace{1mm}
	\subfigure[$\Gamma_{\succ_2}$]{
		\begin{tikzpicture}[yscale=0.75, xscale=0.75]
			\node[p-graph-node] (A2) at (0,3) {$A_2$};
			\node[p-graph-node] (A3) at (1,3) {$A_3$};
			\node[p-graph-node] (A1) at (0,2) {$A_1$};

			\draw[p-graph-edge] (A2) to (A1);

			\node at (0, 1) {};
		\end{tikzpicture}
		\label{pic:p-skyline-containment-g2}
	}
        \hspace{1mm}
	\subfigure[$r$]{
		\begin{tikzpicture}
			\node at (0,0) {
\begin{tabular}{l|c|c|c}
        \hspace{-2mm}&\hspace{-2mm} $A_1$\hspace{-2mm} &\hspace{-2mm} $A_2$ \hspace{-2mm}&\hspace{-2mm} $A_3$\\
\hline
$t_1$ \hspace{-2mm}&\hspace{-2mm} $2$\hspace{-2mm} &\hspace{-2mm} $1$ \hspace{-2mm} &\hspace{-2mm}  $0$ \\
$t_2$ \hspace{-2mm}&\hspace{-2mm} $1$\hspace{-2mm} &\hspace{-2mm} $2$ \hspace{-2mm} &\hspace{-2mm}  $0$ \\
$t_3$ \hspace{-2mm}&\hspace{-2mm} $1$\hspace{-2mm} &\hspace{-2mm} $0$ \hspace{-2mm} &\hspace{-2mm}  $2$ \\
$t_4$ \hspace{-2mm}&\hspace{-2mm} $1$\hspace{-2mm} &\hspace{-2mm} $0$ \hspace{-2mm} &\hspace{-2mm}  $0$
\end{tabular}
};
		\end{tikzpicture}
		\label{pic:p-skyline-containment-r}
	}
 \label{pic:p-skyline-containment}
 \caption{Containment of p-skyline relations}
\end{figure}
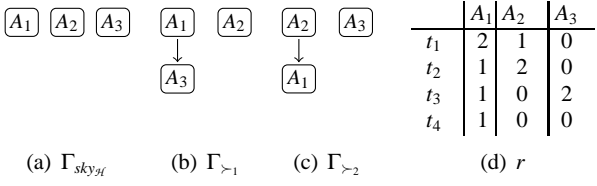

\begin{example}\label{ex:dominance-testing-p-sky}
 Let $\attrset = \{A_1, A_2, A_3\}$, and for every attribute, 
 larger values are preferred.  Consider the relations
\[\begin{array}{l} 
  \skyline_\prefset \ =\  \succ_{A_1} \parcomp \succ_{A_2} \parcomp \succ_{A_3}\\
  \succ_1           \ =\  (\succ_{A_1} \pricomp \succ_{A_3}) \parcomp \succ_{A_2}\\
  \succ_2           \ =\  (\succ_{A_2} \pricomp \succ_{A_1}) \parcomp \succ_{A_3}\\
\end{array}\]
whose p-graphs are shown in Figures
\ref{pic:p-skyline-containment-skyline},
\ref{pic:p-skyline-containment-g1}, and
\ref{pic:p-skyline-containment-g2}, respectively. Theorems
\ref{thm:cnf-subset} and \ref{thm:p-graph-uniqueness} imply that
$\skyline_\prefset \ \subset \ \succ_1$, $\skyline_\prefset \ \subset
\ \succ_2$, $\succ_1 \ \not \subseteq \ \succ_2$, and $\succ_2 \ \not
\subseteq \ \succ_1$. Take the relation instance $r$ shown in Figure
\ref{pic:p-skyline-containment-r}. Then $\winnow_{\skyline_\prefset}(r) =
\{t_1, t_2, t_3\}$, $\winnow_{\succ_1}(r) = \{t_1, t_2\}$, and
$\winnow_{\succ_2}(r) = \{t_2, t_3\}$.
\end{example}

In Theorem \ref{thm:pref-equiv}, we show how one can directly test
tuple dominance.  The dominance is expressed in terms of
\emph{containment constraints} on attribute sets.  This formulation is
essential for our approach to preference elicitation (section
\ref{sec:p-skyline-elicitation}).

Given two tuples $o, o' \in \univ$, a p-skyline relation $\succ$ and its p-graph $\Gamma_\succ$,
let 
\begin{itemize}
  \item $Diff(o,o')$ be the attributes in which $o$ differs from $o'$: 
\[Diff(o,o') = \{A\in\attrset\; |\; o_1.A \neq o_2.A\},\]
  \item $Top_\succ(o,o')$ be the topmost members of $Diff(o,$ $o')$:
\begin{align*}
Top_\succ(o,o') = \{A\; |\; & A \in Diff(o,o') \wedge \\
                            & \neg \exists B \in Diff(o,o').\ B\in\parents{\Gamma_\succ}(A)\},
\end{align*}
  \item $\betterin(o,o')$ be the attributes
in which $o$ is better than $o'$: 
\[\betterin(o_1, o_2) = \{A\in\attrset\; |\; o_1.A \attr_{A} o_2.A\}.\]
\end{itemize}

\begin{theorem}\label{thm:pref-equiv}\thmname{p-skyline dominance testing}
  Let $o, o' \in \univ$ s.t.   $o \neq o'$ and $\succ\  \in\  \formset_\prefset$.  Then the following conditions are equivalent:
 \begin{enumerate}
  \item $o \succ o'$;
  \item $\betterin(o, o') \supseteq Top_\succ(o, o')$;
  \item $\children{\Gamma_\succ}(\betterin(o, o')) \supseteq \betterin(o', o)$.
 \end{enumerate}
\end{theorem}

%The intuition behind Theorem \ref{thm:pref-equiv} is as follows. 
%The second condition 
%testing dominance implies that $o$ is preferred to $o'$ iff
%$o$ is preferred to $o'$ according to all the most important
%attributes in which these tuples are different. Method 3 of testing
%dominance says that $o$ is preferred to $o'$ iff for every
%attribute in which $o'$ is better than $o$, there is a more important
%attribute in which $o$ is better than $o'$.
%Notice that to check tuple dominance by Theorem \ref{thm:pref-equiv}, 
%one needs to check containment of certain attribute sets: no preference
%formulas are involved.

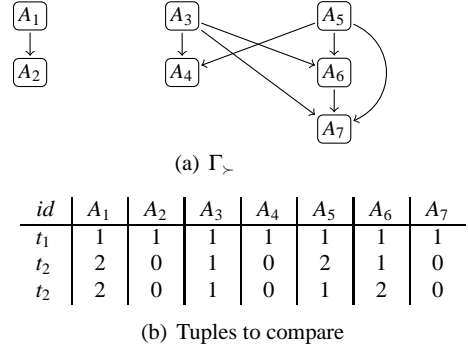
\begin{figure}
	\subfigure[$\Gamma_{\succ}$]{
		\begin{tikzpicture}[yscale=0.75]
			\node[p-graph-node] (A1) at (-1,3) {$A_1$};
			\node[p-graph-node] (A2) at (-1,2) {$A_2$};
			\node[p-graph-node] (A3) at (1,3) {$A_3$};
			\node[p-graph-node] (A4) at (1,2) {$A_4$};
			\node[p-graph-node] (A5) at (3,3) {$A_5$};
			\node[p-graph-node] (A6) at (3,2) {$A_6$};
			\node[p-graph-node] (A7) at (3,1) {$A_7$};
		
			\draw[p-graph-edge] (A1) to (A2);
			\draw[p-graph-edge] (A3) to (A4);
			\draw[p-graph-edge] (A3) to (A6);
			\draw[p-graph-edge] (A3) to (A7);
			
			\draw[p-graph-edge] (A5) to (A4);
			\draw[p-graph-edge] (A5) to (A6);
			\draw[p-graph-edge-left] (A5) to (A7);
		
			\draw[p-graph-edge] (A6) to (A7);
		\end{tikzpicture}
		\label{pic:dominance-testing-p-sky-graph}
	}
	\hspace{1cm}
	\subfigure[Tuples to compare]{
		\begin{tikzpicture}
		\node at (0,0){
			\begin{tabular}{l|c|c|c|c|c|c|c}
			$id$ & $A_1$ & $A_2$ & $A_3$ & $A_4$ & $A_5$ & $A_6$ & $A_7$\\
			\hline
			$t_1$ & 1 & 1 & 1 & 1 & 1 & 1 & 1 \\
			$t_2$ & 2 & 0 & 1 & 0 & 2 & 1 & 0 \\
			$t_2$ & 2 & 0 & 1 & 0 & 1 & 2 & 0 \\
			\end{tabular}
		};
		
		\end{tikzpicture}
		\label{pic:dominance-testing-p-sky-tuples}
	}

 \label{pic:dominance-testing-p-sky}
 \caption{Theorem \ref{thm:pref-equiv} for dominance testing}
\end{figure}

\begin{example}\label{ex:dominance-testing-p-sky-2}
 Let $\attrset = \{A_1, \ldots, A_7\}$, and for every attribute, 
 larger values are preferred.  Let a p-skyline relation $\succ$ be
 represented by the p-graph shown in Figure
 \ref{pic:dominance-testing-p-sky-graph}. Consider the tuples $t_1$,
 $t_2$, $t_3$ shown in Figure
 \ref{pic:dominance-testing-p-sky-tuples}. $\betterin(t_1,$ $t_2) =
 \{A_2, A_4, A_7\}$, $\betterin(t_2,$ $t_1) = \{A_1, A_5\}$,
 $Diff(t_1,t_2) = \{A_1,$ $A_2, A_4, A_5,$ $A_7\}$, and $Top_\succ(t_1,t_2) =
 \{A_1, A_5\}$.  Thus, $t_2 \succ t_1$, $t_1 \not \succ
 t_2$, $\betterin(t_1,$ $t_3) = \{A_2, A_4, A_7\}$, $\betterin(t_3,$ $t_1) 
 = \{A_1, A_6\}$, $Diff(t_1,t_3) = \{A_1,$ $A_2, $ $A_4,$ $A_6,$ $A_7\}$,
 and $Top_\succ(t_1, t_3)$ $= \{A_1, $ $A_4,$ $A_6\}$. So $t_3 \not \succ t_1$
 and $t_1 \not \succ t_3$.
\end{example}

In Theorem \ref{thm:graph-props}, we showed that p-graphs satisfy
\spoenvelopex, where the property \envelope was formulated in terms of
single p-graph nodes. However, it is often necessary to deal with
\emph{sets} of nodes. The next theorem generalizes the
\envelope property to disjoint  sets of nodes.

\begin{theorem}\label{thm:general-envelope}\thmname{\genenvelope}
  Let $\succ$ be a p-skyline relation with the p-graph
  $\Gamma_\succ$, and $\set{A}, \set{B}, \set{C}, \set{D}$,
  disjoint node sets of $\Gamma_\succ$.  Let the subgraphs of
  $\Gamma_\succ$ induced by those node sets be singletons or unions of
  at least two disjoint subgraphs. Then
	\begin{align*}
	 \edgeof{\set{A}}{\set{B}}{\Gamma_{\succ}} \ \wedge &
         \edgeof{\set{C}}{\set{D}}{\Gamma_{\succ}} \wedge
         \edgeof{\set{C}}{\set{B}}{\Gamma_{\succ}} \Rightarrow \\ &
         \edgeof{\set{C}}{\set{A}}{\Gamma_{\succ}} \vee
         \edgeof{\set{A}}{\set{D}}{\Gamma_{\succ}} \vee
         \edgeof{\set{D}}{\set{B}}{\Gamma_{\succ}}
	\end{align*}
\end{theorem}

Unlike \envelope which holds for every combination of four different
nodes, the property of \genenvelope holds for 
node subsets of a special form. That form  is quite general. For
instance, $\formvars(\succ)$ induces disjoint subgraphs if $\succ$ is
defined as Pareto accumulation of p-skyline relations. Theorem
\ref{thm:general-envelope} is used in the following section.

\begin{figure}
	\begin{center}
		\begin{tikzpicture}[yscale=0.5]
			\node[p-graph-node] (A1) at (-0.25,3) {$A_1$};
 			\node[p-graph-node] (A7) at (4.5,2.5) {$A_7$};
			\node[p-graph-node] (A2) at (1.25,3) {$A_2$};
			\node[p-graph-node] (A3) at (0.5,1) {$A_4$};
			\node[p-graph-node] (A4) at (3,3) {$A_3$};
			\node[p-graph-node] (A5) at (2.25,1) {$A_5$};
			\node[p-graph-node] (A6) at (3.755,1) {$A_6$};
		
			\draw[p-graph-edge] (A1) to (A3);
			\draw[p-graph-edge] (A1) to (A5);
			\draw[p-graph-edge] (A1) to (A6);

			\draw[p-graph-edge] (A2) to (A3);
			\draw[p-graph-edge] (A2) to (A5);
			\draw[p-graph-edge] (A2) to (A6);

			\draw[p-graph-edge] (A4) to (A5);
			\draw[p-graph-edge] (A4) to (A6);
			\draw[p-graph-edge] (A4) to (A3);

		\end{tikzpicture}
	\end{center}
 \caption{The \genenvelope property}
 \label{pic:general-envelope}
\end{figure}

\begin{example}\label{ex:general-envelope}
 Let $\attrset = \{A_1, \ldots, A_7\}$. Consider the p-graph
 $\Gamma_{\succ}$ (Figure \ref{pic:general-envelope}) of
  \[\succ = ((\atomic_{A_1} \parcomp \atomic_{A_2} \parcomp
 \atomic_{A_3}) \pricomp (\atomic_{A_4} \parcomp \atomic_{A_5}
 \parcomp \atomic_{A_6})) \parcomp \atomic_{A_7} \] Let $\set{A} =
 \{A_1\}$, $\set{B} = \{A_4\}$, $\set{C} = \{A_2, A_3\}$, $\set{D} =
 \{A_5, A_6\}$. Then the p-graph satisfies  \genenvelope because
 \[\edgeof{\set{A}}{\set{B}}{\Gamma_{\succ}} \wedge
 \edgeof{\set{C}}{\set{D}}{\Gamma_{\succ}} \wedge
 \edgeof{\set{C}}{\set{B}}{\Gamma_{\succ}} \wedge
 \edgeof{\set{A}}{\set{D}}{\Gamma_{\succ}} \]
\end{example}

\subsection{Minimal extensions}\label{sec:p-skyline-min-ext}

We conclude this section by studying the notion of \emph{minimal extension}
of a p-skyline relation. This notion is central for our approach
to preference elicitation (section \ref{sec:p-skyline-elicitation}). Intuitively,
we will construct a p-skyline relation that incorporates user feedback
using an iterative process that starts from the skyline relation and extends
it repeatedly in a minimal way.

\begin{definition}\label{def:min-ext}
{\bf (p-extension)} 
  For a p-skyline relation $\succ \ \in \formset_\prefset$, a
  p-skyline relation $\succ_{ext}\ \in \formset_\prefset$ is a
  \emph{p-extension of $\succ$}
  if $\succ \ \subset
  \ \succ_{ext}$.  The p-extension $\succ_{ext}$ is \emph{minimal} if
  there exists no $\succ'\ \in \formset_\prefset$
  such that $\succ \ \subset \ \succ' \ \subset \ \succ_{ext}$.
\end{definition}

Theorem \ref{thm:cnf-subset} implies that for every p-skyline relation
$\succ$, a p-extension $\succ_{ext}$ of $\succ$, if it exists, may be obtained by
constructing an extension $\Gamma_{\succ_{ext}}$ of the p-graph
$\Gamma_{\succ}$.  Hence, the problem of constructing a minimal
p-extension of a p-skyline relation can be reduced to the problem of
finding a minimal set of edges that when added to $\Gamma_\succ$ form
a graph satisfying
\spoenvelope. However, it is not clear how to find such a minimal set
of edges efficiently: adding a single edge to a graph may not be
enough due to violation of \spoenvelope, as shown in the following
example.

\begin{example}
Take the relation $\succ$ from Example \ref{ex:general-envelope}
(Figure \ref{pic:general-envelope}), and add the edge  $(A_6,A_7)$
to its p-graph. Then to preserve \spo, we need to add
the edges $(A_1,$ $A_7),$ $(A_2,$ $A_7)$, and $(A_3,$ $A_7)$. The resulting graph satisfies \spoenvelope.
However, if instead of the 
edge $(A_6,$ $A_7)$, we add the edge $(A_3,$ $A_7)$, 
then for preserving \envelope, it is enough to add
$(A_1,$ $A_7)$ and $(A_2,$ $A_7)$ (other extension possibilities exist too). 
The resulting graph satisfies \spoenvelope.
\end{example}

The method of constructing all minimal p-extensions we propose in this paper
operates directly on normalized p-ex\-pressions represented as syntax
trees.  In particular, we show a set of transformation rules of syntax
trees such that every unique application of a rule from this set
results in a unique minimal p-extension of the original p-skyline
relation.  If \emph{all} minimal p-extensions of a p-skyline relation
are needed, then one needs to apply to the syntax tree \emph{every}
rule in every possible way.

\begin{figure}[ht]
\begin{center}
	\subfigure[$Rule_1(T_\succ, C_i, C_{i+1})$]{
	\begin{tikzpicture}[>=stealth, yscale=0.5, xscale=0.35]
		\tikzstyle{nonleaf} = [draw=black,circle,inner sep=0pt]
		\tikzstyle{unknown} = []
		\tikzstyle{leaf} = [draw=black,inner sep=1pt]

		\node at (0,4.5) {{\scriptsize Original tree part}};

		\node[nonleaf] (PAR1) at (0,3.5) {{\scriptsize $\parcomp$}};
		\node[unknown] (C1) at (-2.1,2.5) {{\scriptsize $C_1$}};
		\node at (-1,2.5) {{\tiny $\ldots$}};
		\node[nonleaf] (Ci) at (-0.5,1.5) {{\scriptsize $\pricomp$}};
		\node[unknown] (Cj) at (0.6,2.5) {{\scriptsize $C_{i+1}$}};
		\node at (1.4,2.5) {{\tiny $\ldots$}};
		\node[unknown] (Ck) at (2.4,2.5) {{\scriptsize $C_k$}};
		\draw[-] (PAR1) -- (C1);
		\draw[-] (PAR1) -- (Ci);
		\draw[-] (PAR1) -- (Cj);
		\draw[-] (PAR1) -- (Ck);

		\node[unknown] (N1) at (-1.2,0.5) {{\scriptsize $N_1$}};
		\node[unknown] at (-0.5,0.5) {{\scriptsize $\ldots$}};
		\node[unknown] (Nm) at (0.2,0.5) {{\scriptsize $N_m$}};

		\draw[-] (Ci) -- (N1);
		\draw[-] (Ci) -- (Nm);

		% node labels		
 		\node[unknown] (LR) at (-4,3.5) {{\scriptsize $R$}};
 		\draw[-, dashed] (LR) -- (PAR1);
 		\node[unknown] (LCi) at (-4,1.5) {{\scriptsize $C_i$}};
 		\draw[-, dashed] (LCi) -- (Ci);

		% resulting tree

		\node at (12,4.5) {{\scriptsize Transformed tree part}};

		\node[nonleaf] (PAR1) at (12,3.5) {{\scriptsize $\parcomp$}};
		\node[unknown] (C1) at (9.6,2.5) {{\scriptsize $C_1$}};
		\node at (10.4,2.5) {{\tiny $\ldots$}};
		\node[unknown] (Cim1) at (11.2,2.5) {{\scriptsize $C_{i-1}$}};
		\node[unknown] (Cip2) at (12.9,2.5) {{\scriptsize $C_{i+2}$}};
		\node at (13.7,2.5) {{\tiny $\ldots$}};
		\node[unknown] (Ck) at (14.7,2.5) {{\scriptsize $C_k$}};
		\node[nonleaf] (PRI1) at (12,1.5) {{\scriptsize $\pricomp$}};
		\node[unknown] (N1) at   (11.2,0.5) {{\scriptsize $N_1$}};
		\node[nonleaf] (R3) at (12.9,0.5) {{\scriptsize $\parcomp$}};
		\node[unknown] (Ci1) at (13.8,-0.5) {{\scriptsize $C_{i+1}$}};
		\node[nonleaf] (R4) at (12,-0.5) {{\scriptsize $\pricomp$}};
		\node[unknown] (Nm) at   (12.9,-1.5) {{\scriptsize $N_m$}};
		\node[unknown] at (12,-1.5) {{\scriptsize $\ldots$}};
		\node[unknown] (N2) at (11.1,-1.5) {{\scriptsize $N_2$}};

		\draw[-] (PAR1) -- (C1);
		\draw[-] (PAR1) -- (Cim1);
		\draw[-] (PAR1) -- (Cip2);
		\draw[-] (PAR1) -- (Ck);
		\draw[-] (PAR1) -- (PRI1);
		\draw[-] (PRI1) -- (N1);
		\draw[-] (PRI1) -- (R3);
		\draw[-] (R3) -- (Ci1);
		\draw[-] (R3) -- (R4);
		\draw[-] (R4) -- (N2);
		\draw[-] (R4) -- (Nm);

		% node labels		
 		\node[unknown] (LR') at (18,3.5) {{\scriptsize $R'$}};
 		\draw[-, dashed] (LR') -- (PAR1);
 		\node[unknown] (LCi') at (18,1.5) {{\scriptsize $C_i'$}};
 		\draw[-, dashed] (LCi') -- (PRI1);
 		\node[unknown] (LR1') at (18,0.5) {{\scriptsize $R_1'$}};
 		\draw[-, dashed] (LR1') -- (R3);
 		\node[unknown] (LR2') at (8,-0.5) {{\scriptsize $R_2'$}};
 		\draw[-, dashed] (LR2') -- (R4);

	\end{tikzpicture}	
	}
\end{center}

\begin{center}
	\subfigure[$Rule_2(T_\succ, C_i, C_{i+1})$]{
	\begin{tikzpicture}[>=stealth, yscale=0.5,xscale=0.35]
		\tikzstyle{nonleaf} = [draw=black,circle,inner sep=0pt]
		\tikzstyle{unknown} = []
		\tikzstyle{leaf} = [draw=black,inner sep=1pt]

		\node at (0,4.5) {{\scriptsize Original tree part}};

		\node[nonleaf] (PAR1) at (0,3.5) {{\scriptsize $\parcomp$}};
		\node[unknown] (C1) at (-2.1,2.5) {{\scriptsize $C_1$}};
		\node at (-1,2.5) {{\tiny $\ldots$}};
		\node[nonleaf] (Ci) at (-0.5,1.5) {{\scriptsize $\pricomp$}};
		\node[unknown] (Cj) at (0.6,2.5) {{\scriptsize $C_{i+1}$}};
		\node at (1.4,2.5) {{\tiny $\ldots$}};
		\node[unknown] (Ck) at (2.4,2.5) {{\scriptsize $C_k$}};
		\draw[-] (PAR1) -- (C1);
		\draw[-] (PAR1) -- (Ci);
		\draw[-] (PAR1) -- (Cj);
		\draw[-] (PAR1) -- (Ck);

		\node[unknown] (N1) at (-1.2,0.5) {{\scriptsize $N_1$}};
		\node[unknown] at (-0.5,0.5) {{\tiny $\ldots$}};
		\node[unknown] (Nm) at (0.2,0.5) {{\scriptsize $N_m$}};

		\draw[-] (Ci) -- (N1);
		\draw[-] (Ci) -- (Nm);
		
		% node labels		
 		\node[unknown] (LR) at (-4,3.5) {{\scriptsize $R$}};
 		\draw[-, dashed] (LR) -- (PAR1);
 		\node[unknown] (LCi) at (-4,1.5) {{\scriptsize $C_i$}};
 		\draw[-, dashed] (LCi) -- (Ci);

		% resulting tree

		\node at (12,4.5) {{\scriptsize Transformed tree part}};

		\node[nonleaf] (PAR1) at (12,3.5) {{\scriptsize $\parcomp$}};
		\node[unknown] (C1) at (9.6,2.5) {{\scriptsize $C_1$}};
		\node at (10.4,2.5) {{\tiny $\ldots$}};
		\node[unknown] (Cim1) at (11.2,2.5) {{\scriptsize $C_{i-1}$}};
		\node[unknown] (Cip2) at (12.9,2.5) {{\scriptsize $C_{i+2}$}};
		\node at (13.7,2.5) {{\tiny $\ldots$}};
		\node[unknown] (Ck) at (14.7,2.5) {{\scriptsize $C_k$}};
		\node[nonleaf] (PRI1) at (12,1.5) {{\scriptsize $\pricomp$}};
		\node[unknown] (Nm) at   (12.9,0.5) {{\scriptsize $N_m$}};
		\node[nonleaf] (R3) at (11.2,0.5) {{\scriptsize $\parcomp$}};
		\node[unknown] (Ci1) at (10.3,-0.5) {{\scriptsize $C_{i+1}$}};
		\node[nonleaf] (R4) at (12,-0.5) {{\scriptsize $\pricomp$}};
		\node[unknown] (Nm1) at   (12.9,-1.5) {{\scriptsize $N_{m-1}$}};
		\node[unknown] at (12,-1.5) {{\tiny $\ldots$}};
		\node[unknown] (N1) at (11.1,-1.5) {{\scriptsize $N_1$}};

		\draw[-] (PAR1) -- (C1);
		\draw[-] (PAR1) -- (Cim1);
		\draw[-] (PAR1) -- (Cip2);
		\draw[-] (PAR1) -- (Ck);
		\draw[-] (PAR1) -- (PRI1);
		\draw[-] (PRI1) -- (Nm);
		\draw[-] (PRI1) -- (R3);
		\draw[-] (R3) -- (Ci1);
		\draw[-] (R3) -- (R4);
		\draw[-] (R4) -- (N1);
		\draw[-] (R4) -- (Nm1);

		% node labels		
 		\node[unknown] (LR') at (18,3.5) {{\scriptsize $R'$}};
 		\draw[-, dashed] (LR') -- (PAR1);
 		\node[unknown] (LCi') at (18,1.5) {{\scriptsize $C_i'$}};
 		\draw[-, dashed] (LCi') -- (PRI1);
 		\node[unknown] (LR1') at (8,0.5) {{\scriptsize $R_1'$}};
 		\draw[-, dashed] (LR1') -- (R3);
 		\node[unknown] (LR2') at (18,-0.5) {{\scriptsize $R_2'$}};
 		\draw[-, dashed] (LR2') -- (R4);
	\end{tikzpicture}	
	}
\end{center}

\begin{center}
	\subfigure[$Rule_3(T_\succ, C_i, C_{i+1})$]{
	\begin{tikzpicture}[>=stealth, yscale=0.5, xscale=0.35]
		\tikzstyle{nonleaf} = [draw=black,circle,inner sep=0pt]
		\tikzstyle{unknown} = []
		\tikzstyle{leaf} = [draw=black,inner sep=2pt, rounded corners = 2pt]

		\node at (0,4.5) {{\scriptsize Original tree part}};

		\node[nonleaf] (PAR1) at (0,3.5) {{\scriptsize $\parcomp$}};
		\node[unknown] (N1) at (-2.7,2.5) {{\scriptsize $C_1$}};
		\node at (-1.7,2.5) {{\tiny $\ldots$}};
		\node[leaf] (Ni) at (-0.8,2.5) {{\scriptsize $C_{i}$}};
		\node[leaf] (Nj) at (0.9,2.5) {{\scriptsize $C_{i+1}$}};
		\node at (2,2.5) {{\tiny $\ldots$}};
		\node[unknown] (Nk) at (3,2.5) {{\scriptsize $C_k$}};
		\draw[-] (PAR1) -- (N1);
		\draw[-] (PAR1) -- (Ni);
		\draw[-] (PAR1) -- (Nj);
		\draw[-] (PAR1) -- (Nk);

		% node labels		
 		\node[unknown] (LR) at (-4,3.5) {{\scriptsize $R$}};
 		\draw[-, dashed] (LR) -- (PAR1);

		% resulting tree		

		\node at (12,4.5) {{\scriptsize Transformed tree part}};

		\node[nonleaf] (PAR1) at (12,3.5) {{\scriptsize $\parcomp$}};
		\node[unknown] (N1) at (9.6,2.5) {{\scriptsize $C_1$}};
		\node at (10.4,2.5) {{\tiny $\ldots$}};
		\node[unknown] (Nim1) at (11.2,2.5) {{\scriptsize $C_{i-1}$}};
		\node[unknown] (Nip2) at (12.9,2.5) {{\scriptsize $C_{i+2}$}};
		\node at (13.7,2.5) {{\tiny $\ldots$}};
		\node[unknown] (Nk) at (14.7,2.5) {{\scriptsize $C_k$}};
		\node[nonleaf] (PRI1) at (12,1.5) {{\scriptsize $\pricomp$}};
		\node[leaf] (Ni) at   (11.2,0.5) {{\scriptsize $C_i$}};
		\node[leaf] (Nip1) at (12.9,0.5) {{\scriptsize $C_{i+1}$}};

		\draw[-] (PAR1) -- (N1);
		\draw[-] (PAR1) -- (Nim1);
		\draw[-] (PAR1) -- (Nip2);
		\draw[-] (PAR1) -- (Nk);
		\draw[-] (PAR1) -- (PRI1);
		\draw[-] (PRI1) -- (Ni);
		\draw[-] (PRI1) -- (Nip1);

		% node labels		
 		\node[unknown] (LR') at (18,3.5) {{\scriptsize $R'$}};
 		\draw[-, dashed] (LR') -- (PAR1);
 		\node[unknown] (LR1') at (18,1.5) {{\scriptsize $R_1'$}};
 		\draw[-, dashed] (LR1') -- (PRI1);

	\end{tikzpicture}	
	}
\end{center}

\begin{center}
	\subfigure[$Rule_4(T_\succ, C_i, C_{i+1}, s, t)$]{
	\begin{tikzpicture}[>=stealth, yscale=0.5, xscale=0.35]
		\tikzstyle{nonleaf} = [draw=black,circle,inner sep=0pt]
		\tikzstyle{unknown} = []
		\tikzstyle{leaf} = [draw=black,inner sep=1pt]

		\node at (0,4.5) {{\scriptsize Original tree part}};

		\node[nonleaf] (PAR1) at (0,3.5) {{\scriptsize $\parcomp$}};
		\node[unknown] (C1) at (-2.4,2.5) {{\scriptsize $C_1$}};
		\node at (-1.3,2.5) {{\tiny $\ldots$}};
		\node at (1.3,2.5) {{\tiny $\ldots$}};
		\node[unknown] (Ck) at (2.4,2.5) {{\scriptsize $C_k$}};

		\node[nonleaf] (R1) at (-1.4,1.5) {{\scriptsize $\pricomp$}};
		\node[unknown] (N1) at (-2.3,0.5) {{\scriptsize $N_1$}};
		\node[unknown]      at (-1.4,0.5) {{\tiny $\ldots$}};
		\node[unknown] (Nm) at (-0.5,0.5) {{\scriptsize $N_m$}};

		\node[nonleaf] (R2) at (1.4,1.5) {{\scriptsize $\pricomp$}};
		\node[unknown] (M1) at (0.5,0.5) {{\scriptsize $M_1$}};
		\node[unknown]      at (1.4,0.5) {{\tiny $\ldots$}};
		\node[unknown] (Mn) at (2.3,0.5) {{\scriptsize $M_n$}};

		\draw[-] (R1) -- (N1);
		\draw[-] (R1) -- (Nm);
		\draw[-] (R2) -- (M1);
		\draw[-] (R2) -- (Mn);

		\draw[-] (PAR1) -- (C1);
		\draw[-] (PAR1) -- (R1);
		\draw[-] (PAR1) -- (R2);
		\draw[-] (PAR1) -- (Ck);

		% node labels		
 		\node[unknown] (LR) at (-4,3.5) {{\scriptsize $R$}};
 		\draw[-, dashed] (LR) -- (PAR1);
 		\node[unknown] (LCi) at (-4,1.5) {{\scriptsize $C_i$}};
 		\draw[-, dashed] (LCi) -- (R1);
 		\node[unknown] (LCi1) at (4,1.5) {{\scriptsize $C_{i+1}$}};
 		\draw[-, dashed] (LCi1) -- (R2);

		% resulting tree

		\node at (12,4.5) {{\scriptsize Transformed tree part}};

		\node[nonleaf] (PAR1) at (12,3.5) {{\scriptsize $\parcomp$}};
		\node[unknown] (C1) at (9.6,2.5) {{\scriptsize $C_1$}};
		\node[unknown] (Ci1) at (11.3,2.5) {{\scriptsize $C_{i-1}$}};
		\node at (10.4,2.5) {{\tiny $\ldots$}};
		\node[unknown] (Ci2) at (12.7,2.5) {{\scriptsize $C_{i+2}$}};
		\node at (13.6,2.5) {{\tiny $\ldots$}};
		\node[unknown] (Ck) at (14.4,2.5) {{\scriptsize $C_k$}};

		\node[nonleaf] (T1) at (12,1.5) {{\scriptsize $\pricomp$}};
		\node[nonleaf] (R2) at (15,0.5) {{\scriptsize $\parcomp$}};
		\node[nonleaf] (R21) at (13.5,-0.5) {{\scriptsize $\pricomp$}};
		\node[nonleaf] (R22) at (16.5,-0.5) {{\scriptsize $\pricomp$}};
		\node[unknown] (Ni1) at (12.6,-1.5) {{\scriptsize $N_{s+1}$}};
		\node[unknown]      at (13.5,-1.5) {{\tiny $\ldots$}};
 		\node[unknown] (Nm) at (14.4,-1.5) {{\scriptsize $N_m$}};
 		\node[unknown] (Mj1) at (15.6,-1.5) {{\scriptsize $M_{t+1}$}};
 		\node[unknown]      at (16.5,-1.5) {{\tiny $\ldots$}};
		\node[unknown] (Mn) at (17.4,-1.5) {{\scriptsize $M_n$}};
		\draw[-] (T1) -- (R2);
		\draw[-] (R2) -- (R21);
		\draw[-] (R2) -- (R22);
		\draw[-] (R21) -- (Ni1);
		\draw[-] (R21) -- (Nm);
		\draw[-] (R22) -- (Mj1);
		\draw[-] (R22) -- (Mn);

		\node[nonleaf] (R1) at (9,0.5) {{\scriptsize $\parcomp$}};
		\node[nonleaf] (R12) at (10.5,-0.5) {{\scriptsize $\pricomp$}};
		\node[nonleaf] (R11) at (7.5,-0.5) {{\scriptsize $\pricomp$}};
		\node[unknown] (M1) at (11.4,-1.5) {{\scriptsize $M_t$}};
		\node[unknown]      at (10.5,-1.5) {{\tiny $\ldots$}};
 		\node[unknown] (Mj) at (9.6,-1.5) {{\scriptsize $M_1$}};
 		\node[unknown] (N1) at (8.4,-1.5) {{\scriptsize $N_s$}};
 		\node[unknown]      at (7.5,-1.5) {{\tiny $\ldots$}};
		\node[unknown] (Ni) at (6.6,-1.5) {{\scriptsize $N_1$}};
		\draw[-] (T1) -- (R1);
		\draw[-] (R1) -- (R11);
		\draw[-] (R1) -- (R12);
		\draw[-] (R11) -- (N1);
		\draw[-] (R11) -- (Ni);
		\draw[-] (R12) -- (M1);
		\draw[-] (R12) -- (Mj);

		\draw[-] (PAR1) -- (C1);
		\draw[-] (PAR1) -- (Ci1);
		\draw[-] (PAR1) -- (T1);
		\draw[-] (PAR1) -- (Ci2);
		\draw[-] (PAR1) -- (Ck);

		% node labels		
 		\node[unknown] (LR') at (18,3.5) {{\scriptsize $R'$}};
 		\draw[-, dashed] (LR') -- (PAR1);
 		\node[unknown] (LCi') at (18,1.5) {{\scriptsize $C_i'$}};
 		\draw[-, dashed] (LCi') -- (T1);
 		\node[unknown] (LR2') at (18,0.5) {{\scriptsize $R_2'$}};
 		\draw[-, dashed] (LR2') -- (R2);
 		\node[unknown] (LR1') at (6,0.5) {{\scriptsize $R_1'$}};
 		\draw[-, dashed] (LR1') -- (R1);
 		\node[unknown] (LR3') at (6,-0.5) {{\scriptsize $R_3'$}};
 		\draw[-, dashed] (LR3') -- (R11);
 		\node[unknown] (LR4') at (11.5,0.5) {{\scriptsize $R_4'$}};
 		\draw[-, dashed] (LR4') -- (R12);
 		\node[unknown] (LR5') at (12.5,0.5) {{\scriptsize $R_5'$}};
 		\draw[-, dashed] (LR5') -- (R21);
 		\node[unknown] (LR6') at (18,-0.5) {{\scriptsize $R_6'$}};
 		\draw[-, dashed] (LR6') -- (R22);
	\end{tikzpicture}	
	}
\end{center}

	\begin{center}
	\begin{tikzpicture}{xscale=0.7}
		\tikzstyle{nonleaf} = [draw=black,circle,inner sep=0pt]
		\tikzstyle{unknown} = []
		\tikzstyle{leaf} = [draw=black,inner sep=2pt, rounded corners = 2pt]		
		\node[leaf] at (0, 0) {{\scriptsize $C_i$}};
		\node       at (1.3, 0) {{ - leaf node}};

		\node[unknown] at (0, -.5) {{\scriptsize $C_i$}};
		\node          at (2.0, -.5) {{ - leaf or non-leaf node}};

	\end{tikzpicture} 
	\end{center}
\vspace{-5mm}
\caption{Syntax tree transformation rules} 
\label{pic:trans-rules} 
\end{figure}
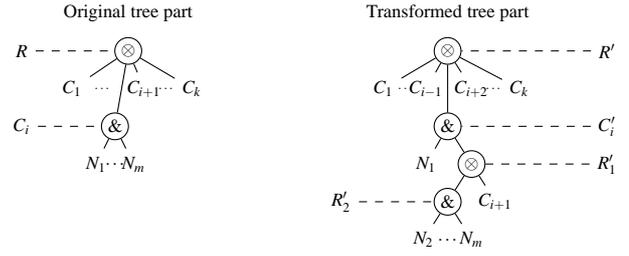
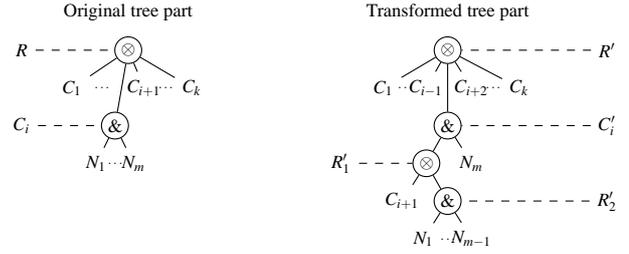
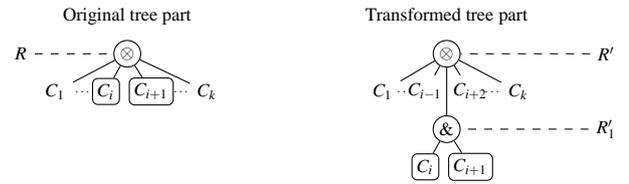
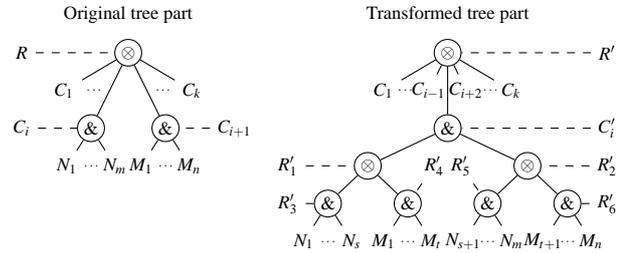

The transformation rules are shown in Figure \ref{pic:trans-rules}. On
the left hand side, we show a part of the syntax tree of an original
p-skyline relation. On the right hand side, we show how this
part is modified in the resulting relation. We assume
that the rest of the syntax tree is left unchanged. All the
transformation rules operate on two children $C_i$ and $C_{i+1}$ of a
$\parcomp$-node of the syntax tree. For simplicity, these nodes are
shown as consecutive children. However, in general $C_i$ and $C_{i+1}$
may be any pair of children nodes of the same $\parcomp$-node. Their
order is unimportant due to the associativity of $\parcomp$.

Let us denote the original relation as $\succ$ and the relation
obtained as the result of applying one of the transformation rules as
$\succ_{ext}$.  Observation \ref{obs:new-edges} shows
that all the rules only \emph{add} edges to
the p-graph of the original preference relation and hence extend the
p-skyline relation.

\begin{observation}\label{obs:new-edges}
If $T_{\succ_{ext}}$ is obtained from $T_{\succ}$ using some of 
$Rule_1,$ $\ldots,$ $Rule_4$, then $\edgesof{\Gamma_{\succ}} \subset
\edgesof{\Gamma_{\succ_{ext}}}$. Moreover,
\begin{itemize}
 \item if $T_{\succ_{ext}}$ is a result of $Rule_1(T_\succ, C_i, C_{i+1})$,
   then
\[\edgesof{\Gamma_{\succ_{ext}}} = \edgesof{\Gamma_{\succ}} \cup \{(X,Y)\ |\ X\in
   \formvars({N_1}), Y \in \formvars({C_{i+1}})\}\]
 \item if $T_{\succ_{ext}}$ is a result of $Rule_2(T_\succ, C_i, C_{i+1})$,
   then
\[\edgesof{\Gamma_{\succ_{ext}}} = \edgesof{\Gamma_{\succ}} \cup \{(X,Y)\ |\ X\in
   \formvars({C_{i+1}}), Y \in \formvars({N_m})\}\]
 \item if $T_{\succ_{ext}}$ is a result of $Rule_3(T_\succ, C_i, C_{i+1})$,
   then
\[\edgesof{\Gamma_{\succ_{ext}}} = \edgesof{\Gamma_{\succ}} \cup (C_i, C_{i+1})\]
 \item if $T_{\succ_{ext}}$ is a result of $Rule_4(T_{\succ}, C_i, C_{i+1},
   s, t)$ for $s \in [1, $ $n-1], t \in [1, m-1]$, then
$\edgesof{\Gamma_{\succ_{ext}}} = \edgesof{\Gamma_{\succ}}  \cup $
\begin{align*}
\{(X,Y)\ |\ X
\in \bigcup_{p \in 1\ldots s}\formvars({N_p}), Y \in \bigcup_{q \in
  t+1\ldots n} \formvars({M_q})\} \ \cup \\ \{(X,Y)\ |\ X \in
\bigcup_{p \in 1\ldots t}\formvars({M_p}), Y \in \bigcup_{q \in
  s+1\ldots m} \formvars({N_q})\} \\
\end{align*}
\end{itemize}
\end{observation}

\vspace{-1cm} We note that every $\pricomp$- and $\parcomp$-node in a
syntax tree has to have at least two children nodes.  This is because the
operators $\pricomp$ and $\parcomp$ must have at least two
arguments. However, as a result of a transformation rule application,
some $\pricomp$- and $\parcomp$-nodes may end up with only one child node.
These nodes are:

\begin{enumerate}
 \item $R'$ if $k = 2$ for $Rule_1, Rule_2, Rule_3, Rule_4$;
 \item $R_2'$ if $m = 2$ for $Rule_1, Rule_2$;
 \item $R_3'$ or $R_5'$ if $s = 1$ or $s = m-1$, respectively, for $Rule_4$;
 \item $R_4'$ or $R_6'$ if $t = 1$ or $t = n-1$, respectively, for $Rule_4$.
\end{enumerate}

\begin{figure}[ht]
%        \vspace{-5mm}
	\begin{center}
	\begin{tikzpicture}[>=stealth, xscale=0.6, yscale=0.5]
		\tikzstyle{nonleaf} = [draw=black,circle,inner sep=0pt]
		\tikzstyle{unknown} = []
		\tikzstyle{leaf} = [draw=black,inner sep=2pt, rounded corners = 2pt]

		\node at (0, 2) {{\scriptsize Before single-child }};
		\node at (0, 1.5) {{\scriptsize node elimination}};
		\node	       (R) at (0,1) {};
		\node[nonleaf] (A) at (0,0) {{\scriptsize $\ \delta\ $}};

		\node[unknown] (B) at (0,-1) {{\scriptsize $N$}};
		\draw (R) -- (A);
		\draw (A) -- (B);

		\node at (8, 2) {{\scriptsize After single-child }};
		\node at (8, 1.5) {{\scriptsize node elimination}};
		\node	       (R) at (8,1) {};
		\node[unknown] (B) at (8,-1) {{\scriptsize $N$}};

		\draw (R) -- (B);
	\end{tikzpicture}	
	\end{center}
	\vspace{-5mm}
	\caption{Single-child node elimination ($\delta \in \{ \pricomp, \parcomp \}$)}
	\label{pic:single-child-node-elim} 
        \vspace{-3mm}
\end{figure}
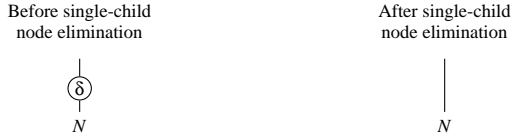

In such cases, we remove the nodes with a single child and connect the
child directly to the parent (Figure
\ref{pic:single-child-node-elim}).

\begin{theorem}\label{thm:min-ext-rules}\thmname{minimal p-extension}
 Let $\succ\ \in \ \formset_\prefset$, and $T_{\succ}$ be a normalized
 syntax tree of $\succ$.  Then $\succ_{ext}$ is a \emph{minimal 
   p-extension} of $\succ$ iff the syntax tree
 $T_{\succ_{ext}}$ of $\succ_{ext}$ is obtained from $T_{\succ}$ by a
 single application of a rule from $Rule_1, \ldots, $ $Rule_4$, followed
by a single-child node elimination if necessary.
\end{theorem}

Theorem \ref{thm:min-ext-rules} has two important corollaries
describing properties of minimal p-extensions.

\begin{corollary}\label{cor:polynomial-time-to-ext}
 For a p-skyline relation $\succ$ with a normalized syntax
 tree $T_{\succ}$, a syntax tree $T_{\succ_{ext}}$ of each of its minimal
 p-extensions $\succ_{ext}$ may be constructed in time
 $\mathcal{O}(|\attrset|)$.
\end{corollary}

In Corollary \ref{cor:polynomial-time-to-ext}, we assume the
adjacency-list representation of syntax trees. The total number of
nodes in a tree is linear in the number of its leaf nodes
\cite{cormen}, which is $|\attrset|$. Thus the number of edges in
$T_{\succ}$ is $\mathcal{O}(|\attrset|)$.  The transformation of
$T_{\succ}$ using every rule requires removing $\mathcal{O}(|\attrset|)$
and adding $\mathcal{O}(|\attrset|)$ edges.

\begin{corollary}\label{cor:number-of-min-ext}
 For a p-skyline relation $\succ$, the number of its minimal
 p-extensions is $\mathcal{O}(|\attrset|^4)$.
\end{corollary}

The justification for Corollary \ref{cor:number-of-min-ext} is as
follows. The set of minimal-extension rules is complete due to Theorem
\ref{thm:min-ext-rules}.  Every rule operates on two nodes $C_i$ and
$C_{i+1}$ of the syntax tree. Hence, the number of such node pairs is
$\mathcal{O}(|\attrset|^2)$. $Rule_4$ also relies on some partitioning
of the sequence of child nodes of $C_{i}$ and $C_{i+1}$. The total
number of such partitionings is $\mathcal{O}(|\attrset|^2)$. Thus, the
total number of different rule applications is
$\mathcal{O}(|\attrset|^4)$.
Consequently, the number of minimal
p-extensions is \emph{polynomial} in the number of attributes. This
differs from the number of \emph{all} p-extensions of a p-skyline
relation, which is $\Omega(|\attrset|!)$.

The last property related to p-extensions that we consider here
is as follows. By Theorem
\ref{thm:cnf-subset}, a p-extension of a p-skyline relation is
obtained by adding edges to its p-graph. However, the total number of
edges in a p-graph is at most $\mathcal{O}(|\attrset|^2)$. Hence, the
next Corollary holds.
\begin{corollary}\label{cor:min-ext-length}
  Let $S$ be a sequence of p-skyline relations 
\[\succ_1, \ldots, \succ_k \ \in\  \formset_\prefset\]
such that for every $i \in [1, k-1]$, $\succ_{i+1}$ is a 
p-extension of $\succ_i$. Then $|S| = \mathcal{O}(|\attrset|^2)$.
\end{corollary}

\newcommand{\favdisprob}{\texttt{DF-PSKY\-LINE}\xspace}
\newcommand{\funcfavdisprob}{\texttt{FDF-PSKY\-LINE}\xspace}
\newcommand{\funcfavdisprobx}{\texttt{FDF\-PSKY\-LINE}\xspace}
\newcommand{\funcoptfavdisprob}{\texttt{OPT-FDF-PSKY\-LINE}\xspace}
\newcommand{\funcoptfavdisprobx}{\texttt{OPT- FDF-PSKY\-LINE}\xspace}

\newcommand{\favprob}{{\tt DF$^+$-PSKY\-LINE}\xspace}
\newcommand{\funcfavprob}{{\tt FDF$^+$-PSKY\-LINE}\xspace}
\newcommand{\optfavprob}{{\tt OPT-FDF$^+$-PSKY\-LINE}\xspace}
\newcommand{\optfavprobx}{{\tt OPT- FDF$^+$-PSKY\-LINE}\xspace}

\vspace{-5mm}
\section{Elicitation of p-skyline relations}\label{sec:p-skyline-elicitation}

In Section \ref{sec:pskylines}, we proposed a class of preference
relations called \emph{p-skyline relations}.  In this section, we
introduce a method of constructing p-skyline relations based on
user-provided feedback.

\vspace{-5mm}
\subsection{Feedback-based elicitation}

As we showed in the previous section, the p-skyline framework is a
generalization of the skyline framework.  The main difference between
those frameworks is that in the p-skyline framework one can express
varying attribute importance. On the other hand, one of the main distinguishing properties of
the skyline framework is the simplicity of representing
preferences. Namely, the user needs to provide only a set of attribute
preferences to specify a preference relation.  For p-skylines, an
additional piece of information, the relative importance of the
attributes (in the form of, e.g., a p-graph or a p-expression), has to be also provided
by the user. But how can relative attribute importance be specified? It seems
impractical to ask the user to compare distinct attributes pairwise for
importance: even though some relationships can be deduced by
transitivity, the number of comparisons may still be too large.  
Another  issue is even more serious: the users themselves may be not
fully aware of their own preferences.

In this section, we propose an alternative approach to elicitation of
attribute importance relationships, based on \emph{user feedback}. We
use the following scenario. A fixed, finite set of tuples is stored in
a database relation $\oset \subseteq \U$.  All the tuples have the
same set of attributes $\attrset$.  We assume that, in addition to
$\attrset$, a corresponding set of attribute preference relations
$\prefset$ is given.  The user partitions $\oset$ into three disjoint
subsets: the set $G$ of tuples she confidently likes (\emph{superior}
examples), the set $W$ of tuples she confidently dislikes
(\emph{inferior} examples), and the set of remaining tuples about
which she is not sure. The output of our method is a p-skyline
relation $\succ$ (with the set of relevant attributes $\attrset$),
according to which all tuples in $G$ are superior and all tuples in
$W$ are inferior. A tuple $o\in\oset$ is \emph{superior} if $\oset$ does not
contain any tuples preferred to $o$, according to $\succ$.  A tuple
$o\in\oset$ is \emph{inferior} if there is at least one superior example in
$\oset$, which is preferred to $o$. The last assumption is justified
by a general principle that the user considers something bad because
she knows of a better alternative.

Formally: given $\attrset$, $\prefset$, $\oset$, $G$, and $W$, we want
to construct a p-expression inducing a p-skyline relation
$\succ\in\formset_\prefset$ such that
\begin{enumerate}
\item $G \subseteq \winnow_{\succ}(\oset)$, i.e.,
the tuples in $G$ are among the most preferred tuples in $\oset$,
according to $\succ$, and 
\item  for every tuple $o'$ in $W$, there is a
tuple $o$ in $G$ such that $o \succ o'$, i.e., $o'$ is an inferior
example.  
\end{enumerate}
Such a p-skyline relation $\succ$ is called \emph{favoring
  $G$ and disfavoring $W$ in $\oset$}. We may also skip ``in $\oset$''
when the context is clear.

The first problem we consider is the existence of a 
p-skyline relation favoring $G$ and disfavoring $W$ in $\oset$.

\smallskip

{\bf Problem \favdisprob.} {\it Given a set of attributes $\attrset$, a set of attribute preference
  relations $\prefset$, a set of superior examples $G$ and a set of inferior
  examples $W$ in a set $\oset$, determine if there exists a p-skyline relation
 $\succ\ \in \formset_\prefset$ favoring $G$ and
  disfavoring $W$ in $\oset$.  }

\smallskip

In most real life scenarios, knowing that a favoring/ disfavoring
p-skyline relation \emph{exists} is not sufficient. One needs to know
the \emph{contents} of  such a relation.

\smallskip

{\bf Problem \funcfavdisprob.} {\it Given a set of attributes $\attrset$, a set of attribute preference
  relations $\prefset$, a set of superior examples $G$ and a set of inferior
  examples $W$ in a set $\oset$, construct a p-skyline relation
 $\succ\ \in \formset_\prefset$ favoring $G$ and
  disfavoring $W$ in $\oset$.  }

\smallskip

We notice that \funcfavdisprob is the \emph{functional version}
\cite{1994-papadimitriou} of \favdisprob. Namely, gi\-ven subsets $G$
and $W$ of $\oset$, an instance of \funcfavdisprob outputs ``no'' if
there is no $\succ \in \formset_\prefset$ favoring $G$ and disfavoring
$W$ in $\oset$.  Otherwise, it outputs \emph{some} p-skyline relation
$\succ \in \formset_\prefset$ favoring $G$ and disfavoring $W$ in
$\oset$.

\begin{example}\label{ex:existence}
 Let the set $\oset$ consist of the following tuples describing cars
 for sale:

\begin{center}
\begin{tabular}{c||c|c|c} 
      & make & price & year \\
 \hline
  $t_1$ & ford & 30k & 2007\\
  $t_2$ & bmw & 45k & 2008 \\
  $t_3$ & kia & 20k & 2007 \\
  $t_4$ & ford & 40k & 2008 \\
  $t_5$ & bmw & 50k & 2006 \\
\end{tabular}
\end{center}

Assume also Mary wants to buy a car and her preferences over
automobile attributes are as follows.
\vspace{1mm}

\begin{description}
  \item[$>_{make}$:] {\it BMW} is better than {\it Ford}, {\it Ford}
    is better than {\it Kia}.
  \item[$>_{year}$:] higher values of $year$ (i.e., newer cars) are
    preferred.
  \item[$>_{price}$:] lower values of $price$ (i.e., cheaper cars) are
    preferred.
\end{description}

Let $G = \{t_4\}$, $W = \{t_3\}$. We elicit a p-skyline relation
$\succ$ favoring $G$ and disfavoring $W$. First, $>_{make}$ cannot be
more important than all other attribute preferences, since then $t_2$
and $t_5$ dominate $t_4$ and thus $t_4$ is not superior.  Moreover,
$>_{price}$ cannot be more important than the other attribute
preferences, because then $t_3$ and $t_1$ dominate $t_4$. However, if
$>_{year}$ is more important than the other attribute preferences,
then $t_4$ dominates $t_1,t_3,t_5$ and $t_2$ does not dominate $t_4$
in $>_{year}$.  At the same time, both $t_2$ and $t_4$ are the best
according to $>_{year}$, but $t_2$ dominates $t_4$ in
$>_{make}$. Therefore, $>_{make}$ should not be more important
than $>_{price}$. Thus, for example, the following p-skyline relation
\footnote{Here we again replace attribute preference relations
by atomic preference relations.} favors $G$ and disfavors $W$ in $\oset$
\[\succ_1 \ = \ \succ_{year} \pricomp ( \succ_{price} \parcomp
\succ_{make} )\]
The set of the best tuples in $\oset$ according to
$\succ_1$ is $\{ t_2, t_4\}$.
\end{example}

Generally, there may be zero, one or more p-skyline relations favoring
$G$ and disfavoring $W$ in $\oset$.  When more than one such relation
exists, we pick a \emph{maximal} one (in the set-theoretic sense).
Larger preference relations imply more dominated tuples and fewer most
preferred ones. Consequently, the result of $\winnow_{\succ}(\oset)$ is
likely to get more manageable due to its decreasing size. Moreover,
maximizing $\succ$ corresponds to minimizing $\winnow_\succ(\oset) - G$,
which implies more precise correspondence of $\succ$ to the real user
preferences.
%A maximal p-skyline
%relation favoring $G$ and disfavoring $W$ in $\oset$ is called
%\emph{optimal}. 
Thus, the next problem considered here is constructing maximal
p-skyline relations favoring $G$ and disfavoring $W$.

\smallskip

%{\bf Problem \optfavdisprob.} {\it Given  a set of attributes $\attrset$, a set of attribute
%preference relations $\prefset$, sets of superior $G$ and inferior $W$
%examples of a set $\oset$, determine if there exists a maximal p-skyline relation
%$\succ\ \in \formset_\prefset$ favoring $G$ and disfavoring $W$ in
%$\oset$.  }

%\smallskip

{\bf Problem \funcoptfavdisprob.} {\it  Given a set of attributes $\attrset$, a set of attribute
preference relations $\prefset$, a sets of superior examples $G$ and a set of inferior examples $W$
in  a set $\oset$, construct a maximal p-skyline relation
$\succ\ \in \formset_\prefset$ favoring $G$ and disfavoring $W$ in
$\oset$.  }

\smallskip

\begin{example}\label{ex:maximal}
 Take $G$, $W$, and $\succ_1$ from Example \ref{ex:existence}. Note
 that to make $t_4$ dominate $t_2$, we need to make $price$
 more important than $year$. As a result, the relation
 \[\succ_2 \ = \ \succ_{year} \pricomp \succ_{price} \pricomp
 \succ_{make}\] 
also favors $G$ and disfavors $W$ in $\oset$ but the
 set of best tuples in $\oset$ according to $\succ_2$ is
 $\{t_4\}$. Moreover, $\succ_2$ is \emph{maximal}.  The justification
 is that no other p-skyline relation favoring $G$ and disfavoring $W$
 contains $\succ_2$ since the p-graph of $\succ_2$ is a total order of
 the attributes $\{year, price, make\}$ and thus $\succ_2$ is a maximal SPO.
\end{example}

Even though the notion of maximal favoring/disfavoring reduces the
space of alternative p-skyline relations, there may still be more than
one maximal favoring/disfavoring p-skyline relation, given $\attrset$, $\prefset$, $G$, $W$,
and $\oset$.

\subsection{Negative and positive constraints}

We formalize now the kind of reasoning from Examples \ref{ex:existence}
and \ref{ex:maximal} using \emph{constraints on attribute sets}.
The constraints guarantee that the constructed p-skyline
relation favors $G$ and disfavors $W$ in $\oset$.

Consider the notion of \emph{favoring} $G$ in $\oset$ first.  For a
tuple $o' \in G$ to be in the set of the most preferred tuples of
$\mathcal{O}$, $o'$ must not be dominated by any tuple in
$\mathcal{O}$.  That is,
\begin{align}\label{eq:superior}
\forall o \in \mathcal{O}, o' \in G \ .\ o \not \succ o'
\end{align}

Using Theorem \ref{thm:pref-equiv}, we can rewrite \eqref{eq:superior}
as
\begin{align}\label{eq:superior2}
\forall o \in \mathcal{O}, o' \in G
\ .\ \children{\Gamma_{\succ}}(\betterin(o, o')) \not \supseteq
\betterin(o', o),
\end{align}
where $\betterin(o_1, o_2) = \{A\in\attrset\; |\; o_1.A \attr_{A} o_2.A\}$.  Note
that no tuple can be preferred to itself by irreflexivity of
$\succ$. Thus, a p-skyline relation favoring $G$ in $\oset$
should satisfy $(|\mathcal{O}|-1) \cdot |G|$ \emph{negative}
constraints $\tau$ in the form:
\[\tau: \children{\Gamma_{\succ}}(\clhs{\tau}) \not \supseteq
\crhs{\tau}\]
where $\clhs{\tau} = \betterin(o, o'), \crhs{\tau} =
\betterin(o', o)$. We denote this set of constraints as $\negsystem(G,
\oset)$.

\begin{example}\label{ex:negative}
  Take Example \ref{ex:existence}. Then some p-skyline relation $\succ
  \in \formset_\prefset$ favoring $G = \{t_3\}$ in $\oset$ has to
  satisfy each negative constraint below
\smallskip

\begin{center}
 \begin{tabular}{|l|l|}
\hline $t_1 \not \succ t_3$ & $\children{\Gamma_{\succ}}(\{make\})
\not \supseteq \{price\}$ \\ \hline $t_2 \not \succ t_3$ &
$\children{\Gamma_{\succ}}(\{make,year\}) \not \supseteq \{price\}$
\\ \hline $t_4 \not \succ t_3$ &
$\children{\Gamma_{\succ}}(\{make,year\}) \not \supseteq \{price\}$
\\ \hline $t_5 \not \succ t_3$ & $\children{\Gamma_{\succ}}(\{make\})
\not \supseteq \{price,year\}$ \\ \hline
 \end{tabular}
\end{center}
\end{example}

\medskip

Now consider the notion of \emph{disfavoring} $W$ in $\oset$. According to
the definition, a p-skyline relation $\succ$ favoring $G$ disfavors
$W$ in $\oset$ iff the following holds
\begin{equation}
\forall o' \in W \ \exists o \in G\ . \ o \succ o'.
\end{equation}
Following Theorem \ref{thm:pref-equiv}, it can be rewritten as a set
of \emph{positive constraints} $\mathcal{P}(W, G)$
\begin{equation}
\forall o' \in W \ \bigvee_{o_i \in G}
\children{\Gamma_\succ}(\betterin(o_i, o')) \supseteq \betterin(o',
o_i).
\end{equation}

Therefore, in order for $\succ$ to disfavor $W$ in $\oset$, it has to
satisfy $|W|$ positive constraints.

\begin{example}\label{ex:positive}
  Take Example \ref{ex:existence}. Then every p-skyline relation $\succ
  \ \in \ \formset_\prefset$ favoring $G = \{t_1, t_3\}$ and
  disfavoring $W = \{t_4\}$ in $\oset$ has to satisfy the constraint
\[t_1 \succ t_4 \vee t_3 \succ t_4\]
which is equivalent to the following positive constraint
\begin{align*}
\children{\Gamma_{\succ}}(\{price\}) \supseteq \{year\} \vee 
\children{\Gamma_{\succ}}(\{price\}) \supseteq \{year, make\},
\end{align*}
which in turn is equivalent to 
\begin{align*}
\children{\Gamma_{\succ}}(\{price\}) \supseteq \{year, make\}.
\end{align*}
\end{example}

Notice that positive and negative constraints are formulated in terms
of relative importance of the attributes captured by the p-graph of the
constructed p-skyline relation. Since p-sky\-line relations are
uniquely identified by p-graphs (Theorem
\ref{thm:p-graph-uniqueness}), we may refer to
\emph{a p-skyline relation satisfying/not satisfying  a system of positive/negative constraints}. Formally, a p-skyline
relation \emph{satisfies} a system of (positive or negative)
constraints iff it satisfies \emph{every constraint} in the system.

Let us summarize the kinds of constraints we have considered so far. To construct
a p-skyline relation $\succ$ favoring $G$ and disfavoring $W$ in
$\mathcal{O}$, we need to construct a p-graph $\Gamma_\succ$ 
that satisfies \spoenvelope to guarantee that $\succ$ be a
p-skyline relation, $\negsystem(G, \oset)$ to guarantee favoring $G$
in $\oset$, and $\possystem(W, G)$ to guarantee disfavoring $W$ in
$\oset$. 
By Theorem \ref{thm:cnf-subset}, the p-graph of a maximal $\succ$ is
\emph{maximal} among all graphs satisfying \spoenvelope,
$\negsystem(G, \oset)$, and $\possystem(W, G)$.

\subsection{Using superior and inferior examples }

In this section, we study the computational complexity of the problems of existence of a
favoring/disfavoring p-skyline relation and of constructing a
favoring/disfavoring p-skyline relation.

\begin{theorem}\label{thm:np-compl}
 \favdisprob is NP-complete.
\end{theorem}

Now consider the problems of constructing favoring/dis\-fa\-vor\-ing p-skyline
relations. First, we consider the problem of constructing \emph{some} p-skyline
relation favoring $G$ and disfavoring $W$ in $\oset$. 
Afterwards we address the problem of constructing a \emph{maximal}
p-skyline relation. The results shown below are based on the
following proposition.

\begin{proposition}\label{prop:polytime-comput}
 Let $\succ$ be a p-skyline relation, $\oset$ a finite set of tuples,
 and $G$ and $W$ disjoint subsets of $\oset$. Then the next two operations
 can be done in polynomial time:
 \begin{enumerate}
  \item verifying if $\succ$ is maximal favoring $G$ and disfavoring $W$ in $\oset$;
  \item constructing a maximal p-skyline relation $\succ_{ext}$ that favors
    $G$ and disfavors $W$ in $\oset$, and is a p-extension of $\succ$ favoring
  $G$ and disfavoring $W$ in $\oset$.
 \end{enumerate}
\end{proposition}

\begin{theorem}\label{thm:func-exist-fnp}
 \funcfavdisprob is \fnpcomplete
\end{theorem}

Surprisingly, the problem of constructing a maximal favoring/disfavoring
p-skyline relation is not harder then the problem of constructing some
favoring/disfavoring p-skyline relation.

\begin{theorem}\label{thm:computation-hard}
 \funcoptfavdisprob is \fnpcomplete
\end{theorem}

\subsection{Using only superior examples}
In view of Theorems \ref{thm:np-compl}, \ref{thm:func-exist-fnp}, and
\ref{thm:computation-hard}, we consider now restricted versions of the
favoring/disfavoring p-skyline relation problems, where we assume 
no inferior examples ($W = \emptyset$).
Denote as \favprob, \funcfavprob, and \optfavprobx the subclasses of
\favdisprob, \funcfavdisprobx, and \funcoptfavdisprob in which the sets of 
inferior examples $W$ are empty.
We show now that these problems are easier than their general
counterparts: they can all be solved in polynomial time.

\medskip

Consider \favprob first. We showed in Corollary \ref{cor:skyline-superset} 
that the set of the best objects according to the skyline
preference relation is the largest among all p-skyline
relations. Hence, the next proposition holds.

\begin{proposition}\label{prop:existence-simple}
  There exists a p-skyline relation $\succ \ \in
  \formset_\prefset$ favoring $G$ in $\oset$ iff
\[G \subseteq \winnow_{\skyline_\prefset}(\oset).\]
\end{proposition}

Proposition \ref{prop:existence-simple} implies that to solve
\favprob, one needs to run a skyline algorithm over $\oset$ and check
if the result contains $G$. This clearly can be done in polynomial
time. 

\funcfavprob can also be solved in polynomial time: if
$G \subseteq \winnow_{\skyline_\prefset}(\oset)$, then $\skyline_\prefset$
is a relation favoring $G$ and disfavoring $W$ in $\oset$. Otherwise,
there is no such a relation.

Now consider \optfavprob. To specify a p-skyline relation
$\succ$ favoring $G$ in $\oset$, we need to construct the
corresponding graph $\Gamma_{\succ}$ which satisfies $\negsystem(G,
\oset)$ and \spoenvelope.  Furthermore, to make the relation $\succ$
maximal favoring $G$ in $\oset$, $\Gamma_\succ$ has to be a
\emph{maximal} graph satisfying these constraints. In the next
section, we present an algorithm for constructing maximal p-skyline
relations.

\subsubsection{Syntax tree transformation}\label{sec:tree-transformation}

Our approach to constructing maximal favoring p-skyline relations
favoring $G$ is based on iterative transformations of
normalized syntax trees. We assume that the provided set of superior
examples $G$ satisfies Proposition \ref{prop:existence-simple}, i.e.,
$G \subseteq \winnow_{\skyline_\prefset}(\oset)$.  The idea beyond our
approach is as follows. First, we generate the set of negative
constraints $\negsystem(G, \oset)$.  The p-skyline relation we start
with is $\skyline_\prefset$ since it is the least p-skyline relation
favoring $G$ in $\oset$.  In every iteration of the algorithm, we pick
an attribute preference relation in $\prefset$ and apply a fixed set of
transformation rules to the syntax tree of the
current p-skyline relation. As a result, we obtain a ``locally maximal'' p-skyline relation
satisfying \emph{the given set $\negsystem(G, \oset)$ of negative
constraints}.  Recall that a negative constraint in $\negsystem(G,
\oset)$ represents the requirement that no tuple in $G$ is
 dominated by a tuple in $\oset$.  Eventually, this technique
produces a maximal p-skyline relation satisfying $\negsystem(G,
\oset)$.

Let us describe now what we mean by ``locally maximal''.
\begin{definition}\label{def:parially-max}
 Let $M$ be a nonempty subset of $\attrset$.  A p-skyline
 relation $\succ\ \in \formset_\prefset$ that favors $G$ in $\oset$
 such that $\edgesof{\Gamma_\succ} \subseteq $ $M \times M$ is
 \emph{$M$-favoring $G$ in $\oset$}.
% A maximal relation among all of
% them is called \emph{maximal}.
\end{definition}

We note that, similarly to a maximal favoring p-skyline relation, a
maximal $M$-favoring p-skyline relation is often  not unique for given $G$, $\oset$,
and $M$.
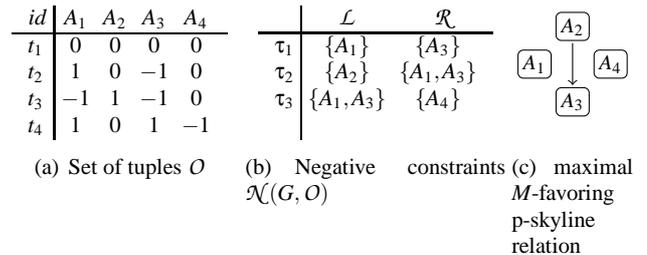
\begin{figure}[ht]
 \subfigure[Set of tuples $\oset$]{
	\begin{tikzpicture}
	 \node at (0, 1.5) {
		\begin{tabular}{l|cccc}
		 	$id$ \hspace{-2mm} & \hspace{-2mm} $A_1$ \hspace{-2mm} & \hspace{-2mm} $A_2$ \hspace{-2mm} & \hspace{-2mm} $A_3$ \hspace{-2mm} & \hspace{-2mm} $A_4$\\
		\hline
			$t_1$ \hspace{-2mm} & \hspace{-2mm} $0$ \hspace{-2mm} & \hspace{-2mm} $0$ \hspace{-2mm} & \hspace{-2mm} $0$ \hspace{-2mm} & \hspace{-2mm} $0$ \\
			$t_2$ \hspace{-2mm} & \hspace{-2mm} $1$ \hspace{-2mm} & \hspace{-2mm} $0$ \hspace{-2mm} & \hspace{-2mm} $-1$ \hspace{-2mm} & \hspace{-2mm} $0$ \\
			$t_3$ \hspace{-2mm} & \hspace{-2mm} $-1$ \hspace{-2mm} & \hspace{-2mm} $1$ \hspace{-2mm} & \hspace{-2mm} $-1$ \hspace{-2mm} & \hspace{-2mm} $0$\\
			$t_4$ \hspace{-2mm} & \hspace{-2mm} $1$ \hspace{-2mm} & \hspace{-2mm} $0$ \hspace{-2mm} & \hspace{-2mm} $1$ \hspace{-2mm} & \hspace{-2mm} $-1$ \\
		\end{tabular}
	  };
	\end{tikzpicture}
	\label{pic:m-favor-set}
  }
  \subfigure[Negative constraints $\negsystem(G, \oset)$]{
	\begin{tikzpicture}
	 \node at (0, 1.76) {
		\begin{tabular}{l|cc}
		 	     & $\clhs{}$ & $\crhs{}$ \\
		\hline
			$\tau_1$ \hspace{-2mm} & \hspace{-2mm} $\{A_1\}$ \hspace{-2mm} & \hspace{-2mm} $\{A_3\}$\\
			$\tau_2$ \hspace{-2mm} & \hspace{-2mm} $\{A_2\}$ \hspace{-2mm} & \hspace{-2mm} $\{A_1, A_3\}$ \\
			$\tau_3$ \hspace{-2mm} & \hspace{-2mm} $\{A_1, A_3\}$ \hspace{-2mm} & \hspace{-2mm} $\{A_4\}$\\
                        & & 
		\end{tabular}
	  };
	\end{tikzpicture}
	\label{pic:m-favor-cons}
  }
  \subfigure[maximal $M$-favoring p-skyline relation]{
		\begin{tikzpicture}[xscale=0.5]
%		  \tikzstyle{leafnode} = [draw=black,rounded corners,inner sep=2pt]
		  \node[p-graph-node] (A1) at (0, 2) {$A_1$};
		  \node[p-graph-node] (A2) at (1, 2.5) {$A_2$};
		  \node[p-graph-node] (A3) at (1, 1.5) {$A_3$};
		  \node[p-graph-node] (A4) at (2, 2) {$A_4$};
		  \draw[->] (A2) to (A3);
                  \node[] at (0,1){$ $};
		\end{tikzpicture}
	\label{pic:m-favor-graph}
  }
  \caption{Example \ref{ex:m-favor}}
  \label{pic:m-favor}
   \vspace{-5mm}
\end{figure}

\begin{example}\label{ex:m-favor}
 Let $\attrset = \{A_1, A_2, A_3, A_4\}$ and $\prefset = \{\attr_{A_1},
 \attr_{A_2}, \attr_{A_3}, \attr_{A_4}\}$, where a greater value
 of the corresponding attribute is preferred, according to every
 $\attr_{A_i}$.  Let the set of objects $\oset$ be as shown in
 Figure \ref{pic:m-favor-set} and $G = \{t_1\}$.  Then the set of
 negative constraints $\negsystem(G, \oset)$ is shown in Figure
 \ref{pic:m-favor-cons}.  Consider the p-skyline relation $\succ$
 represented by the p-graph $\Gamma_{\succ}$ shown in Figure
 \ref{pic:m-favor-graph}. It is a maximal $\{A_1, A_2,
 A_3\}$-favoring relation because $\Gamma_{\succ}$ satisfies all the
 constraints in $\negsystem(G, \oset)$ and every additional edge
 from one attribute to another attribute in $\{A_1, A_2, A_3\}$
 violates $\negsystem(G, \oset)$. In particular, the edge
 $(A_1,A_3)$ violates $\tau_1$ and the edge $(A_2,A_1)$
 violates $\tau_2$. Every other edge between $A_1$, $A_2$ and $A_3$ induces
 one of the two edges above.

At the same time, $\succ$ is not a maximal $\attrset$-favoring
relation because, for example, the edge $(A_4,A_1)$ may be added
to $\Gamma_{\succ}$ without violating $\negsystem(G, \oset)$.
\end{example}

By Definition \ref{def:parially-max}, the edge set of the p-graph of
every maximal $M$-favoring relation is maximal among all the p-graphs
of $M$-favoring relations.  Note that if $M$ is a singleton, the edge
set of a p-graph $\Gamma_{\succ}$ of a maximal $M$-favoring relation
$\succ$ is empty, i.e., $\succ = \skyline_\prefset$. If $M =
\attrset$, then a maximal p-skyline relation $M$-favoring $G$ in
$\oset$ is also a maximal p-skyline relation favoring $G$ in $\oset$.
Thus, if we had a method of transforming a maximal $M$-favoring
p-skyline relation to a maximal $(M \cup \{A\})$-favoring p-skyline
relation for each attribute $A$, we could construct a maximal favoring
p-skyline relation iteratively. A useful property of such a
transformation process is shown in the next proposition.

\begin{proposition}\label{prop:edge-diff}
 Let a relation $\succ\ \in \formset_\prefset$ be a maximal
 $M$-fa\-vor\-ing relation, and a p-extension $\succ_{ext}$ of
 $\succ$ be $(M \cup \{A\})$-favoring. Then every edge in
 $\edgesof{\Gamma_{\succ_{ext}}} - \edgesof{\Gamma_{\succ}}$ starts or
 ends in $A$.
\end{proposition}

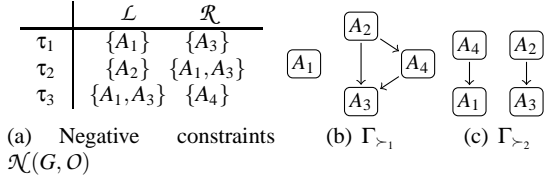
\begin{figure}
\begin{center}
  \subfigure[Negative constraints $\negsystem(G, \oset)$]{
	\begin{tikzpicture}
	 \node at (0, 1.76) {
		\begin{tabular}{l|cc}
		 	     & $\clhs{}$\hspace{-2mm} & \hspace{-2mm}$\crhs{}$ \\
		\hline
			$\tau_1$ & $\{A_1\}$\hspace{-2mm} & \hspace{-2mm}$\{A_3\}$\\
			$\tau_2$ & $\{A_2\}$\hspace{-2mm} & \hspace{-2mm}$\{A_1, A_3\}$ \\
			$\tau_3$ & $\{A_1, A_3\}$\hspace{-2mm} & \hspace{-2mm}$\{A_4\}$
		\end{tabular}
	  };
	\end{tikzpicture}
	\label{pic:m-favor-opt-cons}
  }
  \subfigure[$\Gamma_{\succ_1}$]{
		\begin{tikzpicture}[xscale=0.75]
		  \node[p-graph-node] (A1) at (0, 2) {$A_1$};
		  \node[p-graph-node] (A2) at (1, 2.5) {$A_2$};
		  \node[p-graph-node] (A3) at (1, 1.5) {$A_3$};
		  \node[p-graph-node] (A4) at (2, 2) {$A_4$};
		  
		  \draw[->] (A2) to (A3);
		  \draw[->] (A2) to (A4);
		  \draw[->] (A4) to (A3);
		\end{tikzpicture}
	\label{pic:m-favor-opt-pgraph1}
  }
  \subfigure[$\Gamma_{\succ_2}$]{
		\begin{tikzpicture}[scale=0.75]
		  \node[p-graph-node] (A1) at (0, 1.5) {$A_1$};
		  \node[p-graph-node] (A2) at (1, 2.5) {$A_2$};
		  \node[p-graph-node] (A3) at (1, 1.5) {$A_3$};
		  \node[p-graph-node] (A4) at (0, 2.5) {$A_4$};
		  
		  \draw[->] (A2) to (A3);
		  \draw[->] (A4) to (A1);
		\end{tikzpicture}
	\label{pic:m-favor-opt-pgraph2}
  }
  \caption{Example \ref{ex:m-favor-opt}}
  \label{pic:m-favor-opt}
\end{center}
       \vspace{-5mm}
\end{figure}

\begin{example}\label{ex:m-favor-opt}
 Consider $\negsystem(G, \oset)$ from Example \ref{ex:m-favor} 
 (also depicted in Figure \ref{pic:m-favor-opt-cons}), and the maximal $\{A_1,
 A_2, A_3\}$-favoring relation $\succ$. Several different maximal
 $\attrset$-favoring p-skyline relations containing $\succ$ exist. 
 Two of them are $\succ_1$ and $\succ_2$ whose p-graphs
 are shown in Figures \ref{pic:m-favor-opt-pgraph1} and
 \ref{pic:m-favor-opt-pgraph2}.
\end{example}

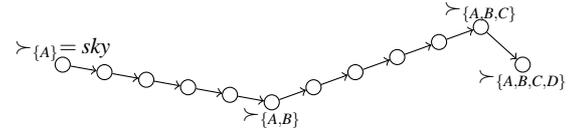
\begin{figure}[ht]
%   \vspace{-0.75cm}
   \begin{tikzpicture}[xscale=0.55,yscale=0.1]
        \tikzstyle{cirnode} = [draw=black,circle,inner sep=2pt]
        \node[cirnode] (A0 ) at ( 0, 0) {};
        \node[cirnode] (A1 ) at ( 1, -1) {};
        \node[cirnode] (A2 ) at ( 2, -2) {};
        \node[cirnode] (A3 ) at ( 3, -3) {};
        \node[cirnode] (A4 ) at ( 4, -4) {};
        \node[cirnode] (A5 ) at ( 5, -5) {};
        \node[cirnode] (A6 ) at ( 6, -3) {};
        \node[cirnode] (A7 ) at ( 7, -1) {};
        \node[cirnode] (A8 ) at ( 8,  1) {};
        \node[cirnode] (A9 ) at ( 9, 3) {};
        \node[cirnode] (A10) at (10, 5) {};
        \node[cirnode] (A11) at (11, 0) {};

        \draw[->](A0 ) to (A1 );
        \draw[->](A1 ) to (A2 );
        \draw[->](A2 ) to (A3 );
        \draw[->](A3 ) to (A4 );
        \draw[->](A4 ) to (A5 );
        \draw[->](A5 ) to (A6 );
        \draw[->](A6 ) to (A7 );
        \draw[->](A7 ) to (A8 );
        \draw[->](A8 ) to (A9 );
        \draw[->](A9 ) to (A10);
        \draw[->](A10) to (A11);

        \node[] at (0,2){$\succ_{\{A\}} = \skyline$};
        \node[] at (5,-7){$\succ_{\{A,B\}}$};
        \node[] at (10,7){$\succ_{\{A,B,C\}}$};
        \node[] at (11,-2){$\succ_{\{A,B,C,D\}}$};
   \end{tikzpicture}

   \caption{A path to a maximal $\attrset$-favoring p-skyline
   relation. The path starts from the maximal singleton-favoring p-skyline
   relation: the skyline relation. Every step is a minimal p-extension. The path goes
   through maximal $M$-favoring p-skyline relations ($\succ_{\{A\}}, 
        \succ_{\{A,B\}},\ldots$) for incrementally increasing $M$.
        The path ends with a maximal $M$-favoring p-skyline relation for
        $M = \attrset$.}
   \label{pic:optmimal-path}
   \vspace{-0.5cm}
\end{figure}

In section \ref{sec:p-skyline-min-ext}, we showed four syntax tree
transformation rules , $Rule_1$ -- $Rule_4$, for extending  p-skyline
relations in a minimal way. Although a maximal $(M \cup
\{A\})$-favoring p-skyline relation is a p-extension of a maximal
$M$-favoring p-skyline relation, it is not necessary a minimal
p-extension in general.  However, an important property of that set of
rules is its completeness, i.e., every minimal p-extension can be
constructed using them. Hence, a maximal $(M \cup
\{A\})$-favoring p-skyline relation can be produced from a maximal
$M$-favoring p-skyline relation by
\emph{iterative application of the minimal extension rules}.
This process is illustrated by Figure \ref{pic:optmimal-path}.

We use the following idea for constructing maximal $(M \cup
\{A\})$-favoring relations.  We start with a maximal $M$-favoring
p-skyline relation $\succ_0$ and apply the transformation rules to
$T_{\succ_0}$ in every possible way guaranteeing that the new edges in
the p-graph go only from or to $A$. In other words, we construct all
minimal $(M \cup \{A\})$-favoring p-extensions of $\succ_0$. We
construct such p-extensions until we find the first one which does not
violate $\negsystem(G, \oset)$. When we find it (denote it as
$\succ_1$), we repeat all the steps above but for $\succ_1$.  This
process continues until for some $\succ_m$, every of its constructed
minimal p-extension violates $\negsystem(G, \oset)$. Since in every
iteration we construct all minimal $(M \cup \{A\})$-favoring
p-extensions, $\succ_m$ is a maximal $(M \cup \{A\})$-favoring
p-extension of $\succ_0$.

There is subtle point here. We can limit ourselves to {\em minimal}
p-extensions because if a minimal p-extension violates $\negsystem(G,
\oset)$, so do all non-minimal p-extensions containing it. Also, if
there exists a p-extension satisfying $\negsystem(G, \oset)$, so does
some minimal one. In fact, each p-extension of a p-skyline relation
can be obtained through a finite sequence of minimal p-extensions.
Those properties are characteristic of \emph{negative} constraints.
The properties do not hold for \emph{positive} constraints and thus
our approach cannot be directly generalized to such constraints.

An important condition to apply Theorem \ref{thm:min-ext-rules} is
that the input syntax tree for every transformation rule be
normalized. At the same time, syntax trees returned by the
transformation rules are not guaranteed to be normalized. Therefore,
we need to normalize a tree before applying transformation rules to
it.

\medskip

Consider the rules $Rule_1$ -- $Rule_4$ which can be used to construct an $(M \cup
\{A\})$-favoring p-skyline relation from an $M$-favoring one. By Proposition
\ref{prop:edge-diff}, such rules may \emph{only} add to the p-graph the edges
that go to $A$ or from $A$.  According to Observation
\ref{obs:new-edges}, $Rule_1$ adds edges going to the node $A$ if
$C_{i+1} = A$ or $N_1 = A$. Similarly, $Rule_2$ adds edges going from
$A$ if $C_{i+1} = A$ or $N_m = A$. $Rule_3$ adds edges going from or
to $A$ if $C_{i} = A$ or $C_{i+1} = A$ correspondingly. However,
$Rule_4$ can only be applied to a pair of $\pricomp$-nodes. Hence, as
we showed in section \ref{sec:p-skyline-min-ext}, $Rule_4$ adds edges
going from at least two nodes to at least two different nodes of a
p-graph. Hence, every application of $Rule_4$ violates Proposition
\ref{prop:edge-diff}. We conclude that \emph{$Rule_1, Rule_2,$ and $Rule_3$
  are sufficient to construct every maximal $(M \cup \{A\})$-favoring
  p-skyline relation}.

\subsubsection{Efficient constraint checking}\label{sec:eff-cons-checking}

Before going into the details of the algorithm of p-skyline relation
elicitation, we consider an important step of the algorithm: testing
if a p-extension of a p-skyline relation satisfies a set of negative
constraints. We propose now an efficient method for this task.

Recall that a negative constraint is of the form
\[\tau : \children{\Gamma_{\succ}}(\clhs{\tau}) \not \supseteq
\crhs{\tau}.\] 
It can be visualized as two layers of nodes
$\clhs{\tau}$ and $\crhs{\tau}$.  For a p-skyline relation
$\succ\ \in \formset_\prefset$ satisfying $\tau$, its p-graph
$\Gamma_{\succ}$ may contain edges going between the nodes of the
layers $\clhs{\tau}$ and $\crhs{\tau}$. However, in order for $\succ$
to satisfy $\tau$, there should be at least one member of
$\crhs{\tau}$ with no incoming edges from $\clhs{\tau}$.

The method of efficient checking of negative constraints against a
p-graph that we propose here is based on the fact that the edge set of the
p-graph of a transformed p-skyline relation monotonically increases.
Therefore, while we transform a p-skyline relation $\succ$, we can
simply drop the elements of $\crhs{\tau}$ which already have incoming
edges from $\clhs{\tau}$.  If we do so after every transformation of
the p-skyline relation $\succ$, the negative constraint $\tau$ will be
violated by $\Gamma_{\succ}$ only if $\crhs{\tau}$ is empty.  The next
proposition says that such a modification of negative constraints is
valid.

\begin{proposition}\label{prop:cons-modif}
 Let a relation $\succ\ \in \formset_\prefset$ satisfy a system of
 negative constraints $\negsystem$.  Construct the system of negative
 constraints $\negsystem'$ from $\negsystem$ in which every constraint $\tau' \in
 \negsystem'$ is created from a constraint $\tau$ of $\negsystem$ in the
 following way:

\begin{itemize}
 \item $\clhs{\tau'} = \clhs{\tau}$
 \item $\crhs{\tau'} = \crhs{\tau} - \{B \in \crhs{\tau}\ |\ \exists A
   \in \clhs{\tau} \ .\ \edgeof{A}{B}{\Gamma_\succ}\}$
\end{itemize}

\noindent Then every p-extension $\succ'$ of $\succ$
satisfies $\negsystem$ iff $\succ'$ satisfies
$\negsystem'$.
\end{proposition}

A constraint $\tau'$ constructed from $\tau$ as shown in Proposition
\ref{prop:cons-modif} is called a \emph{minimal negative constraint
  w.r.t. $\succ$}. The corresponding system of negative constraints
  $\negsystem'$ is called a \emph{system of minimal negative
  constraints w.r.t. $\succ$}.

Minimization of a system of negative constraints is illustrated in the
next example.

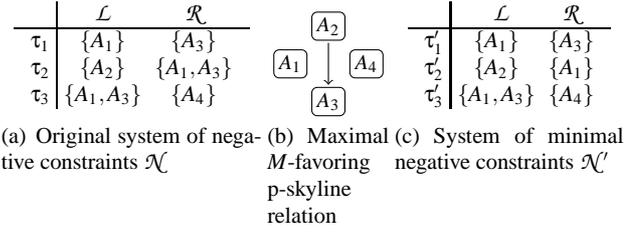
\begin{figure}[ht]
\begin{center}
  \subfigure[Original system of negative constraints $\negsystem$]{
	\begin{tikzpicture}
	 \node at (0, 1.76) {
		\begin{tabular}{l|cc}
		 	     & \hspace{-2mm} $\clhs{}$ \hspace{-2mm} & \hspace{-2mm}  $\crhs{}$ \\
		\hline
			$\tau_1$ \hspace{-2mm} & \hspace{-2mm} $\{A_1\}$ \hspace{-2mm}  & \hspace{-2mm} $\{A_3\}$\\
			$\tau_2$ \hspace{-2mm} & \hspace{-2mm} $\{A_2\}$ \hspace{-2mm} & \hspace{-2mm} $\{A_1, A_3\}$ \\
			$\tau_3$ \hspace{-2mm} & \hspace{-2mm} $\{A_1, A_3\}$ \hspace{-2mm} & \hspace{-2mm} $\{A_4\}$
		\end{tabular}
	  };
	\end{tikzpicture}
	\label{pic:min-cons-syst-initial}
  }
  \subfigure[Maximal $M$-favoring p-skyline relation]{
		\begin{tikzpicture}[xscale=0.5]
		  \node[p-graph-node] (A1) at (0, 2) {$A_1$};
		  \node[p-graph-node] (A2) at (1, 2.5) {$A_2$};
		  \node[p-graph-node] (A3) at (1, 1.5) {$A_3$};
		  \node[p-graph-node] (A4) at (2, 2) {$A_4$};
		  \draw[->] (A2) to (A3);
		\end{tikzpicture}
	\label{pic:min-cons-syst-graph}
  }
  \subfigure[System of minimal negative constraints $\negsystem'$]{
	\begin{tikzpicture}
	 \node at (0, 1.76) {
		\begin{tabular}{l|cc}
		 	     & $\clhs{}$ \hspace{-2mm} & \hspace{-2mm} $\crhs{}$ \\
		\hline
			$\tau_1'$ \hspace{-2mm} & \hspace{-2mm} $\{A_1\}$ \hspace{-2mm} & \hspace{-2mm} $\{A_3\}$\\
			$\tau_2'$ \hspace{-2mm} & \hspace{-2mm} $\{A_2\}$ \hspace{-2mm} & \hspace{-2mm} $\{A_1\}$ \\
			$\tau_3'$ \hspace{-2mm} & \hspace{-2mm} $\{A_1, A_3\}$ \hspace{-2mm} & \hspace{-2mm} $\{A_4\}$
		\end{tabular}
	  };
	\end{tikzpicture}
	\label{pic:min-cons-syst-minimal}
  }
\end{center}
  \caption{Example \ref{ex:min-cons-syst}}
  \label{pic:min-cons-syst}
%  \vspace{-5mm}
\end{figure}

\begin{example}\label{ex:min-cons-syst}
 Consider the system of negative constraints $\negsystem$ and the
 p-skyline relation $\succ$ from Example \ref{ex:m-favor} (they are
 shown in Figures \ref{pic:min-cons-syst-initial} and
 \ref{pic:min-cons-syst-graph} correspondingly). The result
 $\negsystem'$ of minimization of $\negsystem$ w.r.t $\succ$ is shown
 in Figure \ref{pic:min-cons-syst-minimal}. Only the constraint
 $\tau_2'$ is different from $\tau_2$ because
 $\edgeof{A_2}{A_3}{\Gamma_{\succ}}$ and $A_2 \in \clhs{\tau_2}$, $A_3
 \in \crhs{\tau_2}$.
\end{example}

The next proposition summarizes the constraint checking rules over a
system of minimal negative constraints.

\begin{proposition}\label{prop:cons-checking}
 Let a relation $\succ\ \in \formset_\prefset$ satisfy
 a system of negative constraints $\negsystem$, and $\negsystem$ be
 minimal w.r.t. $\succ$. Let $\succ'$ be a
 p-extension of $\succ$ such that every edge in $\edgesof{\Gamma_{\succ'}} -
 \edgesof{\Gamma_{\succ}}$ starts or ends in $A$.  Denote the
 \emph{new} parents and children of $A$ in $\Gamma_{\succ'}$ as $P_A$
 and $C_A$ correspondingly. Then $\succ'$ violates $\negsystem$ iff
 there is a constraint $\tau \in \negsystem$ such that
 
 \begin{enumerate}
   \item $\crhs{\tau} = \{A\} \wedge P_A \cap \clhs{\tau} \neq \emptyset, \mbox{ or}$
   \item $A \in \clhs{\tau} \wedge \crhs{\tau} \subseteq C_A$
 \end{enumerate}
\end{proposition}

Proposition \ref{prop:cons-checking} is illustrated in the next example.

\begin{figure}[ht]
\begin{center}
		\begin{tikzpicture}
		  \node[p-graph-node] (A1) at (0, 2) {$A_1$};
		  \node[p-graph-node] (A2) at (1, 2.5) {$A_2$};
		  \node[p-graph-node] (A3) at (1, 1.5) {$A_3$};
		  \node[p-graph-node] (A4) at (2, 2) {$A_4$};
		  \draw[->] (A2) to (A3);
		  \draw[->] (A2) to (A4);;
		\end{tikzpicture}
\end{center}
  \caption{$\Gamma_{\succ'}$ from Example \ref{ex:min-cons-syst-sat}}
  \label{pic:ex:min-cons-syst-sat}
\end{figure}
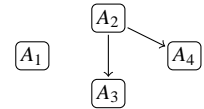

\begin{example}\label{ex:min-cons-syst-sat}
 Take the system of minimal negative constraints $\negsystem'$
 w.r.t. $\succ$ from Example \ref{ex:min-cons-syst}. Construct
 a p-extension $\succ'$ of $\succ$ such that every edge
 in $\edgesof{\Gamma_{\succ'}} - \edgesof{\Gamma_{\succ}}$ starts or
 ends in $A_4$.  Consider possible edges going to $A_4$. Use
 Proposition \ref{prop:cons-checking} to check if a new edge violates
 $\negsystem'$.  The edge $(A_1,A_4)$ is not allowed in $\Gamma_{\succ'}$
 because then $A_1 \in \clhs{\tau_3'}$ and $\{A_4\} =
 \crhs{\tau_3'}$  (and thus the constraint $\tau_3'$ is violated).
 The edge $(A_3,A_4)$ is not allowed in ${\Gamma_{\succ'}}$ 
 because $A_3 \in \clhs{\tau_3'}$ and $\{A_4\} =
 \crhs{\tau_3'}$.  However, the edge
 $(A_2,A_4)$ is allowed in $\Gamma_{\succ'}$. The p-graph of the
 resulting $\succ'$ is shown in Figure
 \ref{pic:ex:min-cons-syst-sat}. One can analyze the edges going from
 $A_4$ in a similar fashion.
\end{example}

\subsubsection{p-skyline elicitation}

In this section, we show an algorithm for p-skyline relation elicitation
which exploits the ideas developed in the previous sections. 

The function {\tt elicit} (Algorithm \ref{alg:elicit}) is the main
function of the algorithm. It takes four arguments: the set of
superior examples $G$, the entire set of tuples $\oset$, the set of
attribute preferences $\prefset$, and the set of all relevant
attributes $\attrset$.  It returns a normalized syntax tree of a
maximal p-skyline relation favoring $G$ in $\oset$.  Following
Proposition
\ref{prop:existence-simple}, we require $G$ to be a subset of
$\winnow_{\skyline_\prefset}(\oset)$. First, we construct the set of
negative constraints $\negsystem$ for the superior tuples $G$. We
start with $\skyline_\prefset$ as the initial p-skyline relation
favoring $G$ in $\oset$. After that, we take the set $M$ consisting of
a single attribute. In every iteration, we enlarge it and construct a
maximal $M$-favoring p-skyline relation. As a result, the function
returns a maximal p-skyline relation favoring $G$ in $\oset$. The
construction of a maximal $(M \cup \{A\})$-favoring relation from a
maximal $M$-favoring relation is performed in the {\tt repeat}/ {\tt until}
loop (lines 5-8). Here we use the function {\tt push} which constructs
a minimal $(M \cup \{A\})$-favoring p-extension of the relation
represented by the syntax tree $T$. It returns $true$ if $T$ has been
(minimally) extended to a relation not violating $\negsystem$, and
further p-extensions are feasible (though they may still violate
$\negsystem$). Otherwise, it returns $false$. The syntax tree $T$
passed to {\tt push} has to be normalized. Hence, after extending the
relation, we normalize its syntax tree (line 7) using the
normalization procedure sketched in Section
\ref{sec:syntax-tree}. The {\tt repeat/until} loop terminates when all
minimal extensions of $T$ violate $\negsystem$.

\begin{algorithm}
  \caption{{\tt elicit}($G$, $\oset$, $\prefset$, $\attrset$)}
  \begin{algorithmic}[1]
	\REQUIRE{$G \subseteq \winnow_{\skyline_\prefset}(\oset)$}
	\STATE $\negsystem = \negsystem(G, \oset)$
	\STATE $T =\  $a normalized syntax tree of $\skyline_\prefset$
	\STATE $M =\  $set containing an arbitrary attribute from $\attrset$
	\FOR{each attribute $A$ in $\attrset - M$}
		\REPEAT
			\STATE $r = $ push($T$, $M$, $A$, $\negsystem$);
			\STATE {\tt normalizeTree}(root of $T$);
		\UNTIL{$r$ is false}
		\STATE $M = M \cup \{A\}$
	\ENDFOR
	\RETURN $T$
   \end{algorithmic}
   \label{alg:elicit}
   \vspace{-1mm}
\end{algorithm}

Let us now take a closer look at the function {\tt push} (Algorithm
\ref{alg:main2}).  It takes four arguments: a set $M$ of attributes, a
normalized syntax tree $T$ of an $M$-favoring p-skyline relation
$\succ$, the current attribute $A$, and a system of negative
constraints $\negsystem$ minimal w.r.t. $\succ$.  It returns $true$ if
a transformation rule $q\in\{Rule_1, Rule_2, Rule_3\}$ has been applied to $T$ without violating
$\negsystem$, and $false$ if no transformation rule can be applied to $T$
without violating $\negsystem$. When {\tt push} returns true,
$\negsystem$ and $T$ have been changed. Now $\negsystem$ is minimal
w.r.t. the p-skyline relation represented by the modified syntax tree,
and $T$ has been modified by the rule $q$ and is normalized.

The goal of {\tt push} is to find an appropriate transformation rule which
adds to the current p-graph edges going from $M$ to $A$ or vice
versa. The function has two branches: the first for the parent of the
node $A$ in the syntax tree $T$ being a $\pricomp$-node (i.e., we may
apply $Rule_1$ where $N_1$ is $A$ or $Rule_2$ where $N_m$ is $A$), and
the second for it being $\parcomp$-node (i.e., we may apply $Rule_1$
or $Rule_2$ where $C_{i+1}$ is $A$, or $Rule_3$ where $C_i$ or
$C_{i+1}$ is $A$). In the first branch (line 2-14), we distinguish
between applying $Rule_1$ (line 3-8) and $Rule_2$ (line 9-14). It is
easy to notice that, with the parameters specified above, the rules
are exclusive, but the application patterns are similar. First, we
find an appropriate child $C_{i+1}$ of $R$ (lines 4 and 10). (It is
important for $\formvars(C_{i+1})$ to be a subset of $M$ because we
want to add edges going from $M$ to $A$ or from $A$ to $M$.)  Then we
check if the corresponding rule application does not violate
$\negsystem$ using the function {\tt checkConstr} (lines 5 and 11), as
per Proposition \ref{prop:cons-checking}.  If the rule application
does not violate $\negsystem$, we apply the corresponding rule to $T$
(lines 6 and 12) and minimize $\negsystem$ w.r.t. the p-skyline
relation which is the result of the transformation (Proposition
\ref{prop:cons-modif}) using the function {\tt minimize}.

The second branch of {\tt push} is similar to the first one and
different only in the transformation rules applied.  So it is
easy to notice that {\tt push} checks every possible rule application
not violating $\negsystem$, and adds to the p-graph only edges
going from $A$ to the elements of $M$ or vice versa.

In our implementation of the algorithm, all sets of attributes are
represented as bitmaps of fixed size $|\attrset|$. Similarly, every
negative constraint $\tau$ is represented as a pair of bitmaps
corresponding to $\clhs{\tau}$ and $\crhs{\tau}$.  With every node
$C_i$ of the syntax tree, we associate a variable storing
$\formvars(C_i)$.  Its value is updated whenever the children list of
$C_i$ is changed.

\begin{theorem}\label{thm:main2-runtime}
 The function {\tt elicit} returns a syntax tree of a maximal
 p-skyline relation favoring $G$ in $\oset$. Its running time is
 $O(|\negsystem| \cdot |\attrset|^3)$.
\end{theorem}

\newcommand{\algfor}{{\bf for\ }}
\newcommand{\algif}{{\bf if\ }}
\newcommand{\algthen}{{\bf then}}
\newcommand{\algelse}{{\bf else}}
\newcommand{\algelsif}{{\bf else if\ }}
\newcommand{\algreturn}{{\bf return\ }}
\newcommand{\algrepeat}{{\bf repeat\ }}
\newcommand{\alguntil}{{\bf until\ }}

\newcommand{\tabone}{\hspace{3mm}}
\newcommand{\tabtwo}{\hspace{6mm}}
\newcommand{\tabthree}{\hspace{9mm}}
\newcommand{\tabfour}{\hspace{12mm}}
\newcommand{\tabfive}{\hspace{15mm}}

\begin{algorithm}
  \caption{{\tt push}($T$, $M$, $A$, $\negsystem$)}
  \begin{algorithmic}[1]
\REQUIRE{$T$ is normalized}
\STATE	\algif the parent of $A$ in $T$ is of type $\pricomp$
\STATE	\tabone $C_i := $ parent of $A$ in $T$; $R := $ parent of $C_i$ in $T$;
\STATE	\tabone	\algif $R$ is defined, and $A$ is the first child of $C_i$ 
\STATE	\tabtwo	\algfor {each child $C_{i+1}$ of $R$ s.t. $\formvars(C_{i+1}) \subseteq M$}
\STATE	\tabthree \algif checkConstr($\negsystem$, $A$, $\emptyset$, $\formvars(C_{i+1})$)
\STATE	\tabfour  apply $Rule_1(T, C_i, C_{i+1})$
\STATE	\tabfour  $\negsystem := minimize(\negsystem, \formvars(A), \formvars(C_{i+1}))$
\STATE	\tabfour  \algreturn $true$
\STATE	\tabone	\algelsif $R$ is defined, and $A$ is the last child of $C_i$
\STATE	\tabtwo	\algfor {each child $C_{i+1}$ of $R$ s.t. $\formvars(C_{i+1}) \subseteq M$}
\STATE	\tabthree \algif checkConstr($\negsystem$, $A$, $\formvars(C_{i+1})$, $\emptyset$)
\STATE	\tabfour  apply $Rule_2(T, C_i, C_{i+1})$
\STATE	\tabfour  $\negsystem := minimize(\negsystem, \formvars(C_{i+1}), \formvars(A))$
\STATE	\tabfour  \algreturn $true$
\STATE	\algelse \ \ // the parent of $A$ in $T$ is of type $\parcomp$
\STATE	\tabone   $R :=$ parent of $A$ in $T$;
\STATE 	\tabone \algfor each child $C_i$ of $R$ s.t. $\formvars(C_i) \subseteq M$
\STATE	\tabtwo	\algif	$C_i$ is of type $\pricomp$
\STATE 	\tabthree $N_1 := $ first child of $C_i$, $N_m := $ last child of $C_i$
\STATE	\tabthree \algif checkConstr($\negsystem$, $A$, $\formvars(N_1)$, $\emptyset$)
\STATE	\tabfour apply $Rule_1(T, C_i, A)$
\STATE 	\tabfour $\negsystem := $minimize($\negsystem$, $\formvars(N_1)$, $\formvars(A)$)
\STATE 	\tabfour \algreturn $true$
\STATE 	\tabthree \algelsif checkConstr($\negsystem$, $A$, $\emptyset$, $\formvars(N_m)$)
\STATE 	\tabfour apply $Rule_2(T, C_i, A)$
\STATE 	\tabfour $\negsystem :=$ minimize($\negsystem$, $\formvars(A)$, $\formvars(N_m)$)
\STATE 	\tabfour \algreturn $true$
\STATE	\tabtwo	\algelse \ \ // $C_i$ is a leaf node, since $T$ is normalized
\STATE	\tabthree \algif checkConstr($\negsystem$, $A$, $\formvars(C_i)$, $\emptyset$)
\STATE 	\tabfour apply $Rule_3(T, C_i, A)$
\STATE 	\tabfour $\negsystem := $minimize($\negsystem$, $\formvars(C_i)$, $\formvars(A)$)
\STATE 	\tabfour \algreturn $true$
\STATE 	\tabthree \algelsif checkConstr($\negsystem$, $A$, $\emptyset$, $\formvars(C_i)$
\STATE 	\tabfour apply $Rule_3(T, A, C_i)$
\STATE 	\tabfour $\negsystem := $minimize($\negsystem$, $\formvars(A)$, $\formvars(C_i)$)
\STATE 	\tabfour \algreturn $true$
\STATE	\algreturn $false$
  \end{algorithmic}
\label{alg:main2}
   \vspace{-1mm}
\end{algorithm}

\begin{algorithm}[ht]
 \caption{{\tt checkConstr}($\negsystem$, $A$, $P_A$, $C_A$)}
 \begin{algorithmic}
   \FOR{each $\tau \in \negsystem$}
	\IF{$\crhs{\tau} = \{A\} \wedge P_A \cap \clhs{\tau} \neq \emptyset$ {\bf or} $A \in \clhs{\tau} \wedge \crhs{\tau} \subseteq C_A$}
		\RETURN $false$
	\ENDIF
   \ENDFOR
   \RETURN $true$
 \end{algorithmic}
 \label{alg:checkConstr}
%   \vspace{-1mm}
\end{algorithm}

\begin{algorithm}[ht]
  \caption{{\tt minimize}($\negsystem$, $U$, $D$)}\label{alg:minimizeSystem}
  \begin{algorithmic}[1]
	\FOR{each constraint $\tau$ in $\negsystem$}
		\IF{$U \cap \clhs{\tau} \neq \emptyset$}
			\STATE $\crhs{\tau} \leftarrow \crhs{\tau} - D$
		\ENDIF
	\ENDFOR
	\RETURN $\negsystem$
   \end{algorithmic}
%   \vspace{-1mm}
\end{algorithm}

The order in which the
attributes are selected and added to $M$ in {\tt elicit} is
arbitrary. Moreover, the order of rule application in {\tt push}
may be also changed. That is, we currently try to apply $Rule_1$ (line
21) first and $Rule_2$ (line 25) afterwards. However, one can apply the rules
in the opposite order. The same observation applies to $Rule_3(T, A, C_i)$ and
$Rule_3(T, C_i, A)$ (lines 30 and 34, respectively).  If the algorithm
is changed along those lines, the generated p-skyline relation may be
different. However, even if the p-skyline relation is different, it will still be a
maximal p-skyline relation favoring $G$ in $\oset$.  Note also that
due to the symmetry of $\parcomp$, the order of children nodes of a
$\parcomp$-node may be different in normalized p-skyline trees of
equivalent p-skyline relations. Hence, the order in which the leaf
nodes are stored in the normalized syntax tree of $\skyline_\prefset$
(line 2 of {\tt elicit}) also affects the resulting p-skyline
relation.

\tikzstyle{smallfont}=[font=\fontsize{8}{8}]
\tikzstyle{nonleaf}=[smallfont, draw=black,circle,inner sep=1pt]
\tikzstyle{unknown}=[smallfont]
\tikzstyle{leaf}=[smallfont, draw=black,inner sep=1pt]

  \begin{figure}[ht]
	\centering
        \subfigure[]{
	\label{pic:ex:constraints-0}
\begin{tabular}{|l|l|}
\hline $\tau_1: t_1 \not \succ t_3$ &
$\children{\Gamma_{\succ}}(\{\mbox{\tt make}\}) \not \supseteq
\{\mbox{\tt price}\}$ \\ \hline $\tau_2: t_2 \not \succ t_3$ &
$\children{\Gamma_{\succ}}(\{\mbox{\tt make,year}\}) \not \supseteq
\{\mbox{\tt price}\}$ \\ \hline $\tau_3: t_4 \not \succ t_3$ &
$\children{\Gamma_{\succ}}(\{\mbox{\tt make,year}\}) \not \supseteq
\{\mbox{\tt price}\}$ \\ \hline $\tau_4: t_5 \not \succ t_3$ &
$\children{\Gamma_{\succ}}(\{\mbox{\tt make}\}) \not \supseteq
\{\mbox{\tt price,year}\}$ \\ \hline
\end{tabular}
        }
	\subfigure[]{
	\label{pic:ex:discover-1}
	\begin{tikzpicture}[>=stealth, yscale=0.4, xscale=0.6]

		\node[nonleaf] (PAR) at (0,1.3) {$\parcomp$};
		\node[leaf]    (P) at (-1.5,0) {{\tt price}};
		\node[leaf]    (M) at (0,0) {{\tt make}};
		\node[leaf]    (Y) at ( 1.5,0) {{\tt year}};

		\draw[-] (PAR) -- (P);
		\draw[-] (PAR) -- (M);
		\draw[-] (PAR) -- (Y);
	\end{tikzpicture}
	}
	\subfigure[]{
	\label{pic:ex:discover-2}
	\begin{tikzpicture}[>=stealth, yscale=0.4, xscale=0.6]
		\node[nonleaf] (PAR) at (1,4.3) {$\parcompsymbol$};
		\node[nonleaf] (PRI) at (0,3) {$\pricompsymbol$};
		\node[leaf]    (Y) at (2,3) {{\tt year}};

		\node[leaf]    (P) at (-1,2) {{\tt price}};
		\node[leaf]    (M) at ( 1,2) {{\tt make}};

		\draw[-] (PAR) -- (PRI);
		\draw[-] (PAR) -- (Y);
		\draw[-] (PRI) -- (P);
		\draw[-] (PRI) -- (M);
	\end{tikzpicture}
	}
	\subfigure[]{
	\label{pic:ex:discover-3}
	\begin{tikzpicture}[>=stealth, yscale=0.4, xscale=0.6]
		\node[nonleaf] (PRI) at (-1,4.3) {$\pricompsymbol$};
		\node[nonleaf] (PAR) at (0,3) {$\parcompsymbol$};
		\node[leaf]    (P) at (-2,3) {{\tt price}};

		\node[leaf]    (M) at (-1,2) {{\tt make}};
		\node[leaf]    (Y) at ( 1,2) {{\tt year}};

		\draw[-] (PRI) -- (PAR);
		\draw[-] (PAR) -- (M);
		\draw[-] (PRI) -- (P);
		\draw[-] (PAR) -- (Y);
	\end{tikzpicture}
	}
	\subfigure[]{
	\label{pic:ex:discover-4}
	\begin{tikzpicture}[>=stealth, yscale=0.4, xscale=0.6]
		\node[nonleaf] (PRI) at (0,1.3) {$\pricompsymbol$};
		\node[leaf]    (P) at (-1.5,0) {{\tt price}};
		\node[leaf]    (Y) at ( 1.5,0) {{\tt year}};
		\node[leaf]    (M) at ( 0,0) {{\tt make}};

		\draw[-] (PRI) -- (Y);
		\draw[-] (PRI) -- (P);
		\draw[-] (PRI) -- (M);
	\end{tikzpicture}
	}
	\caption{Example \ref{ex:discover}}
	\label{pic:ex:discover}
\vspace{-6mm}
  \end{figure}
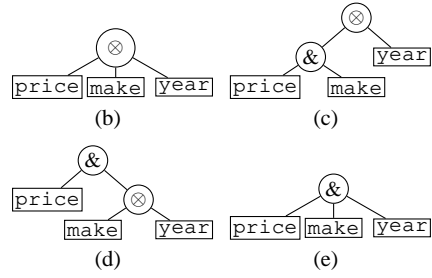

\begin{example}\label{ex:discover}
 Take $\oset$ and $\prefset$ from Example \ref{ex:existence}, and $G$
 from Example \ref{ex:negative}. Then the corresponding system of
 negative constraints $\negsystem = \negsystem(G, \oset)$ (Example
 \ref{ex:negative}) is shown in Figure \ref{pic:ex:constraints-0}.
Consider the attributes in the following order:
{\em make}, {\em price}, {\em year}.
Run {\tt elicit}. The tree $T$ (line 2) is shown in Figure
\ref{pic:ex:discover-1}. The initial value of $M$ is $\{\mbox{{\em
    make}}\}$. First, call {\em push}$(T,$ $\{${\em make}$\},$ $\mbox{{\em price}},
\negsystem)$. The parent of {\em price} is a $\parcompsymbol$-node
(Figure \ref{pic:ex:discover-1}), so we go to line 16 of {\tt push},
where $R$ is set to the $\parcompsymbol$-node (Figure
\ref{pic:ex:discover-1}). After $C_i$ is set to the node {\em make} in
line 17, we go to line 29 because it is a leaf node. The {\tt
  checkConstr} test in line 29 fails because $\negsystem$ prohibits
the edge $\mbox{({\em make},\ {\em price})}$. Hence, we go
to line 33 where the {\tt checkConstr} test succeeds. We apply
$Rule_3(T, $ $\mbox{{\em price}}, C_i)$, {\tt push} returns $true$, and
the resulting syntax tree $T$ is shown in Figure
\ref{pic:ex:discover-2}. Next time we call {\tt push}$(T, \{${\em make}$\},$
$\mbox{{\em price}}, \negsystem)$ in the line 6 of {\tt elicit}, we get
to the line 4 of {\tt push}. Since $\mbox{{\em year}} \not \in M$, we
immediately go to line 37 and return $false$.  In {\tt elicit} $M$
is set to $\{\mbox{{\em make}}, \mbox{{\em price}}\}$ and {\tt
push}$(T, \{${\em make}$,${\em price}$\}, \mbox{{\em year}},
\negsystem)$ is called. There we go to line 16 ($R$ is set to the
$\parcompsymbol$-node in Figure
\ref{pic:ex:discover-2}), $C_i$ is set to the $\pricompsymbol$-node
(Figure \ref{pic:ex:discover-2}), we apply $Rule_1(T, C_i, \mbox{{\em
    year}})$ (the resulting tree $T$ is shown in Figure
\ref{pic:ex:discover-3}), and $true$ is returned. When $push(T, \{${\em make}$,$
{\em price}$\},\mbox{{\em year}}, \negsystem)$ is called the next time, we first go to
line 16, $R$ is set to the $\parcompsymbol$-node (Figure
\ref{pic:ex:discover-3}), and $C_i$ to the node {\em make}. Then
$Rule_3(T, C_i, \mbox{{\em year}})$ is applied (line 30) resulting in
the tree $T$ shown in Figure \ref{pic:ex:discover-4}, and $true$ is
returned. Now {\tt push}$(T, \{${\em make}$, ${\em price}$\}, year, \negsystem)$ 
gets called once again from {\tt elicit} and returns $false$; and thus the tree in
Figure \ref{pic:ex:discover-4} is the final one. According to the
corresponding p-skyline relation, $t_3$ dominates all other tuples in
$\oset$.
\end{example}

The final p-skyline relation constructed in Example \ref{ex:discover}
is a prioritized accumulation of all the attribute preference
relations. This is because $\negsystem$ effectively contained only
one constraint (all constraints are implied by $\tau_2$, as shown below). When more constraints are involved, an elicited
p-skyline relation may also have occurrences of Pareto
accumulation.

\subsection{Reducing the size of systems of negative constraints}\label{sec:reduction}

As we showed in Theorem \ref{thm:main2-runtime}, the running time of the
function {\tt elicit} linearly depends on the size of the system of
negative constraints $\negsystem$.  If $\negsystem = \negsystem(G,
\oset)$, then $\negsystem$ contains $(|\oset|-1)\cdot |G|$
constraints.  A natural question which arises here is whether we
really need all the constraints in $\negsystem$ to elicit a maximal
p-skyline relation satisfying $\negsystem$. In particular, \emph{can
we replace $\negsystem$ with an equivalent subset of $\negsystem$?}

We define equivalence of systems of negative constraints in a
natural way.

\begin{definition}
 Given two systems of negative constraints $\negsystem_1$ and
 $\negsystem_2$, and two negative constraints $\tau_1$, $\tau_2$:
 \begin{itemize}
   \item $\negsystem_1$ (resp. $\tau_1$) \emph{implies} $\negsystem_2$
     (resp. $\tau_2$) iff every $\succ \in \formset_\prefset$
     satisfying $\negsystem_1$ (resp. $\tau_1$) also satisfies
     $\negsystem_2$ (resp. $\tau_2$);
   \item $\negsystem_1$ (resp. $\tau_1$) \emph{strictly implies}
     $\negsystem_2$ (resp. $\tau_2$) iff every $\succ \in
     \formset_\prefset$ satisfying $\negsystem_1$ (resp. $\tau_1$)
     also satisfies $\negsystem_2$ (resp. $\tau_2$), but
     $\negsystem_2$ (resp. $\tau_2$) does not imply $\negsystem_1$
     (resp. $\tau_1$);

   \item $\negsystem_1$ (resp. $\tau_1$) \emph{is equivalent} to
     $\negsystem_2$ (resp. $\tau_2$) iff $\negsystem_1$
     (resp. $\tau_1$) implies $\negsystem_2$ (resp. $\tau_2$) and vice
     versa.
  \end{itemize}
\end{definition}

In particular, a subset of $\negsystem(G, \oset)$ from Example
\ref{ex:discover} that is equivalent to $\negsystem(G, \oset)$ is
$\negsystem' = \{\tau_2\}$: first, $\negsystem'$ clearly implies
$\negsystem(G, \oset)$; second, $\{\tau_3\}$ is trivially implied by
$\{\tau_2\}$, $\{\tau_1\}$ is implied by $\{\tau_2\}$ (if {\em price}
is not a child of either {\em make} or {\em year}, it is not a
child of {\em make}), and $\{\tau_4\}$ is implied by $\{\tau_2\}$ (if
{\em price} is a child of neither {\em make} nor {\em year},
then both {\em price} and {\em year} cannot be  children of {\em make}).

Below we propose a number of methods for computing an equivalent 
subset of a system of negative constraints.

\subsubsection{Using $\skyline_{\prefset}(\oset)$ instead of $\oset$}
\label{sec:using-sky}

The first method of reducing the size of a system of negative constraints
is based on the following observation. Recall that each negative constraint
is used to show that a tuple should not be preferred to a superior
example. We also know that the relation $\skyline_\prefset$ is the
least p-skyline relation. By definition of the winnow operator, for
every $o' \in (\oset - \winnow_{\skyline_\prefset}(\oset))$ there is a
tuple $o \in \winnow_{\skyline_\prefset}(\oset)$ s.t. $o$ is preferred
to $o'$ according to $\skyline_\prefset$. Since $\skyline_\prefset$ is
the least  p-skyline relation, the same $o$ is preferred to $o'$
according to every p-skyline relation.  Thus, to guarantee favoring
$G$ in $\oset$, the system of negative constraints needs to contain
only the constraints showing that the tuples in
$\winnow_{\skyline_\prefset}(\oset)$ are not preferred to the superior
examples. Hence, the following proposition holds.

\begin{proposition}\label{prop:red-rule-1}
  Given $G \subseteq \winnow_{\skyline_\prefset}(\oset)$, $\negsystem(G,
  \oset)$ is \emph{equivalent} to 
   $\negsystem(G, \winnow_{\skyline_{\prefset}}(\oset))$.
\end{proposition}

Notice that $\negsystem(G, \winnow_{\skyline_{\prefset}}(\oset))$ contains
$(|\winnow_{\skyline_\prefset}(\oset)|-1) \cdot |G|$ negative
constraints. Proposition
\ref{prop:red-rule-1} also imply an important result: \emph{if a user
  considers a tuple $t$ superior based on the comparison with
  $\winnow_{\skyline_{\prefset}}(\oset)$, comparing $t$ with the tuples in $(\oset
  -\winnow_{\skyline_{\prefset}}(\oset))$ does not add any new
  information}.

\subsubsection{Removing redundant constraints}
\label{sec:rem-red}
The second method of reducing the size of a negative constraint system
is based on determining the implication of distinct negative
constraints in a system.  Let two $\tau_1, \tau_2 \in \negsystem$ be
such that $\clhs{\tau_2} \subseteq \clhs{\tau_1}$, $\crhs{\tau_1}
\subseteq \crhs{\tau_2}$. It is easy to check that $\tau_1$ implies
$\tau_2$.  Thus, the constraint $\tau_2$ is \emph{redundant} and may be
deleted from $\negsystem$.  This idea can also be expressed as
follows:
\begin{align*}
\tau \mbox{ implies }& \tau' \mbox{ iff } \clhs{\tau'}
\subseteq \clhs{\tau} \wedge (\attrset - \crhs{\tau'}) \subseteq
(\attrset - \crhs{\tau}).
\end{align*}

Let us represent $\tau$ as a bitmap representing $(\attrset -
\crhs{\tau})$ appended to a bitmap representing $\clhs{\tau}$. We
assume that a bit is set to $1$ iff the corresponding attribute is in
the corresponding set. Denote such a representation as $bitmap(\tau)$.
\begin{example}
  Let $\clhs{\tau} = \{A_1, A_3, A_5\}$, $\crhs{\tau} = \{A_2\}$,
  $\clhs{\tau'} = \{A_1, $ $A_5\}$, $\crhs{\tau'} = \{A_2, A_4\}$. Let
  $\attrset = \{A_1, \ldots, A_5\}$. As a result, $bit\-map(\tau) =
  10101\ 10111$ and $bitmap(\tau') = 10001\ 10101$.
\end{example}

Consider $bitmap(\tau)$ as a vector with $2\cdot |\attrset|$
dimensions. From the negative constraint implication rule, it follows
that $\tau$ strictly implies $\tau'$ iff $bitmap(\tau)$ and
$bitmap(\tau')$ satisfy the \emph{Pareto improvement principle}, i.e.,
the value of every dimension of $bitmap(\tau)$ is greater or
  equal to the corresponding value in $bitmap(\tau)$, and there is at
  least one dimension whose value in $bitmap(\tau)$ is greater than in
  $bitmap(\tau')$. Therefore, the set of all non-redundant
constraints in $\negsystem$ corresponds to the \emph{skyline} of the
set of bitmap representations of all constraints in
$\negsystem$. Moreover, $bitmap(\tau)$ can have only two values in
every dimension: $0$ or $1$.  Thus, algorithms for computing skylines over 
low-cardinality domains (e.g. \cite{DBLP:conf/vldb/MorsePJ07}) can be used
to compute the set of non-redundant constraints.

\subsubsection{Removing redundant sets of constraints}

The method of determining redundant constraints in the previous
section is based on distinct constraint implication. A more powerful
version of this method would compute and discard \emph{redundant
  subsets of $\negsystem$} rather then redundant distinct
constraints. However, as we show in this section, that problem appears
to be significantly harder.

\newcommand{\negsystimplsubset}{\texttt{SUBSET-EQUIV}\xspace}

\smallskip 
{\bf Problem \negsystimplsubset.}{\it\ Given systems of negative
constraints $\negsystem_1$ and $\negsystem_2$ s.t. $\negsystem_2
\subseteq \negsystem_1$, check if $\negsystem_2$ is equivalent to
$\negsystem_1$.}

To determine the complexity of \negsystimplsubset, we use a helper
problem.

\newcommand{\negsystimpl}{\texttt{NEG-SYST-IMPL}\xspace}
\newcommand{\negsystimplx}{\texttt{NEG-SYST- IMPL}\xspace}

{\bf Problem \negsystimpl.}{\it\ 
  Given two systems of negative constraints $\negsystem_1$ and
  $\negsystem_2$, check if $\negsystem_1$ implies $\negsystem_2$.
}

\smallskip 

It turns out that the problems \negsystimpl and \negsystimplsubset are
intractable in general.

\begin{theorem}\label{thm:negsystimpl}
  \negsystimpl is co-NP complete
\end{theorem}

\begin{theorem}\label{thm:negsystimplsubset}
  \negsystimplsubset is co-NP complete
\end{theorem}

We notice that even though the problem of minimizing the size of 
a system of negative constraints is intractable in general, the 
methods of reducing its size we proposed in sections 
\ref{sec:rem-red} and \ref{sec:using-sky} result in a  significant 
decrease in the size of the system. This is illustrated in 
Section \ref{sec:experiments}.

\vspace{-5mm}
\section{Experiments}\label{sec:experiments}

We have performed extensive experimental study of the proposed
framework. The algorithms were implemented in Java. The experiments
were run on Intel Core 2 Duo CPU 2.1 GHz with 2.0GB RAM under Windows
XP.  We used four data sets: one real-life and three synthetic.

\vspace{-3mm}
\subsection{Experiments with real-life data}\label{sec:p-skyline-real-exp}
\vspace{-2mm}

In this subsection, we focus on experimenting with the accuracy of the
{\tt elicit} algorithm and the reduction of winnow result size,
achieved by modeling user preferences using p-skyline relations.  We
use a data set $NHL$ which stores statistics of NHL players
\cite{nhl}, containing 9395 tuples.  We consider three sets of relevant
attributes $\attrset$ containing 12, 9, and 6 attributes. The
size of the corresponding skylines is 568, 114, and 33, respectively.

\vspace{-3mm}
\subsubsection{Precision and recall}
\vspace{-2mm}

The aim of the first experiment is to demonstrate that the
{\tt elicit} algorithm has high accuracy. We use the
following scenario. We assume that the real, hidden preferences of the user 
are modeled as a p-skyline relation  $\succ_{hid}$. We also assume
that the user provides the set of relevant attributes $\attrset$, the
set of corresponding attribute preferences $\prefset$, and a set $G_{hid}$ of
tuples which she likes most in $NHL$ (i.e., $G_{hid}$ are superior
examples and $G_{hid} \subseteq \winnow_{\succ_{hid}}(NHL)$). We use
$G_{hid}$ to construct a maximal p-skyline relation $\succ$ favoring
$G_{hid}$ in $NHL$. To measure the accuracy of {\tt elicit},
we compare the set of the best tuples
$\winnow_{\succ}(NHL)$  with the set of the best tuples
$\winnow_{\succ_{hid}}(NHL)$. The latter is supposed to correctly reflect
user preferences.

To model user preferences, we randomly generate 100 p-skyline
relations $\succ_{hid}$. For each $\winnow_{\succ_{hid}}(NHL)$, we
randomly pick 5 tuples from it, and use the tuples as superior examples
$G_{hid}$ to elicit three different maximal p-skyline relations
$\succ$ favoring $G_{hid}$ in $NHL$.  Out of those three relations,
we pick the one resulting in $\winnow_{\succ}(NHL)$ of the smallest
size.  Then we add 5 more tuples from $\winnow_{\succ_{hid}}(NHL)$ to
$G_{hid}$ and repeat the same procedure. We keep adding tuples to
$G_{hid}$ from $\winnow_{\succ_{hid}}(NHL)$ until $G_{hid}$ reaches
$\winnow_{\succ_{hid}}(NHL)$.

To measure the accuracy of the {\tt elicit} algorithm, we compute the
following three values: 
\begin{enumerate}
\item $precision$ of the p-skyline elicitation
method:
\[precision = \frac{|\winnow_{\succ}(NHL) \cap \winnow_{\succ_{hid}}(NHL)|}{|\winnow_{\succ}(NHL)|},\]
\item $recall$ of the p-skyline elicitation method:
\[recall = \frac{|\winnow_{\succ}(NHL) \cap \winnow_{\succ_{hid}}(NHL)|}{|\winnow_{\succ_{hid}}(NHL)|},\]
\item \emph{$F$-measure} which combines $precision$ and $recall$:
\[F = 2 \cdot \frac{precision\cdot recall}{precision + recall}\]
\end{enumerate}

We plot the average values of those measures in Figures
\ref{pic:real-life-accuracy-6}, \ref{pic:real-life-accuracy-9}, 
and \ref{pic:real-life-accuracy-12}.  As can be observed, $precision$ of the
{\tt elicit} algorithm is high in all experiments.  In
particular, it is greater than $0.9$ in most cases, regardless of the
number of superior examples and
the number of relevant attributes. At the same time, $recall$
starts from a low value when the
number of superior examples is low. This is justified by the fact that
{\tt elicit} constructs a \emph{maximal} relation
favoring $G_{hid}$ in $NHL$.  Thus, when $G_{hid}$ contains few
tuples, it is not sufficient to capture the  preference
relation $\succ_{hid}$, and thus the ratio of false negatives is
rather high. However, when we increase the number of superior
examples, $recall$ consistently grows.

\pgfcreateplotcyclelist{\mycyclelist}{%
    {black,mark=*},
    {black,mark=square},
    {black,mark=o},
    {mark=+},
    {mark=diamond},
    {mark=triangle},
    {mark=diamond*}
}

\newcommand{\figscale}{0.8}
\newcommand{\axisdef}{height=5cm,width=5.5cm, 
		legend columns=3,
		every axis legend/.append style={
	    	at={(0.5,1.03)},
    		anchor=south,
		},
                }

\pgfplotsset{every axis label/.append style={font=\normalsize}}

	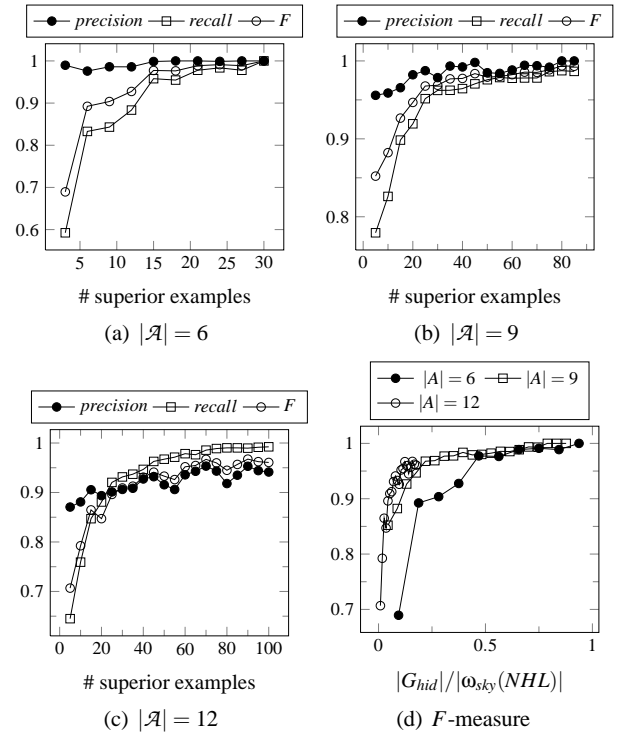
\begin{figure}[ht]
		\centering
          \subfigure[$|\attrset| = 6$]{
	  \label{pic:real-life-accuracy-6}
            \begin{tikzpicture}[scale=\figscale]
              \begin{axis}[
                \axisdef,
		xlabel={\# superior examples},
		cycle list name=\mycyclelist,
		xtick={5,10,...,30},
		ytick={0.5,0.6,0.7,0.8,0.9,1.0},
                ]
                
                \addplot coordinates {
		( 3, 0.989767688451899)
		( 6, 0.97592932429889)
		( 9, 0.986078758782125)
		( 12, 0.98598417137585)
		( 15, 0.998628257887517)
		( 18, 1)
		( 21, 1)
		( 24, 0.998960498960499)
		( 27, 1)
		( 30, 1)
                };
                
                \addplot coordinates {
		( 3, 0.592429246113677)
		( 6, 0.832842702616515)
		( 9, 0.843063019343894)
		( 12, 0.883424257117249)
		( 15, 0.957873478572403)
		( 18, 0.954595313089937)
		( 21, 0.977886139176462)
		( 24, 0.983567144857467)
		( 27, 0.977855477855478)
		( 30, 1)
                };
                
                \addplot coordinates {
		( 3, 0.689433308812614)
		( 6, 0.89237059320009)
		( 9, 0.903704132968351)
		( 12, 0.927557433296869)
		( 15, 0.977387405820013)
		( 18, 0.976381011895954)
		( 21, 0.988765379795383)
		( 24, 0.991126784376543)
		( 27, 0.988757396449704)
		( 30, 1)
                };
                \legend{$precision$, $recall$, $F$};
              \end{axis}
            \end{tikzpicture}
          }
          \subfigure[$|\attrset| = 9$]{
	  \label{pic:real-life-accuracy-9}
            \begin{tikzpicture}[scale=\figscale]              
              \begin{axis}[
                \axisdef ,
		xlabel={\# superior examples},
		cycle list name=\mycyclelist,
		xtick={0,20,...,80},
		ytick={0.5,0.6,0.7,0.8,0.9,1.0},
		minor tick num = 1
                ]

                \addplot coordinates {
		( 5, 0.955850263448606)
		( 10, 0.958801710630136)
		( 15, 0.96561231235404)
		( 20, 0.982203779407812)
		( 25, 0.987473516120962)
		( 30, 0.978591432186879)
		( 35, 0.993411424924932)
		( 40, 0.992270404075888)
		( 45, 0.998043818466354)
		( 50, 0.984896693756538)
		( 55, 0.984007529072976)
		( 60, 0.988363628002344)
		( 65, 0.994407894736842)
		( 70, 0.993829401088929)
		( 75, 0.992003886119794)
		( 80, 1)
		( 85, 1)
                };
                
                \addplot coordinates {
		( 5, 0.779444462338276)
		( 10, 0.826104586641321)
		( 15, 0.898401982607758)
		( 20, 0.919309112108549)
		( 25, 0.951539684601221)
		( 30, 0.962180907955835)
		( 35, 0.962335983462743)
		( 40, 0.96444902760955)
		( 45, 0.970569658812083)
		( 50, 0.975861101859755)
		( 55, 0.978672456620002)
		( 60, 0.977752996829996)
		( 65, 0.978108127605479)
		( 70, 0.978116976959065)
		( 75, 0.986159450243813)
		( 80, 0.987457797264134)
		( 85, 0.986578596022767)
                };
                
                \addplot coordinates {
		( 5, 0.852147394446494)
		( 10, 0.882281248264226)
		( 15, 0.926609078033304)
		( 20, 0.946744778087616)
		( 25, 0.967756931757355)
		( 30, 0.968799281595619)
		( 35, 0.976953416694923)
		( 40, 0.977513802185021)
		( 45, 0.983689596735891)
		( 50, 0.979191093064782)
		( 55, 0.980090372839218)
		( 60, 0.982339884966312)
		( 65, 0.985600810162402)
		( 70, 0.985273089507593)
		( 75, 0.988601420872981)
		( 80, 0.993592786744794)
		( 85, 0.993144613797799)
                };
                \legend{$precision$, $recall$, $F$};
              \end{axis}
            \end{tikzpicture}
          }
          \subfigure[$|\attrset| = 12$]{
	  \label{pic:real-life-accuracy-12}
            \begin{tikzpicture}[scale=\figscale]              
              \begin{axis}[
                \axisdef ,
		xtick={0,20,...,100},
		minor tick num = 1,
		xlabel={\# superior examples},
		cycle list name=\mycyclelist,
                ]

                \addplot coordinates {
		( 5, 0.870574918420044)
		( 10, 0.880946828141015)
		( 15, 0.905526882687219)
		( 20, 0.893635135887654)
		( 25, 0.901202734173826)
		( 30, 0.906199098512529)
		( 35, 0.908541560105889)
		( 40, 0.927318122206667)
		( 45, 0.931871505577833)
		( 50, 0.915573233361804)
		( 55, 0.906032206028987)
		( 60, 0.93565080492885)
		( 65, 0.942851288971352)
		( 70, 0.953303055323547)
		( 75, 0.943285519466036)
		( 80, 0.918072977005943)
		( 85, 0.934968022208866)
		( 90, 0.952998437292273)
		( 95, 0.944181420022897)
		( 100, 0.941373093672142)
                };
                
                \addplot coordinates {
		( 5, 0.644964279378226)
		( 10, 0.759227687349238)
		( 15, 0.847053836609492)
		( 20, 0.880775786774398)
		( 25, 0.920380117047566)
		( 30, 0.931436951295358)
		( 35, 0.936978301734853)
		( 40, 0.946659980482126)
		( 45, 0.962786756222681)
		( 50, 0.967293561423631)
		( 55, 0.971348544392576)
		( 60, 0.978436123101012)
		( 65, 0.976362332225705)
		( 70, 0.985566404743368)
		( 75, 0.988783549503386)
		( 80, 0.990185263942176)
		( 85, 0.990153067194495)
		( 90, 0.989622171999728)
		( 95, 0.991420596921554)
		( 100, 0.992295734188511)
                };
                
                \addplot coordinates {
		( 5, 0.706739180227526)
		( 10, 0.7923567646827)
		( 15, 0.864777963383453)
		( 20, 0.84709502412824)
		( 25, 0.896082662920981)
		( 30, 0.909129624678142)
		( 35, 0.912753042907264)
		( 40, 0.930989618363204)
		( 45, 0.941061150730116)
		( 50, 0.933018537509292)
		( 55, 0.926073817717918)
		( 60, 0.951922909541035)
		( 65, 0.954536974615323)
		( 70, 0.966990956950824)
		( 75, 0.959795312544167)
		( 80, 0.944905713391191)
		( 85, 0.956711138937735)
		( 90, 0.967432973490417)
		( 95, 0.962306512154168)
		( 100, 0.96072104561765)
                };
                \legend{$precision$, $recall$, $F$};
              \end{axis}
            \end{tikzpicture}
          }
        \subfigure[$F$-measure]{
            \label{pic:f-measure}
            \begin{tikzpicture}[scale=\figscale]              
              \begin{axis}[
                \axisdef ,
		legend columns=2,
		minor tick num = 1,
		width=5.5cm,
		xlabel={$|G_{hid}|/|\winnow_{\skyline}(NHL)|$},
		cycle list name=\mycyclelist,
                ]

                \addplot coordinates {

 		( 0.09375, 0.689433308812614)
		( 0.1875, 0.89237059320009)
		( 0.28125, 0.903704132968351)
		( 0.375, 0.927557433296869)
		( 0.46875, 0.977387405820013)
		( 0.5625, 0.976381011895954)
		( 0.65625, 0.988765379795383)
		( 0.75, 0.991126784376543)
		( 0.84375, 0.988757396449704)
		( 0.9375, 1)
                };
                
                \addplot coordinates {
		( 0.043859649122807, 0.852147394446494)
		( 0.087719298245614, 0.882281248264226)
		( 0.131578947368421, 0.926609078033304)
		( 0.175438596491228, 0.946744778087616)
		( 0.219298245614035, 0.967756931757355)
		( 0.263157894736842, 0.968799281595619)
		( 0.307017543859649, 0.976953416694923)
		( 0.350877192982456, 0.977513802185021)
		( 0.394736842105263, 0.983689596735891)
		( 0.43859649122807, 0.979191093064782)
		( 0.482456140350877, 0.980090372839218)
		( 0.526315789473684, 0.982339884966312)
		( 0.570175438596491, 0.985600810162402)
		( 0.614035087719298, 0.985273089507593)
		( 0.657894736842105, 0.988601420872981)
		( 0.701754385964912, 0.993592786744794)
		( 0.745614035087719, 0.993144613797799)
		( 0.789473684210526, 1)
		( 0.833333333333333, 1)
		( 0.87719298245614, 1)
                };
                
                \addplot coordinates {
		( 0.00880281690140845, 0.706739180227526)
		( 0.0176056338028169, 0.7923567646827)
		( 0.0264084507042254, 0.864777963383453)
		( 0.0352112676056338, 0.84709502412824)
		( 0.0440140845070423, 0.896082662920981)
		( 0.0528169014084507, 0.909129624678142)
		( 0.0616197183098592, 0.912753042907264)
		( 0.0704225352112676, 0.930989618363204)
		( 0.0792253521126761, 0.941061150730116)
		( 0.0880281690140845, 0.933018537509292)
		( 0.096830985915493, 0.926073817717918)
		( 0.105633802816901, 0.951922909541035)
		( 0.11443661971831, 0.954536974615323)
		( 0.123239436619718, 0.966990956950824)
		( 0.132042253521127, 0.959795312544167)
		( 0.140845070422535, 0.944905713391191)
		( 0.149647887323944, 0.956711138937735)
		( 0.158450704225352, 0.967432973490417)
		( 0.167253521126761, 0.962306512154168)
		( 0.176056338028169, 0.96072104561765)
                };
                \legend{$|A|=6$, $|A|=9$, $|A|=12$};
              \end{axis}
            \end{tikzpicture}
          }
\vspace{-3mm}
          \caption{Accuracy of p-skyline elicitation}
	  \label{pic:real-life-accuracy}
\vspace{-4mm}
	\end{figure}

In Figure \ref{pic:f-measure}, we plot the values of the $F$-measure with
respect to the share of the skyline used as superior examples.
As one can observe, the value of $F$
starts from a comparatively low value of $0.7$ but quickly reaches
$0.9$ via a small increase of the size of $G_{hid}$. Another important
observation is that the value of $F$ is generally inversely dependent
on the number of relevant attributes (given the same ratio of superior
examples used). This is justified by the following observation. To
construct a p-skyline relation favoring $G_{hid}$ in $NHL$, the
algorithm uses a set of negative constraints $\negsystem$.
Intuitively, the constructed p-skyline relation $\succ$ will match the
original relation $\succ_{hid}$ better if the set $\negsystem$
captures $\succ_{hid}$ sufficiently well. The number of
constraints in $\negsystem$ depends not only on the number of superior
examples but also on the 
skyline size. Since skyline sizes are generally smaller for smaller sets of
$\attrset$, more superior examples are needed for smaller $\attrset$ to
capture  $\succ_{hid}$.

\vspace{-2mm}
\subsubsection{Winnow result size}
\vspace{-2mm}

In Section \ref{sec:intro}, we discussed a well known deficiency of
the skyline framework: skylines are generally of large size for
large sets of relevant attributes $\attrset$.  The goal of the experiments in
this section is twofold. First, we demonstrate that using p-skyline
relations to model user preferences results in
\emph{smaller winnow query results} in comparison to skyline relations.
Second, we show that the reduction of query result size is significant
if the \emph{hidden} user preference relation is a p-skyline relation.
In particular, we show that it is generally hard to find a p-skyline
relation favoring \emph{an arbitrary} subset of the
skyline.

\begin{figure}[ht]
		\vspace{-2mm}
		\centering
          \subfigure[$|\attrset| = 6$]{
            \begin{tikzpicture}[scale=\figscale]
              \begin{axis}[
                \axisdef ,
		minor y tick num = 1,
		ylabel={$winnow\mbox{-}size\mbox{-}ratio$},
		xlabel={\# superior examples},
		cycle list name=\mycyclelist,
		xtick={5,10,...,30},
		width=5.8cm
                ]
                
                \addplot coordinates {
		( 3 , 0.30893349075167256970 )
		( 6 , 0.57531840140535792727 )
		( 9 , 0.62023460410557184848 )
		( 12 , 0.68077324973876697879 )
		( 15 , 0.77104377104377104242 )
		( 18 , 0.80303030303030303030 )
		( 21 , 0.89107289107289107273 )
		( 24 , 0.89680589680589680606 )
		( 27 , 0.95221445221445221515 )
		( 30 , 0.99724517906336088182 )
                };
		\addplot coordinates{
		  (3, 0.515151515151515)
		  (6, 0.742424242424242)
		  (9, 0.818181818181818)
		  (12, 0.901515151515152)
		  (15, 1)
		  (18, 0.992424242424242)
		  (21, 1)
		  (24, 1)
		  (27, 1)
		  (30, 1)
		};
                \legend{$G_{hid}$,$G_{rand}$};
              \end{axis}
            \end{tikzpicture}
          }
          \subfigure[$|\attrset| = 9$]{
            \begin{tikzpicture}[scale=\figscale]              
              \begin{axis}[
                \axisdef ,
		ylabel={$winnow\mbox{-}size\mbox{-}ratio$},
		xlabel={\# superior examples},
		cycle list name=\mycyclelist,
		xtick={0,20,...,80},
		minor tick num = 1,
		width=5.8cm
                ]

                \addplot coordinates {
		( 5 , 0.38332131699110790702 )
		( 10 , 0.42626493770658530351 )
		( 15 , 0.49218750000000000000 )
		( 20 , 0.53735632183908045965 )
		( 25 , 0.57063711911357340702 )
		( 30 , 0.63087719298245614035 )
		( 35 , 0.65179252479023646053 )
		( 40 , 0.69683354728284124912 )
		( 45 , 0.74610136452241715439 )
		( 50 , 0.78018575851393188860 )
		( 55 , 0.79399255715045188772 )
		( 60 , 0.79851973684210526316 )
		( 65 , 0.79495614035087719298 )
		( 70 , 0.81790683605565638246 )
		( 75 , 0.83427318295739348333 )
		( 80 , 0.84143049932523616754 )
		( 85 , 0.86925647451963241404 )
                };
		\addplot coordinates{
		  (5, 0.68640350877193)
		  (10, 0.802631578947368)
		  (15, 0.899122807017544)
		  (20, 0.881578947368421)
		  (25, 0.903508771929824)
		  (30, 0.932017543859649)
		  (35, 0.93859649122807)
		  (40, 0.969298245614035)
		  (45, 0.969298245614035)
		  (50, 0.969298245614035)
		  (55, 0.986842105263158)
		  (60, 1)
		  (65, 1)
		  (70, 1)
		  (75, 1)
		  (80, 1)
		};
                \legend{$G_{hid}$,$G_{rand}$};
              \end{axis}
            \end{tikzpicture}
          }
          \subfigure[$|\attrset| = 12$]{
            \begin{tikzpicture}[scale=\figscale]              
              \begin{axis}[
                \axisdef ,
		ylabel={$winnow\mbox{-}size\mbox{-}ratio$},
		xlabel={\# superior examples},
		cycle list name=\mycyclelist,
		xtick={0,20,...,100},
		minor tick num = 1,
		width=5.8cm
                ]

                \addplot coordinates {
		( 5 , 0.21852766081022215775 )
		( 10 , 0.30709827144686299613 )
		( 15 , 0.36500933310707619208 )
		( 20 , 0.41807808878587983592 )
		( 25 , 0.42996117732033224982 )
		( 30 , 0.48174965284665740933 )
		( 35 , 0.49127372933251684014 )
		( 40 , 0.51072104267395417271 )
		( 45 , 0.51642316586083666180 )
		( 50 , 0.53134336320956039261 )
		( 60 , 0.55513637379834562940 )
		( 70 , 0.56306280304779496655 )
		( 80 , 0.59735915492957746479 )
		( 90 , 0.62445582586427656849 )
		( 100 , 0.63601982263954094947 )
                };

                \addplot coordinates {
		  (5, 0.43075117370892)
		  (10, 0.580985915492958)
		  (15, 0.631455399061033)
		  (20, 0.751173708920188)
		  (25, 0.842136150234742)
		  (30, 0.882629107981221)
		  (35, 0.865023474178404)
		  (40, 0.916079812206573)
		  (45, 0.975352112676056)
		  (50, 0.983714788732394)
		  (60, 0.984154929577465)
		  (70, 0.981807511737089)
		  (80, 0.984154929577465)
		  (90, 0.991512646130105)
		  (100, 0.997831681359931)

                };

                \legend{$G_{hid}$,$G_{rand}$};
              \end{axis}
            \end{tikzpicture}
          }
% \vspace{-3mm}
          \caption{p-skyline size reduction}
	  \label{pic:real-life-size-reduction}
% \vspace{-5mm}
	\end{figure}
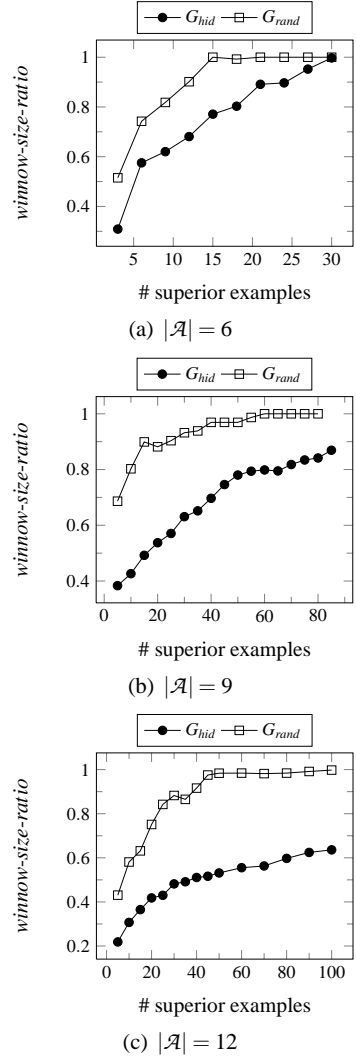

In this experiment, sets of superior examples are generated using two
methods.  First, they are drawn randomly from the set of the best
objects $\winnow_{\succ_{hid}}(NHL)$ according to a hidden p-skyline
relation $\succ_{hid}$, as in the previous 
experiment and denoted $G_{hid}$. Second, they are drawn randomly from
the  skyline $\winnow_{\skyline}(NHL)$ and 
denoted $G_{rand}$. Notice that $G_{rand}$ may not be favored by any
p-skyline relation (besides $\skyline_\prefset$, of course).  We use
these sets to elicit p-skyline relations $\succ$ that favor them. In
Figure \ref{pic:real-life-size-reduction}, we plot
\[winnow\mbox{-}size\mbox{-}ratio = \frac{|\winnow_{\succ}(NHL)|}{|\winnow_{\skyline_\prefset}(NHL)|},\]
which shows the difference in the size of the results of p-skyline and
skyline queries.

Consider the graphs for $G_{hid}$. As the figures suggest, using
p-skyline relations to model user preferences results in a significant
reduction in the size of winnow query result, in comparison to skyline
relations.  It can be observed that using larger sets of relevant
attributes $\attrset$ generally results in smaller values of
$winnow\mbox{-}size\mbox{-}ratio$.  Moreover, for larger relevant
attribute sets, {\it winnow-size-ratio} grows slowly. That is due to
larger skyline size for such sets.

Another important observation is that
$winnow\mbox{-}size\mbox{-}ratio$ is always smaller for superior
examples which correspond to p-skyline relations ($G_{hid}$), in
comparison to superior examples drawn randomly ($G_{rand}$) from the
skyline. The fact that superior examples correspond to a real
p-skyline relation implies that they share some similarity expressed
using the attribute importance relationships.  For a set
of random skyline tuples $G_{rand}$, such similarity exists when it
contains only a few tuples.  Increasing the size of such a set
decreases the similarity of the tuples, which results in a quick
growth of $winnow\mbox{-}size\mbox{-}ratio$.

% Moreover, p-skyline query results are smaller 
% when superior examples are similar (i.e., $G_{hid}$) rather then arbitrary (i.e., $G_{rand}$).

\vspace{-10mm}
\subsection{Experiments with synthetic data}

Here we present experiments with synthetic data. The main goals of the
experiments is to demonstrate that the proposed p-skyline relation
elicitation approach is scalable and allows effective optimizations.  We use
three synthetic data sets here: correlated $S_1$, anti-correlated
$S_2$, and uniform $S_3$. Each of them contains 50000 tuples. We use
three different sets $\attrset$ of 10, 15, and 20 relevant attributes.
For each of those sets, we pick a different set of superior examples
$G$. Sets $G$ are constructed of \emph{similar} tuples,
similarity being measured as Euclidean distance. As before, given a set $G$, we
use {\tt elicit} to construct maximal p-skyline relations $\succ$
favoring $G$. This setup is supposed to model an automated process of
identifying superior objects $G$, in which a user is involved only
indirectly.

\vspace{-2mm}
\subsubsection{Scalability}
\vspace{-2mm}

\begin{figure}[ht]
\centering
\subfigure[against \# superior examples]{
	\begin{tikzpicture}[scale=\figscale]
	\pgfplotsset{every axis legend/.append style={
	    	at={(0.5,1.03)},
    		anchor=south,
		},
		every axis/.append style={
		font=\normalsize
		}
	}

     	\begin{semilogyaxis}[
        	height=5cm,
		width=5.8cm,
		cycle list name=\mycyclelist,
		ylabel={ms},
		xlabel={\# superior examples},
		legend columns=2
      	]

\addplot coordinates {
		  (1, 52)
		  (5, 211)
		  (10, 565)
		  (20, 751)
		  (25, 521)
		  (30, 616)
		  (40, 602)
		  (50, 557)
		  (75, 474)
		  (100, 604)
		  (125, 482)
		  (150, 641)
};

\addplot coordinates {
		  (1, 12)
		  (5, 31)
		  (10, 165)
		  (20, 128)
		  (25, 114)
		  (30, 107)
		  (40, 138)
		  (50, 147)
		  (75, 159)
		  (100, 207)
		  (125, 187)
		  (150, 228)
};

\addplot coordinates {
		  (1, 1.051466)
		  (5, 1.296051)
		  (10, 1.139188)
		  (20, 1.008508)
		  (25, 1.634286)
		  (50, 2.065473)
		  (75, 1.727594)
		  (100, 1.703289)
		  (125, 2.338844)
		  (150, 2.124572)
};
	\legend{anticorr, uniform, corr};
	     \end{semilogyaxis}
	\label{pic:exp-synthetic-performance-examples}
	\end{tikzpicture}
}

\subfigure[against data set size]{
	\begin{tikzpicture}[scale=\figscale]
	\pgfplotsset{every axis legend/.append style={
	    	at={(0.5,1.03)},
    		anchor=south,
		},
		every axis/.append style={
		font=\normalsize
		}
	}

     	\begin{semilogyaxis}[
        	height=5cm,
		width=4cm,
		ylabel={ms},
		cycle list name=\mycyclelist,
		xlabel={data set size},
		legend columns=1
      	]

\addplot coordinates {
(10000, 249.0364560000000000)
(25000, 319.7849170000000000)
(40000, 492.1312625000000000)
(50000, 557.1746586666666700)
};

\addplot coordinates {
(10000, 104)
(25000, 117)
(40000, 139)
(50000, 147)
};

\addplot coordinates {
(10000, 2.138667)
(25000, 3.16287)
(40000, 3.337067)
(50000, 3.565473)
};

\legend{anticorr, uniform, corr};
	     \end{semilogyaxis}
		\label{pic:exp-synthetic-performance-dataset}
	\end{tikzpicture}
}
\subfigure[against $|\attrset|$]{
	\begin{tikzpicture}[scale=\figscale]
	\pgfplotsset{every axis legend/.append style={
	    	at={(0.5,1.03)},
    		anchor=south,
		},
		every axis/.append style={
		font=\normalsize
		}
	}

     	\begin{semilogyaxis}[
        	height=5cm,
		width=4cm,
		ylabel={ms},
		cycle list name=\mycyclelist,
		xlabel={$|\attrset|$},
		legend columns=1
      	]

\addplot coordinates {
	(10, 42.32190475000000000000)
	(15, 201.1510678750000000)
	(20, 503.3810274444444400)
};

\addplot coordinates {
	(10, 1.423219047500000000)
	(15, 67.1510678750000000)
	(20, 147)
};

\addplot coordinates {
	(10, 0.262883)
	(15, 0.578006)
	(20, 3.065473)
};

\legend{anticorr, uniform, corr};
	     \end{semilogyaxis}
		\label{pic:exp-synthetic-performance-attrcount}
	\end{tikzpicture}
}
% \vspace{-4mm}
\caption{Performance of p-skyline elicitation}
\label{pic:exp-synthetic-performance}
\vspace{-5mm}
\end{figure}
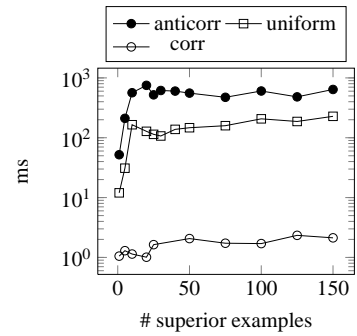
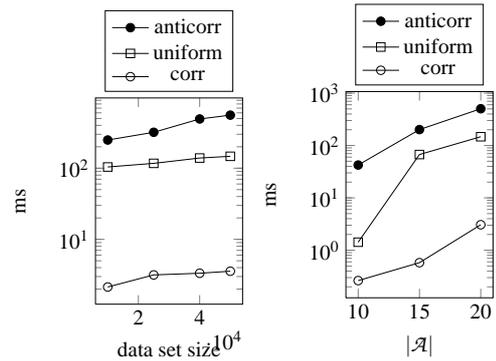

In this section, we show that the {\tt elicit} algorithm is scalable
with respect to various parameters. In Figure
\ref{pic:exp-synthetic-performance}, we plot the dependence of the average
running time of {\tt disco\-ver} on the number of superior examples
$|G|$ used to elicit a p-skyline relation (Figure
\ref{pic:exp-synthetic-performance-examples}, $|S_i| = 50000$,
$|\attrset| = 20$), the size of $S_i$ for $i=1,\ldots,3$ (Figure
\ref{pic:exp-synthetic-performance-dataset}, $|G| = 50$, $|\attrset| =
20$), and the number $|\attrset|$ of relevant attributes (Figure
\ref{pic:exp-synthetic-performance-attrcount}, $|S_i| = 50000$, $|G|
= 50$).  The measured time does not include the time to construct the
system of negative constraints and find the non-redundant constraints
in it.  According to our experiments, the preprocessing time
predominantly depends on the performance of the skyline computation
algorithm.

According to Figure \ref{pic:exp-synthetic-performance-examples}, the
running time of the algorithm increases until the size of $G$ reaches
$30$. After that, it does not vary much. This is due to the fact that
the algorithm performance depends on the number of negative
constraints used.  We use only \emph{non-redundant} constraints for
elicitation. As we show further (Figure
\ref{pic:exp-synthetic-data-exps-constr}), the dependence of the size
of a system of non-redundant constraints on the number of superior
examples has a pattern similar to Figure
\ref{pic:exp-synthetic-performance-examples}.

The growth of the running time with the increase in the data set size
(Figure \ref{pic:exp-synthetic-performance-dataset}) is justified by
the fact that the number of negative constraints depends on skyline
size (Section \ref{sec:reduction}).  For the data sets used in the
experiment, the skyline size grows with the size of the 
data set.  Similarly, the running time of the algorithm grows with
the number of relevant attributes (Figure
\ref{pic:exp-synthetic-performance-attrcount}), due to the increase in
the skyline size.

We conclude that the {\tt elicit} algorithm is efficient
and its running time scales well with respect to the number of
superior examples, the size of the data set, and the number of
relevant attributes used.

\vspace{-2mm}
\subsubsection{Reduction in the number of negative constraints}
\vspace{-2mm}

In this section, we demonstrate that the algorithm {\tt elicit} allows
effective optimizations. Recall that the running time of {\tt elicit}
depends linearly (Theorem \ref{thm:main2-runtime}) on the number of
negative constraints in the system $\negsystem$. Here we show that the
techniques proposed in Section
\ref{sec:reduction} result in a significant reduction in the size of
$\negsystem$.

In Figure \ref{pic:exp-synthetic-data-exps-constr}, we show how the
number of negative constraints depends on the number of superior
examples used to construct them. For every data set, we plot two
values: the number of \emph{unique} negative constraints in
$\negsystem(G, \winnow_{\skyline_\prefset}(S_i))$  for $i=1,\ldots,3$, and the number of
\emph{unique non-redundant} constraints in the corresponding system.  
We note that the reduction in the number of constraints achieved using
the methods we proposed in Section \ref{sec:reduction} is
significant. In particular, for the anti-correlated data set and $G$
of size 150, the total number of constraints in $\negsystem(G, S_i)$
is approximately $7.5\cdot 10^6$. Among them, about $5.5 \cdot 10^6$
are unique in $\negsystem(G,$
$\winnow_{\skyline_\prefset}(S_i))$. However, less than $1\%$ of them
(about $12\cdot 10^3$) are non-redundant.

\vspace{-2mm}
\subsubsection{Winnow result size}
\vspace{-2mm}

\begin{figure}[ht]
\vspace{-3mm}
\centering
	\subfigure[Constraint number reduction]{
	\begin{tikzpicture}[scale=\figscale]
	\pgfplotsset{every axis legend/.append style={
	    	at={(0.5,1.03)},
    		anchor=south,
		},
		every axis/.append style={
		font=\normalsize
		}
	}

     	\begin{semilogyaxis}[
        	height=5cm,
          	width=6cm,
		cycle list name=\mycyclelist,
		xlabel={\# superior examples},
		legend columns=2
      	]

\addplot coordinates {
		  (1, 31231.000000000000)
		  (5, 174556.750000000000)
		  (10, 362938.666666666667)
		  (20, 737821.333333333333)
		  (25, 909672.000000000000)
		  (30, 1091916.000000000000)
		  (40, 1473008.000000000000)
		  (50, 1826106.333333333333)
		  (75, 2757558.500000000000)
		  (100, 3694486.500000000000)
		  (125, 4570431.000000000000)
		  (150, 5500399.500000000000)
};

\addplot coordinates {
		  (1, 2233.0000000000000000)
		  (5, 8128.0000000000000000)
		  (10, 14895.333333333333)
		  (20, 15540.000000000000)
		  (25, 7958.6666666666666667)
		  (30, 18180.500000000000)
		  (40, 12911.5000000000000000)
		  (50, 13675.000000000000)
		  (75, 12338.0000000000000000)
		  (100, 12321.5000000000000000)
		  (125, 13062.0000000000000000)
		  (150, 12134.5000000000000000)
};

\addplot coordinates {
		  (1, 25513.000000000000)
		  (5, 126363)
		  (10, 232938)
		  (20, 403641)
		  (25, 453136)
		  (30,  536310)
		  (40,  764990)
		  (50, 1283113)
		  (75, 1736944)
		  (100, 2436367)
		  (125, 2896583)
		  (150, 3286730)
};

\addplot coordinates {
		  (1, 1660)
		  (5, 6837)
		  (10, 10895)
		  (20, 11170)
		  (25, 9921)
		  (30, 14560)
		  (40, 14357)
		  (50, 15526)
		  (75, 14683)
		  (100, 13762)
		  (125, 11008)
		  (150, 10573)
};

\addplot coordinates {
		 (1, 9942)
		 (5, 62687)
		 (10, 129719)
		 (20, 237100)
		 (25, 284266)
		 (50, 401779)
		 (75, 465588)
		 (100, 508741)
		 (125, 525239)
		 (150, 535364)
};

\addplot coordinates {
		 (1, 307)
		 (5, 104)
		 (10, 136)
		 (20, 175)
		 (25, 139)
		 (50, 188)
		 (75, 64)
		 (100, 72)
		 (125, 80)
		 (150, 83)
};
\legend{$anticorr\mbox{-}un$, 
        $anticorr\mbox{-}nrd$, 
        $uniform\mbox{-}un$, 
        $uniform\mbox{-}nrd$, 
        $corr\mbox{-}un$, 
        $corr\mbox{-}nrd$};

\end{semilogyaxis}
	\end{tikzpicture}
	\label{pic:exp-synthetic-data-exps-constr}
}
\hspace{1cm}
	\subfigure[p-skyline size reduction]{
	\begin{tikzpicture}[scale=\figscale]
	\pgfplotsset{every axis legend/.append style={
	    	at={(0.5,1.03)},
    		anchor=south,
		},
		every axis/.append style={
		font=\normalsize
		}
	}

     	\begin{semilogyaxis}[
        	height=5cm,
          	width=6cm,
		cycle list name=\mycyclelist,
		ylabel={$winnow\mbox{-}size\mbox{-}ratio$},
		xlabel={\# superior examples},
		legend columns=1
      	]

	\addplot coordinates{
                (5,0.00481829513855595)
                (10,0.00572921660753668)
                (20,0.00707162719340301)
                (25,0.00563333013711765)
                (30,0.00711957042861252)
                (40,0.00786269057435996)
                (50,0.00922907277783105)
                (75,0.00877361204334069)
                (100,0.0076229743983124)
                (125,0.010211909099626)
                (150,0.0101879374820213)
	};

	\addplot coordinates{
                (5,0.00624003889894379)
                (10,0.00726653880439774)
                (20,0.00880628866257868)
                (25,0.00861719657473189)
                (30,0.0125611172641076)
                (40,0.0130203409060212)
                (50,0.0154515248926227)
                (75,0.0151543801831492)
                (100,0.0193414192711851)
                (125,0.022582997919987)
                (150,0.0273373132715633)
	};

	\addplot coordinates{
                (5,0.00362960339943343)
                (10,0.0116265344664778)
                (20,0.0402502360717658)
                (25,0.0769298866855524)
                (30,0.131698536355052)
                (40,0.150554768649669)
                (50,0.131728045325779)
                (75,0.465976156751652)
                (100,0.660912417374882)
                (125,0.566336166194523)
                (150,0.753186968838527)
	};
\legend{anticorr, uniform, corr};
	     \end{semilogyaxis}
	\end{tikzpicture}
	\label{pic:exp-synthetic-data-exps-skyline}
	}
\vspace{-3mm}
\caption{Synthetic data experiments}
\label{pic:exp-synthetic-data-exps}
\vspace{-3mm}
\end{figure}
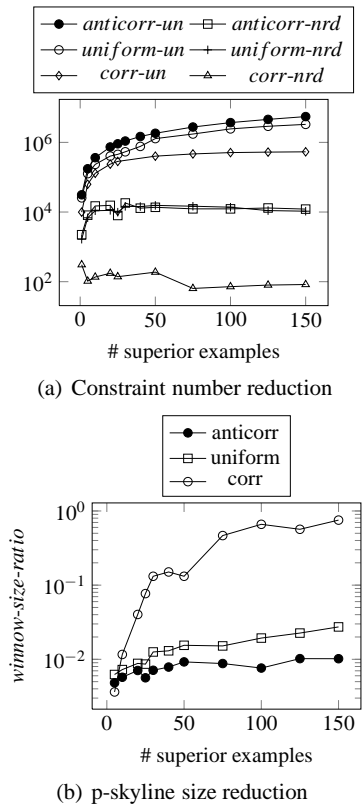

In Section \ref{sec:p-skyline-real-exp}, we showed how the size
of p-skyline query result depends on the number of relevant attributes
and the size of the skyline.  In this section, we show that another
parameter which affects the size of winnow query result is \emph{data
distribution}.  In Figure \ref{pic:exp-synthetic-data-exps-skyline}, we
demonstrate how the size of the p-skyline query result varies with the number
of superior examples. We compare this size with the size of the
corresponding skyline and plot the value of
$winnow\mbox{-}size\mbox{-}ratio$ defined in the previous
section. Here we use anti-correlated, uniform, and 
correlated data sets of $50000$ tuples each. The number of relevant
attributes is 20.  The size of the corresponding skylines is:
$41716$ (anti-correlated), $37019$ (uniform), and $33888$
(correlated). For anti-correlated and  uniform data sets, the
values of $winnow\mbox{-}size\mbox{-}ratio$ quickly reach a certain
bound and then grow slowly with the number of superior examples.  This
bound is approximately $1\%$ of the skyline size (i.e., about 350
tuples) for both data sets. At the same time, the growth of
$winnow\mbox{-}size\mbox{-}ratio$ for correlated data set is
faster. Note that the values of $winnow\mbox{-}size\mbox{-}$ $ratio$ are
generally lower for synthetic data sets, in comparison to the real-life
data set \emph{NHL}.  This is due to the larger set of
relevant attributes and larger skyline sizes in the current
experiment.

We conclude that the experiments that we have carried out show that
incorporating relative attribute importance into skyline relations in
the form of p-skyline relations results in a significant reduction in
query result size. The proposed algorithm {\tt elicit} for eliciting a
maximal p-skyline relation favoring a given set of superior examples
has good scalability in terms of the data set size and the number of
relevant attributes.  The algorithm has high accuracy even for small
sets of superior examples.

\vspace{-5mm}
\section{Related work}\label{sec:relwork}
\vspace{-2mm}
In this section, we discuss related work that has been done 
in the areas covered in the paper: \emph{modeling preferences as skyline relations} and \emph{preference elicitation}.

\vspace{-5mm}
\subsection{Modeling preferences as skyline relations}

The p-skyline framework is based on the preference constructor
approach proposed in \cite{kiesling2002}. That approach was extended in
\cite{Kiessling2005} by relaxing definitions of the accumulation
operators and by using \emph{SV}-re\-la\-ti\-ons, instead of equality, as
indifference relations. \cite{Kiessling2005} showed that such an
extension preserves the SPO properties of the resulting preference
relations. The resulting relations were shown to be \emph{larger} (in
the set theoretic sense) than the relations composed using the
equality-based accumulational operators. However, 
relative importance of attributes implicit in such relations was 
addressed neither in \cite{kiesling2002} nor
\cite{Kiessling2005}.  Containment of preference
relations and minimal extensions were also not considered in these  works.

\cite{Borz01theskyline} proposed the original skyline framework.
That paper introduced an extension of SQL in
which the skyline queries can be formulated.
The paper also proposed a number of algorithms for computing 
skylines. Since then, many algorithms for that task
have been developed 
(\cite{DBLP:conf/vldb/TanEO01,DBLP:conf/vldb/KossmannRR02,Godfrey2003,DBLP:conf/vldb/LeeZLL07,DBLP:journals/vldb/GodfreySG07}
and others).

\cite{godfrey2005} showed that the number of skyline
points in a dataset may be exponential in the number of
attributes. Since then, a number of approaches have been developed
for reducing the size of skylines by computing only \emph{the most 
representative} skyline objects. 

\cite{DBLP:conf/sigmod/ChanJTTZ06} proposed to
compute the set of \emph{k-dominant} skyline points instead of the
entire skyline.  Another variant of the skyline operator was presented
in \cite{DBLP:conf/icde/LinYZZ07}. That operator computes \emph{$k$
most representative} tuples of a skyline.
\cite{DBLP:conf/icde/LinYZZ07} showed that when the number of
attributes involved is greater than two, the problem is {\tt NP}-hard
in general. For such cases,
\cite{DBLP:conf/icde/LinYZZ07} proposed a polynomial time
approximation algorithm. 

More recently, \cite{DBLP:conf/icde/TaoDLP09} proposed the
\emph{distance-based representative skyline} operator. This approach
is based on the observation that if a skyline of a dataset consists of
clusters, then in many cases, a user is interested in seeing only good
representatives from each skyline cluster rather than the entire
skyline (which may be quite large). If interested, the user may drill
down to each cluster further on. The representativeness here
is measured as the maximum of the distance from the cluster center to
each object of the cluster. The authors studied the problem of computing
$k$ most representative skyline objects and proposed an efficient
approximation algorithm for datasets with arbitrary dimensionality.

Another recent work in the area of skyline-size reduction is
\cite{DBLP:conf/sigmod/ZhaoDTT10}. There, the authors proposed the 
\emph{order-based representative skyline} operator. The approach is based on
a well-known fact that an object is in a skyline iff it maximizes some
monotone utility function. As a measure of skyline object similarity,
the authors used the similarity between (possibly infinite) sets of
orders which favor the corresponding objects. The authors developed an
algorithm for computing representatives of clusters of similar
objects. They also proposed a method of eliciting user preferences
which allows to drill down to clusters in an iterative manner.

Another direction of research using the skyline framework concerns
\emph{subspace skyline computation} 
\cite{DBLP:conf/vldb/PeiJET05,DBLP:conf/vldb/YuanLLWYZ05}. 
An interesting problem in this framework is how
to identify the subspaces to whose skylines a given tuple belongs.
\cite{DBLP:conf/vldb/PeiJET05}
showed an approach to that problem, which uses the notion of
\emph{decisive subspace}. A subspace skyline
can be computed using every skyline algorithm. However, to compute $k$
subspace skylines (for $k$ different attribute sets), an
algorithm for efficient computing of \emph{all} subspace skylines at
once
\cite{DBLP:conf/vldb/PeiJET05,DBLP:conf/vldb/YuanLLWYZ05} 
may be more efficient.
\cite{DBLP:conf/vldb/YuanLLWYZ05} introduced the related  notion 
of \emph{skyline cube}. The skyline cube
approach was used in \cite{DBLP:journals/is/LeeYH09} to
find the \emph{most interesting} subspaces given an upper bound on the
size of the corresponding skyline and a total order of attributes,
the latter representing the importance of the attributes to the user.

We notice that the framework based on subspace skylines is, in a
sense, orthogonal to the p-skyline framework proposed here. Both of
them extend the skyline framework. In the subspace skyline framework,
the relative importance of attributes is fixed (i.e., all considered
attributes are of equal importance) while the sets of the relevant
attributes may vary.  In the p-skyline approach, the set of relevant
attributes is fixed while the relative importance of them may vary.
However, given a set of attribute preference relations, all subspace
skylines and the results of all full p-skyline relations are subsets
of the (full-space) skyline (assuming the distinct value property for
subspace skylines).

\cite{1773672} studied the properties of skyline preference relations
and showed that they are the only relations satisfying the introduced 
properties of \emph{rationality},
\emph{transitivity}, \emph{scaling robustness}, and \emph{shifted robustness}. The
authors analyzed these properties and the outcome of their relaxation in
skyline preference relations. They also showed how to adapt existing skyline
computation algorithms to relaxed
skylines. This work is particular interesting in the context of the
current paper, since it gives some insights to possible approaches for
computing
\emph{p-skyline} winnow queries.

\vspace{-5mm}
\subsection{Preference elicitation}

An approach to elicit preferences aggregated using the accumulation
operators was proposed in \cite{Kiessling2003}. Web server logs were
used there to elicit preference relations.  The approach was based on
statistical properties of log data -- more preferable tuples appear
more frequently. The mining process was split into two parts:
eliciting attribute preferences and eliciting accumulation operators
which aggregate the attribute preferences. Attribute preferences to be
elicited were in the form of predefined preference constructors such
as LOWEST, HIGHEST, POS, NEG etc. \cite{Kiessling2003} used a heuristic approach to elicit the way attribute
preferences are aggregated (using {\em Pareto} and {\em prioritized}
accumulation operators). 
The case when more than one different combination of accumulation
operators may be eli\-ci\-ted in the same data was not addressed. Moreover,
no criteria of optimality of elicited preference relations were defined.

A framework for preference elicitation which is complementary to the
approach we have developed here was presented in
\cite{Pei2008}. In that work, preferences are modeled as skyline
relations. Given a set of relevant attributes and a set of attribute
preferences over some of them, the objective is to determine attribute
preferences over the remaining attributes. The elicitation process is
based on user feedback in terms of a set of superior and a set of
inferior examples. The work is focused on eliciting minimal (in terms
of relation size) attribute preference relations.  \cite{Pei2008}
showed that the problem of existence of such relations is NP-complete,
and the computation problem is NP-hard.  Two greedy heuristic
algorithms were provided. The algorithms are not sound, i.e., for some
inputs, the computed preferences may fail to be minimal.  That
approach and the approach we presented here are different in the
following sense. First, \cite{Pei2008} dealt with skyline relations,
and thus all attribute preferences are considered to be equally
important. In contrast, the focus of our work is to elicit differences
in attribute importance.  Second, \cite{Pei2008} focused on eliciting
minimal attribute preferences.  In contrast, we are interested in
constructing maximal tuple preference relations, since such relations guarantee
a better fit to the provided set of superior examples. At the same
time, our work and \cite{Pei2008} complement each other. Namely, when
attribute preferences are not provided explicitly by the user, the
approach of
\cite{Pei2008} may be used to elicit them.

Another approach to preference relation elicitation in the skyline
framework was introduced in \cite{Balke2008}. It proposed to reduce
skyline sizes by revising skyline preference relations by supplying
additional tuple relationships: preference and equivalence. Such
relationships are obtained from user answers to simple questions.  

In quantitative preference frameworks \cite{Fishburn1970}, preferences
are represented as \emph{utility functions}: a tuple $t$ is preferred
to another tuple $t'$ iff $f(t) > f(t')$ for a utility function $f$.
Attribute priorities are often represented here as
\emph{weight coefficients} in polynomial utility functions. A number
of methods have been proposed to elicit utility functions -- some of them are
\cite{Chajewska00makingrational,777132}. Utility functions were shown 
to be effective for reasoning with preferences and querying
data\-bases with preferences (Top-K queries)
\cite{375567,1164167,Bacchus96utilityindependence}. 
Some work has been performed on eliciting utility functions for
preferences represented in other models\cite{777138}.

\cite{1227578} described another mod\-el of preference elicitation in the form of utility
functions. The authors proposed a framework for constructing a utility function
consistent with a set of comparative statements about preferences
(e.g., ``A is better than B'' or ``A is as good as B''). That approach
does not rely on any structure of preference relations.
\cite{Ha99ahybrid} proposed an approach to composing
binary preference relations and \emph{multi-linear} utility functions.
A quantitative framework for eliciting binary preference relations
based on knowledge based artificial neural network (KBANN) was
presented in
\cite{Haddawy03preferenceelicitation}. 
\cite{DBLP:journals/jair/ViappianiFP06} studied the problems
of incremental elicitation of  user preference based on user provided
\emph{example critiques}.

\vspace{-3mm}
\section{Conclusion and future work}
\vspace{-3mm}

In this work, we explored the p-skyline framework which extends
skylines with the notion of attribute importance captured by p-graphs.
We studied the properties of p-skyline relations -- checking
dominance, containment and equality of such relations -- and
showed efficient methods for performing the checks using p-graphs. We
proposed a complete set of transformation rules for efficient
computation of minimal extensions of p-skyline relations.

The main problem studied here was the \emph{elicitation of p-skyline
relations based on user-provided feedback in the form of superior and
inferior examples.} We showed that the problems of existence and
construction of a maximal p-skyline relation favoring and disfavoring
given sets of superior and inferior examples are intractable in
general. For restricted versions of these problems -- when the
provided inferior example sets are empty -- we designed polynomial
time algorithms. We also identified some bottlenecks of constructing
maximal p-skyline relations: the system of negative constraints used
may be quite large in general, which directly affects the algorithm
performance. To tackle that problem, we proposed several optimization
techniques for \emph{reducing} the size of such systems. We also
showed that the problem of \emph{minimization} of such systems is
unlikely to be solvable in polynomial time in general.  We conducted
experimental studies of the proposed elicitation algorithm and
optimization techniques. The study shows that the algorithm
has good scalability in terms of the data set size and the number of
relevant attributes, and high accuracy even for small sets of superior
examples.

At the same time, we note that the our framework has a number
of limitations that can be addressed in future work. 
First, we 
focused on \emph{full} p-skyline relations. An interesting
direction of future work would be to study the properties of
\emph{partial} p-skyline relations (i.e., defined on top of sets $\attrset$ and $\prefset$ of variable size). 

Second, attribute preference relations considered in this work are
limited to \emph{total orders}. There are several reasons for
this limitation:
\begin{itemize}
\item the limitation is natural in many contexts;
\item attribute preferences in skyline relations are also typically total orders
(although there are several papers, e.g.,
\cite{DBLP:conf/sigmod/ChanET05,Balke06exploitingindifference}, in
which this limitation is lifted);
\item some of our results require the assumption that attribute preferences
are total orders, e.g., Theorem \ref{thm:pref-equiv}.
\end{itemize}
It would be interesting to see how our results can be generalized if the 
restriction of attribute preferences to total orders is relaxed.
(To avoid any possible confusion, we emphasize that tuple preference relation
considered in our work are \emph{not} limited to total orders.)

Third, the DIFF attributes,
discussed in the original skyline paper
\cite{Borz01theskyline}, were also not considered in this paper.
This is another possible generalization.

Fourth, the type of user feedback for p-skyline relation elicitation
-- superior and inferior examples -- may not fit some real-life
scenarios. So a potentially promising direction is to adapt the
p-skyline elicitation approach to other types of feedback.  For that,
one should study appropriate classes of attribute set constraints.

Finally, the problem of computing winnow queries with p-skyline
relations is left for future work.

\vspace{-3mm}
{\footnotesize
\bibliographystyle{spbasic}
%\bibliography{preference}

}

\newpage
%$ $
%\newpage

\section*{Appendix: Proofs}

Before going into the proofs, we introduce
\emph{$\wha[0]$-struc\-tu\-res}. A $\wha[0]$-struc\-tu\-re is based on the set
of attributes $\attrset^0$ and a function $\mathcal{W}^0 = \{W_A: A
\in \attrset^0\}$ mapping $\attrset^0$ to subsets of $\attrset^0$.

\begin{definition}\label{def:up-struct}
{\bf ($\wha[0]$-structure)}
  Let $\WSET^0$ and $\attrset^0$ be as discussed above and
  such that for every $A \in \attrset^0$, $A \not \in \W_A$. Then the
  \emph{$\wha[0]$-structure} is a tuple $\wha[0]$, and the
  \emph{relation generated by $\wha[0]$} is 
   \[\succ_{\wha[0]} \ \equiv\ TC \left(\bigcup_{A \in \attrset^0}
  \pskylinephi_{A}\right),\] 
  where 
\[\pskylinephi_{A} \equiv \{(o_1,
  o_2)\ |\ o_1.A >_A o_2.A \} \cap \equivset{\attrset - (\W_A \cup
  \{A\})},\] and $\attr_A$ is the attribute preference relation for $A$ in
  $\prefset$.
\end{definition}

Let a tuple $o$ dominate a tuple $o'$ according to the relation
$\succ_{\wha[0]}$ generated by $\wha[0]$. By Definition
\ref{def:up-struct}, this is possible iff there exist a
sequence of tuples $\tupleseq{o}{o'} = (o_1, o_2, \ldots, o_{m},
o_{m+1})$ such that $o_1 = o, o_{m+1} = o'$, and a sequence of
attributes $\attrseq{o}{o'} = (A_{i_1}, \ldots, A_{i_m})$, all in $\attrset^0$, such that
\[\pskylinephi_{A_{i_1}}(o_1, o_2), \ldots,
\pskylinephi_{A_{i_m}}(o_m, o_{m+1})\] 
Then the pair
($\tupleseq{o}{o'}, \attrseq{o}{o'}$) is called a \emph{\derivingseq}
for $o \succ_{\wha[0]} o'$.  Given a pair of tuples, the corresponding
\derivingseq is not unique in general.

We notice that the $\wha[0]$-structures are an efficient tool
used here to prove some theorems describing properties of
p-skyline relations. 
Now,  Theorem \ref{thm:up-skyline}
can be reformulated as follows:

\medskip

{\bf Theorem \ref{thm:up-skyline}'}{
Every p-skyline relation ${\succ} \in \formset_\prefset$ 
can be represented as a relation $\succ_{\wha}$ generated by a $\wha$-structure
such that for every $A \in \attrset$, $\W_A = \children{\Gamma_{\succ}}(A).$
}

\smallskip

%Below we prove the alternative formulation of the theorem.

\noindent{\bf Proof of Theorem \ref{thm:up-skyline}'.}
We show here that for every $\succ \ \in \ \formset_\prefset$,
\begin{align*}
\succ \ & \equiv \ \succ_{\wham}\ \equiv \ TC\left(
\bigcup_{A \in \formvars(\succ)}
\pskylinephi_A\right)\\ \pskylinephi_{A} & \equiv \{ (o_1, o_2)
\ |\ o_1.A \ {\succ_A} \ o_2.A \} \ \cap \ \equivset{\attrset - (\W_A \cup
  \{A\})}
\end{align*}
where $\WSET_A = \children{\Gamma_\succ}(A)$ for $A \in \formvars(\succ)$.
We prove the theorem by induction on the sizes of $\prefset$ (and $\attrset$).

  \emph{Base step.} Let $\prefset = \{ >_A \}$ and $\attrset =
  \{A\}$. Then $\formset_\prefset$ consists of a single atomic
  p-skyline relation $\succ$ induced by $>_A$. Let $\W_A =
\children{\Gamma_\succ}(A) = \emptyset$. Then
\begin{align*}
  	\succ \ = \ \succ_{\wha}\ & \equiv \ TC(
        \pskylinephi_A )\\ \pskylinephi_A & \equiv \{(o_1,
        o_2)\ |\ o_1.A >_A o_2.A \} \ \cap \equivset{\attrset - (\W_A \cup
          \{A\}).}
\end{align*}
  \emph{Inductive step.}  Now assume that the theorem holds for
  $\prefset$ and $\attrset$ of size up to $n$. Prove that it holds for
  $\prefset$ and $\attrset$ of size $n+1$. Let $\succ \ = \ \succ_1
  \parcomp \succ_2$ (the case of $\succ \ = \ \succ_1 \pricomp
  \succ_2$ is similar). By the definition of induced p-skyline relations,
\begin{align*}
\succ \ \equiv \ ( \succ_1 \cap \ \equivset{\formvars(\succ_2)})
\ \cup\ (\succ_2 \cap \ \equivset{\formvars(\succ_1)}) \ \cup
\ (\succ_1 \cap \succ_2).
\end{align*}
Thus, for two p-skyline relations $\succ_1$ and $\succ_2$ the
inductive assumption implies that $\succ_1$ and $\succ_2$ can be
represented by the structures $\wha[1]$ and $\wha[2]$, 
for $\attrset^1 = \formvars(\succ_1)$ and $\attrset^2 =\formvars(\succ_2)$.
That is,
  \begin{align}
  \succ_1 \ \equiv \ \succ_{\wha[1]}\ \equiv\ TC(\bigcup_{A \in
    \formvars(\succ_1)} \pskylinephi^1_A
  ) \label{eq:pref-tc-succ1}\\ 
  \succ_2 \ \equiv \ \succ_{\wha[2]}\ \equiv\ TC(\bigcup_{A \in \formvars(\succ_2)}
  \pskylinephi^2_A )\label{eq:pref-tc-succ2}
  \end{align}
%   \begin{equation}\label{eq:pref-tc-succ2}
%   
%   \end{equation}
  where 
  \begin{align}\label{eq:pref-tc-phi1}
  \pskylinephi^1_A \equiv \{(o_1, o_2)\ |\ o_1.A >_A o_2.A \}\ \cap
  \equivset{\formvars(\succ_1) - (\W^1_A \cup \{A\})}\\ \pskylinephi^2_A
  \equiv \{(o_1, o_2)\ |\ o_1.A >_A o_2.A \} \ \cap
  \equivset{\formvars(\succ_2) - (\W^2_A \cup \{A\}).}
  \end{align}
Since $\succ$ is a p-skyline relation,
\begin{align}\label{eq:up-skyline-x}
	\formvars(\succ_1) \cap \formvars(\succ_2) = \emptyset.
\end{align}

\eqref{eq:up-skyline-x}, \eqref{eq:pref-tc-succ1}, and \eqref{eq:pref-tc-succ2} imply
  \begin{align} \label{eq:pref-tc-succ-expr4-0}
  \succ \ \equiv \ & TC \left(\bigcup_{A \in \formvars(\succ_1)}
  \pskylinephi^1_A \right) \cap \equivset{\formvars(\succ_2)} \ \cup \notag \\
        & TC \left(\bigcup_{A \in \formvars(\succ_2)} \pskylinephi^2_A
  \right) \cap \equivset{\formvars(\succ_1)} \ \cup \ \notag \\ 
        & TC
  \left(\bigcup_{A \in \formvars(\succ_1)} \pskylinephi^1_A \right)
  \cap TC \left(\bigcup_{A \in \formvars(\succ_2)} \pskylinephi^2_A
  \right)
  \end{align}
or equivalently
  \begin{align} \label{eq:pref-tc-succ-expr4}
  \succ \ \equiv \ & TC \left(\bigcup_{A \in \formvars(\succ_1)}
  \pskylinephi^1_A \cap \equivset{\formvars(\succ_2)} \right) \ \cup \notag \\
        & TC \left(\bigcup_{A \in \formvars(\succ_2)} \pskylinephi^2_A \cap
  \equivset{\formvars(\succ_1)} \right) \ \cup \ \notag \\ & TC
  \left(\bigcup_{A \in \formvars(\succ_1)} \pskylinephi^1_A \right)
  \cap TC \left(\bigcup_{A \in \formvars(\succ_2)} \pskylinephi^2_A
  \right).
  \end{align}

\noindent
  Construct the function $\W$ as follows
  \[ \W_A = \left\{
                   \begin{array}{ll}
                     \W^1_A, & \mbox{if }A \in \formvars(\succ_1)\\
                     \W^2_A, & \mbox{if }A \in \formvars(\succ_2).
                   \end{array}
            \right.
  \]
  Let 
$\attrset = \formvars(\succ_1) \cup \formvars(\succ_2) = \formvars(\succ)$ 
and 
$\succ_{\wha}$ be generated by such $\wha$
  \begin{equation}\label{eq:pref-tc-succ*-def}
  \succ_{\wha}\  \equiv \ TC(\bigcup_{A \in \attrset} \pskylinephi^*_A)
  \end{equation}
  for 
  \begin{equation}\label{eq:pref-tc-phi*}
  \pskylinephi^*_A \equiv \{(o_1, o_2)\ |\ o_1.A \attr_A o_2.A
  \}\ \cap \equivset{\attrset - (\W_A \cup \{A\}).}
  \end{equation}
 
  We prove that $\succ_{\wha}$ is equal to
  $\succ$. Before going into the proof, notice that
  \eqref{eq:pref-tc-succ-expr4} can be rewritten as
\vspace{-3mm}
  \begin{align} \label{eq:pref-tc-succ-expr5}
  \succ \ \equiv \ & TC \left(\bigcup_{A \in \formvars(\succ_1)} \pskylinephi^*_A \right) \cup 
                 TC \left(\bigcup_{A \in \formvars(\succ_2)} \pskylinephi^*_A \right) \cup \notag \\
                 & TC \left(\bigcup_{A \in \formvars(\succ_1)} \pskylinephi^1_A \right) \cap 
                 TC \left(\bigcup_{A \in \formvars(\succ_2)} \pskylinephi^2_A \right).
  \end{align}

  \begin{enumerate}
    \item Let $o \succ_{\wha} o'$. Let $(\tupleseq{o}{o'}, \attrseq{o}{o'})$ be some
	\derivingseq for $o \succ_{\wha)} o'$. W.l.o.g. let 
	$\attrseq{o}{o'} = (A_1, \ldots,$  $A_m)$, 
	$\tupleseq{o}{o'} = 
	(o = o_1, o_2, \ldots, o_m, o_{m+1} = o')$, and 
      \begin{equation}\label{eq:pref-tc-phi*chain}
      \pskylinephi^*_{A_1}(o_1, o_2) , \pskylinephi^*_{A_2}(o_2,o_3),
      \ldots, \pskylinephi^*_{A_m}(o_m, o_{m+1}).
      \end{equation}
	By construction, each attribute $A_i \in \attrseq{o}{o'}$ is
        either in $\formvars(\succ_1)$ or
        \mbox{$\formvars(\succ_2)$}. For every such $A_i$,
        $\pskylinephi^*_{A_i}(o_i, o_{i+1})$ implies $o_i \succ
        o_{i+1}$ by \eqref{eq:pref-tc-succ-expr5}.  Therefore,
        \eqref{eq:pref-tc-phi*chain} implies
      \begin{equation}\label{eq:pref-tc-succ-chain}
        o_1 \succ o_2, o_2 \succ o_3, ..., o_m \succ o_{m+1}.
      \end{equation}
	Transitivity of p-skyline relations implies $o_1 \succ
        o_{m+1}$, i.e.  $o \succ o'$.
   \item Let $o \succ o'$. Then \eqref{eq:pref-tc-succ-expr5} leads to
     three cases
     \begin{enumerate}
       \item $(o,o') \in TC \left(\bigcup_{A \in \formvars(\succ_1)} \pskylinephi^*_A \right)$. 
         Then $o \succ_{\wha} o'$ by \eqref{eq:pref-tc-succ*-def}.
       \item $(o,o') \in TC \left(\bigcup_{A \in \formvars(\succ_2)}
         \pskylinephi^*_A \right)$.  Then $o \succ_{\wha}
         o'$ by the same reasoning.
       \item $(o,o') \in TC \left(\bigcup_{A \in \formvars(\succ_1)}
         \pskylinephi^1_A \right) \cap TC \left(\bigcup_{A \in
           \formvars(\succ_2)} \pskylinephi^2_A \right)$.

         In this case, \eqref{eq:up-skyline-x} implies that there is
         an object $o''$ whose values of $\formvars(\succ_2)$ are
         equal to those of $o$, and the values of $\formvars(\succ_1)$
         are equal to those of $o'$. Then we have
         \begin{align*}
         (o, o'') \in TC \left(\bigcup_{A \in \formvars(\succ_1)}
           \pskylinephi^1_A \right) \cap \equivset{\formvars(\succ_2)}
           \notag \\ (o'', o') \in TC \left(\bigcup_{A \in
             \formvars(\succ_2)} \pskylinephi^1_A \right) \cap
           \equivset{\formvars(\succ_1)}
         \end{align*}
         or equivalently
         \begin{align*}
         (o, o'') \in TC \left(\bigcup_{A \in \formvars(\succ_1)}
           \pskylinephi^1_A \cap \equivset{\formvars(\succ_2)} \right)
           \notag \\ (o'', o') \in TC \left(\bigcup_{A \in
             \formvars(\succ_2)} \pskylinephi^1_A \cap
           \equivset{\formvars(\succ_1)} \right)
         \end{align*}
         
         which implies by \eqref{eq:pref-tc-phi*} and
         \eqref{eq:pref-tc-succ*-def} 
        \[o \succ_{\wha} o'', o'' \succ_{\wha} o'.\]
         The transitivity of $\succ_{\wha}$ implies
         $o \succ_{\wha} o'$.\end{enumerate}
         \end{enumerate}

\noindent
Recall that by Definition \ref{def:p-graph}, 
  \[ \children{\Gamma_\succ}(A) = \left\{
                   \begin{array}{ll}
                     \children{\Gamma_{\succ_1}}, & \mbox{if }A \in \formvars(\succ_1)\\
                     \children{\Gamma_{\succ_2}} & \mbox{if }A \in \formvars(\succ_2).
                   \end{array}
            \right..
  \]
Hence, given the inductive hypothesis, we proved that
  \[ W_A = \children{\Gamma_\succ}(A) = \left\{
                   \begin{array}{ll}
                     W_A^1 = \children{\Gamma_{\succ_1}}, & \mbox{if }A \in \formvars(\succ_1)\\
                     W_A^2 = \children{\Gamma_{\succ_2}} & \mbox{if }A \in \formvars(\succ_2).
                   \end{array}
            \right..
  \]
\qed

\medskip

\noindent {\bf Theorem \ref{thm:graph-props}. }{\it
 A directed graph $\Gamma$ with the set of nodes $\attrset$ is a p-graph 
 of some p-skyline relation iff 
 \begin{enumerate}
  \item $\Gamma$ is an SPO, and
  \item $\Gamma$ satisfies the \envelope property:
	\begin{align*}
	\forall A, B, C, D & \in \attrset, 
        \mbox{all different}\\ \edgeof{A}{B}{\Gamma} & \wedge
        \ \edgeof{C}{D}{\Gamma} \wedge \edgeof{C}{B}{\Gamma}
        \Rightarrow \\
        & \edgeof{C}{A}{\Gamma} \vee \edgeof{A}{D}{\Gamma}
        \vee \edgeof{D}{B}{\Gamma}.
	\end{align*}
\end{enumerate}
}

\smallskip 

To prove the theorem, we introduce the notion of the
\emph{typed partition} of a directed graph.

\begin{definition}\label{def:paritition}
 Let $\Gamma$ be a directed graph, and $\Gamma_1$, $\Gamma_2$ be two
 nonempty subgraphs of $\Gamma$ such that $\nodesof{\Gamma_1} \cap
 \nodesof{\Gamma_2} = \emptyset$ and $\nodesof{\Gamma_1} \cup
 \nodesof{\Gamma_2} = \nodesof{\Gamma}$. Then the pair  $\pair{\Gamma_1}{\Gamma_2}$
is a \emph{\noedgepartition} (respectively \emph{\edgepartition}) of
 $\Gamma$ if
 $\notedgesymof{\nodesof{\Gamma_1}}{\nodesof{\Gamma_2}}{\Gamma}$, respectively
 $\edgeof{\nodesof{\Gamma_1}}{\nodesof{\Gamma_2}}{\Gamma}$.
\end{definition}

\newcommand{\pskylinefork}{fork\xspace}
\newcommand{\pskylineforks}{forks\xspace}

The proof of Theorem \ref{thm:graph-props} is based on Lemmas
\ref{lemma:graph-sep} and \ref{lemma:find-separation}. Lemma
\ref{lemma:graph-sep} establishes relationships between nodes in an
\spoenvelopex graph, while Lemma \ref{lemma:find-separation}
establishes relationships between typed partitions in such a graph.

\begin{figure}
	\centering
	\begin{tikzpicture}[xscale=0.5]
		\tikzstyle{cir} = [draw=black,circle,inner sep=1pt]
		\node[cir] (A) at (0,1) {{\tiny $A$}};
		\node[cir] (B) at (2,1) {{\tiny $B$}};
		\node[cir] (C) at (1,0) {{\tiny $C$}};

		\draw[->, thick] (A) -- (C);
		\draw[->, thick] (B) -- (C);
	\end{tikzpicture}
	\hspace{1cm}
	\begin{tikzpicture}[xscale=0.5]
		\tikzstyle{cir} = [draw=black,circle,inner sep=1pt]
		\node[cir] (A) at (0,0) {{\tiny $A$}};
		\node[cir] (B) at (2,0) {{\tiny $B$}};
		\node[cir] (C) at (1,1) {{\tiny $C$}};

		\draw[->, thick] (C) -- (A);
		\draw[->, thick] (C) -- (B);
	\end{tikzpicture}
	\hspace{1cm}
	\begin{tikzpicture}[xscale=0.5]
		\tikzstyle{cir} = [draw=black,circle,inner sep=1pt]
		\node[cir] (A) at (0,1) {{\tiny $A$}};
		\node[cir] (B) at (0,0) {{\tiny $B$}};

		\draw[->, thick] (A) -- (B);
	\end{tikzpicture}
	\hspace{1cm}
	\begin{tikzpicture}[xscale=0.5]
		\tikzstyle{cir} = [draw=black,circle,inner sep=1pt]
		\node[cir] (A) at (0,0) {{\tiny $A$}};
		\node[cir] (B) at (0,1) {{\tiny $B$}};

		\draw[->, thick] (B) -- (A);
	\end{tikzpicture}
 \caption{Forks of $A$ and $B$}
 \label{pic:forks}
\end{figure}
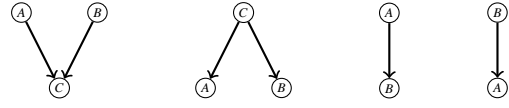

\begin{definition}\label{def:fork}
 Two nodes $A$ and $B$ of a directed graph $\Gamma$ form a
 \emph{\pskylinefork} if $A$ is different from $B$,  and
they conform 
to one of the patterns in Figure \ref{pic:forks}.
The node $C$ of $\Gamma$ has to be different from $A$ and $B$.
\end{definition}

%Figure \ref{pic:forks} shows all possible \pskylineforks of two nodes
%$A$ and $B$ in a graph.

\begin{lemma}\label{lemma:graph-sep}
 Let a directed graph $\Gamma$ with at least two nodes satisfy \spoenvelope. Then $\Gamma$ has
 a \noedgepartition, or every pair of nodes of $\Gamma$ forms a
 \pskylinefork.
\end{lemma}

\noindent{\bf Proof. }
 For the sake of contradiction, assume $\Gamma$ has no
 \noedgepartition, and some pair of different nodes $A$ and $B$ of
 $\Gamma$ does not form a \pskylinefork, i.e.,
\begin{align*}
& \notedgeof{A}{B}{\Gamma} \wedge \notedgeof{B}{A}{\Gamma} \ \wedge \neg \exists C \in \nodesof{\Gamma}\\
& \ \edgeof{A}{C}{\Gamma} \wedge 
\edgeof{B}{C}{\Gamma} \vee 
\edgeof{C}{A}{\Gamma} \wedge \edgeof{C}{B}{\Gamma}.
\end{align*}
Let a subgraph $\Gamma_1$ of $\Gamma$ have the following set of nodes
\[\nodesof{\Gamma_1} = \{A\} \cup \parents{\Gamma}(\{A\} \cup
\children{\Gamma}(A)) \cup \children{\Gamma}(\{A\} \cup
\parents{\Gamma}(A)),\] 
and the subgraph $\Gamma_2$ of $\Gamma$ have
the nodes $\nodesof{\Gamma_2} = \nodesof{\Gamma} -
\nodesof{\Gamma_1}.$ Assuming that $B \in \nodesof{\Gamma_1}$
leads to contradiction by case analysis. So $B \in \nodesof{\Gamma_2}$.
We conclude that both $\Gamma_1$ and $\Gamma_2$ are nonempty.
Also, by case analysis we show that 
$\notedgesymof{\nodes(\Gamma_1)}{\nodes(\Gamma_2)}{\Gamma}$. \qed

\medskip

\begin{lemma}\label{lemma:find-separation}
 A directed graph $\Gamma$ satisfying \spoenvelope with at least two
 nodes has a \edgepartition or a \noedgepartition 
 $\pair{\Gamma_{\succ_1}}{\Gamma_{\succ_2}}$  such that $\Gamma_{\succ_1}$ and $\Gamma_{\succ_2}$ satisfy \ \spoenvelope.
\end{lemma}

\noindent{\bf Proof.}  We assume that no \noedgepartition of $\Gamma$ exists
and show that there exists a \edgepartition.  Since $\Gamma$ is a
finite SPO, there exists a nonempty set $Top \subseteq
\nodesof{\Gamma}$ of all the nodes which have no incoming edges. If
$Top$ is a singleton, then $Top$ dominates every node in
$\nodesof{\Gamma} - Top$, and we get the \edgepartition
$\pair{Top}{\nodesof{\Gamma} - Top}$.  Assume $Top$ is not
singleton. Pick two nodes $T_1, T_2 \in Top$. $T_1$ and $T_2$ have no
incoming edges, and Lemma \ref{lemma:graph-sep} implies that there
exists a node $Z_1$ such that $\edgeof{T_1}{Z_1}{\Gamma} \wedge
\edgeof{T_2}{Z_1}{\Gamma}$. If $|Top|>2$, pick some node $T_k$ ($T_k \neq T_1, T_k
\neq T_2$) from $Top$.  Since $T_k$ has no incoming edges either,
Lemma \ref{lemma:graph-sep} implies that either $T_k$ is a parent of
$Z_1$ or they have a common child (which is also a child of $T_1$ and
$T_2$ by the transitivity of $\Gamma$). Therefore, by picking every node
of $Top$, we can show that there exists at least one node $Z$ which is
a child of all nodes in $Top$. Denote as $M$ the set of all the
nodes dominated by  every node in $Top$.
Above we showed that $M$ contains at least one node.

Now let us show that if a node $X$ is not in $M$ then
$\edgeof{X}{M}{\Gamma}$. Clearly, if $X \in Top$, then
$\edgeof{X}{M}{\Gamma}$. So let $X \not \in Top$. By definition of
$Top$, there is a node $T_1 \in Top$ such that
$\edgeof{T_1}{X}{\Gamma}$. Assume there is a node $Z \in M$ such that
$\notedgeof{X}{Z}{\Gamma}$. By definition of $M$,
$\edgeof{T_1}{Z}{\Gamma}$.  Now pick some node $T$ ($T \neq T_1$) of
$Top$. By definition of $M$, $\edgeof{T}{Z}{\Gamma}$.  Let us apply
\envelope:
\begin{align*}
\edgeof{T}{Z}{\Gamma} \ \wedge \ & \edgeof{T_1}{Z}{\Gamma} \ \wedge
\ \edgeof{T_1}{X}{\Gamma} \Rightarrow \\
& \edgeof{T_1}{T}{\Gamma} \ \vee
\ \edgeof{T}{X}{\Gamma} \ \vee \ \edgeof{X}{Z}{\Gamma}.
\end{align*}
The first and the last disjuncts in the right-hand-side of the
expression contradict the assumptions $\notedgeof{X}{Z}{\Gamma}$ and
$T \in Top$. Therefore, the only choice is
$\edgeof{T}{X}{\Gamma}$. However, $T$ is an arbitrary node in
$Top$. Therefore, $\edgeof{Top}{X}{\Gamma}$ and thus $X \in M$ by
definition of $M$. We conclude that 
$\pair{\nodesof{\Gamma} - M}{M}$ is  a \edgepartition of $\Gamma$

Finally, it is easy to check that every subgraph of an \spoenvelope graph
satisfies \spoenvelope.  \qed

\medskip

\noindent{\bf Proof of Theorem \ref{thm:graph-props}.}  By induction on the
the structure of the p-expression inducing a given p-skyline relation, 
it is easy to show that \spoenvelope is
satisfied by p-graphs.  Now we show that every  directed graph satisfying
\spoenvelopex is a p-graph of some p-skyline relation.  Given such a
graph $\Gamma$, we construct the corresponding p-skyline relation recursively.
If $\Gamma$ contains a single node, then the corresponding p-skyline
relation is the atomic preference relation induced by the attribute
preference relation of the corresponding attribute.  If $\Gamma$
has more than one node, then by Lemma
\ref{lemma:find-separation}, $\Gamma$ has either a \edgepartition or
a \noedgepartition $\pair{\Gamma_{1}}{\Gamma_2}$ into nonempty
subgraphs satisfying
\spoenvelope.
If $\pair{\Gamma_1}{\Gamma_2}$ is a
\edgepartition (\noedgepartition), then the corresponding p-skyline
relation is a prioritized (Pa\-reto, respectively) accumulation of the
p-skyline relations corresponding to $\Gamma_1$ and $\Gamma_2$. This
recursive construction exactly corresponds to the construction of $\W$
shown in Theorem \ref{thm:up-skyline}.  \qed
\medskip
\noindent {\bf Proposition \ref{prop:common-anc}. }{\it
 Let $A$ and $B$ be leaf nodes in a normalized syntax tree $T_{\succ}$
 of a p-skyline relation $\succ \ \in \ \formset_\prefset$. Then
 $\edgeof{A}{B}{\Gamma_{\succ}}$ iff the least common ancestor $C$ of
 $A$ and $B$ in $T_\succ$ is labeled by $\pricomp$, and $A$ precedes
 $B$ in the left-to-right tree traversal.
}
\smallskip

\noindent {\bf Proof of Proposition \ref{prop:common-anc}. }

\noindent
\leftsideproof Let $\succ_{C}$ be a p-skyline relation represented by
the syntax tree with the root node $C$.  Definition \ref{def:p-graph}
implies $ \edgeof{A}{B}{\Gamma_{\succ_C}}$
and 
$\edgesof{\Gamma_{\succ_C}} \subseteq \edgesof{\Gamma_{\succ}}$.

\noindent \rightsideproof Let $\edgeof{A}{B}{\Gamma_{\succ}}$.  If $C$
is of type $\pricomp$ but $B$ precedes $A$ in left-to-right tree
traversal, then  Definition \ref{def:p-graph} implies 
$\edgeof{B}{A}{\Gamma_{\succ_C}}$ and hence
$\edgeof{B}{A}{\Gamma_{\succ}}$, which is a contradiction to \spo of
$\Gamma_{\succ}$.  If $C$ is of type $\parcomp$, then by 
Definition \ref{def:p-graph}, $\notedgesymof{A}{B}{\Gamma_{\succ_C}}$ and
hence $\notedgesymof{A}{B}{\Gamma_{\succ}}$, which contradicts the
initial assumption.  \qed
\medskip
\noindent {\bf Theorem \ref{thm:p-graph-uniqueness}. }{\it
 Two p-skyline relations \mbox{$\succ_1,$} $\succ_2 \in \formset_\prefset$
 are equal iff their p-graphs are identical.
}

\smallskip
\noindent To prove the theorem, we use the next lemma.

\begin{lemma}\label{lemma:phi-subset}
  Assume that $\succ_1$ (resp. $\succ_2$) are p-skyline relations in
  $\formset_\prefset$, generated by  $\wham[1]$ and $\wham[2]$, respectively.
  If for some $A \in \attrset$, $\W^1_A - \W^2_A \neq
  \emptyset$, then there is a pair $o, o' \in \univ$ such that
		\[o \succ_{1} o'\ {\rm and}\ o \not \succ_{2} o'.\]
\end{lemma}

\noindent{\bf Proof. }
We construct two tuples $o$ and $o'$ such that $o \succ_{\wham[1]} o'$
(and thus $o
\succ_1 o')$, and $o \not \succ_{\wham[2]} o'$ (and thus $o
\not \succ_2 o'$).

For every attribute $A_i \in \attrset$, pick two values 
  $v_{A_i}, v_{A_i}' \in \Dom_{A_i}$ such that $v_{A_i} \attr_{A_i} v_{A_i}'$. Construct the tuples $o$ and $o'$ as follows:

  \[ o.A_i = \left\{
                   \begin{array}{ll}
                     v_{A_i}, & \mbox{if }A_i = A,\\
                     v_{A_i}, & \mbox{if }A_i \in \attrset - (\{A\} \cup \W_{A}^1),\\
                     v_{A_i}', & \mbox{otherwise} \ (A_i \in \W_{A}^1)\\
                   \end{array}
            \right.
  \]   
  
  \[ o'.A_i = \left\{
                   \begin{array}{ll}
                     v_{A_i}', & \mbox{if }A_i = A,\\
                     v_{A_i}, & \mbox{if }A_i \in \attrset - (\{A\} \cup \W_{A}^1),\\
                     v_{A_i}, & \mbox{otherwise}\ (A_i \in \W_{A}^1)\\
                   \end{array}
            \right.
  \]   

  By construction, it is clear that
  \[(o, o') \in \{(o_1, o_2)\ |\ o_1 \succ_{A} o_2\} \cap \equivset{\attrset - (\{A\} \cup \W_{A}^1)}\]
  and thus $o \succ_{\wham[1]} o'$ and $o \succ_1 o'$. 
  Now assume $o \succ_{\wham[2]} o'$ (and thus $o \succ_2 o'$), i.e.
	\begin{equation}\label{eq:phi-subset-tc1}
  (o, o') \in TC \left(\bigcup_{A_i \in \attrset} \pskylinephi_{A_i}\right)
	\end{equation}
  where 
	\begin{equation}\label{eq:phi-subset-tc2}
  \pskylinephi_{A_i}\equiv \{(o_1, o_2)\ |\ o_1 {\succ_{A_i}} o_2\} \cap \equivset{\attrset - (\{A_i\} \cup \W_{A_i}^2)}.
	\end{equation}

  \eqref{eq:phi-subset-tc1} implies that there should exist
  a \derivingseq ($\tupleseq{o}{o'}, \attrseq{o}{o'}$) for 
  $o \succ_{\wham[2]} o'$. That is,
  $\tupleseq{o}{o'} = (o_1 = o, o_2, \ldots, $ $o_{m},
o_{m+1} = o')$ is a sequence of tuples, and
 $\attrseq{o}{o'} = (A_{i_1}, \ldots,$ $A_{i_m})$ is a sequence of 
attributes such that
\begin{equation}\label{eq:phi-subset-chain0}
\pskylinephi_{A_{i_1}}(o_1, o_2), \ldots, \pskylinephi_{A_{i_m}}(o_m, o_{m+1}).\end{equation} 

Note that by \eqref{eq:phi-subset-tc2}, $o_{i_k}$ may be worse than 
$o_{i_{k+1}}$ in the values of $\W_{A_{i_k}^2}$ only.

First, we prove that $\attrseq{o}{o'} \subseteq \W_{A}^2 \cup \{A\}$. 
For the sake of contradiction, assume $M = \attrseq{o}{o'} - (\W_{A}^2
\cup \{A\})$ is nonempty.  Pick an element $A_{top} \in M$ which
has no ancestors from $M$ in $\Gamma_{\succ_2}$ (such an element
exists due to acyclicity of $\Gamma_{\succ_2}$).  Since
$\pskylinephi_{A_{{top}}}$ is in the chain
\eqref{eq:phi-subset-chain0}, we get
\[o.A_{top} \attr_{A_{{top}}} o'.A_{top}.\]
By construction of
$o$, $o'$ that implies $A_{top} = A$, which is a contradiction. Thus,
$\attrseq{o}{o'} \subseteq \W_{A}^2 \cup \{A\}$.
  
Second, we prove $o \not \succ_{\wham[2]} o'$. 
For that, pick $B \in \W_{A}^1 - \W_{A}^2$. By construction of 
$o$ and $o'$, $o'.B \attr_{B} o.B$. That implies that there is a pair
of tuples $o_k,o_{k+1}$ in $\tupleseq{o}{o'}$ in which the value
of $B$ is changed from a less preferred to a more preferred one.
That is possible only if $B \in W^2_C$ for some attribute $C \in 
\attrseq{o}{o'} \subseteq W_A^2 \cup \{A\}$. By Theorem \ref{thm:up-skyline},
$B \in \children{\Gamma_{\succ_2}}(C)$ and $C \in
\children{\Gamma_{\succ_2}}(A) \cup \{A\}$. By transitivity of
$\Gamma_{\succ_2}$ (Theorem \ref{thm:graph-props}), $B \in
\children{\Gamma_{\succ_2}}(A)$  (i.e., $B \in W_{A}^2$), which contradicts
the definition of $B$. Hence, $o \not \succ_{\wham[2]} o'$.\qed

Now we go back to the proof of Theorem  \ref{thm:p-graph-uniqueness}.
\smallskip
\noindent{\bf Proof of Theorem \ref{thm:p-graph-uniqueness}. }

\rightsideproof Every two  p-skyline relations which have the same
p-graph are represented by the same structure $\wha$, by the
definition of p-graph. Therefore, the p-skyline relations are equal.

\leftsideproof Pick two equal p-skyline relations $\succ_1$
and $\succ_2$. Let the structures $\wham[1]$, $\wham[2]$ and the
p-graphs $\Gamma_{\succ_1}$, $\Gamma_{\succ_2}$ represent $\succ_1$
and $\succ_2$, respectively. Clearly, the node sets of
$\Gamma_{\succ_1}$ and $\Gamma_{\succ_2}$ are equal to $\attrset$. If
their edge sets are different, then the functions $\W^1$ and $\W^2$
are different. Pick $A \in \attrset$ such that $\W^1_A \neq \W^2_A$.
Without loss of generality, we can assume $\W^1_A - \W^2_A \neq
\emptyset$. Lemma \ref{lemma:phi-subset} implies that $\succ_1$ and
$\succ_2$ are not equal, which is a contradiction.  \qed

\noindent {\bf Theorem \ref{thm:cnf-subset}. }{\it
  For p-skyline relations $\succ_1, \succ_2 \ \in
  \formset_\prefset$,\ 
  $\succ_1 \ \subset\ \succ_2 \ \ \Leftrightarrow
  \ \ \edgesof{\Gamma_{\succ_1}} \subset \edgesof{\Gamma_{\succ_2}}.$
}

\smallskip
\noindent{\bf Proof.}

\noindent
\leftsideproof Let the
structures $\wham[1]$ and $\wham[2]$ generate relations 
$\succ_{\wham[1]}$ and $\succ_{\wham[2]}$ equal to $\succ_1$ and
$\succ_2$, correspondingly. $\edgesof{\Gamma_{\succ_1}} \subset
\edgesof{\Gamma_{\succ_2}}$ implies that for all $A \in \attrset$,
$\W^1_A \subseteq \W^2_A$. Hence, $\succ_{\wham[1]} \ \subseteq \ $
$\succ_{\wham[2]}$ and $\succ_1 \ \subseteq \ \succ_2$.
Theorem \ref{thm:p-graph-uniqueness} implies $\succ_1 \ \subset \
\succ_2$.
 
\noindent
 \rightsideproof Let $\edgesof{\Gamma_{\succ_1}} \not \subset
 \edgesof{\Gamma_{\succ_2}}$.  If $\edgesof{\Gamma_{\succ_1}} =
 \edgesof{\Gamma_{\succ_2}}$, then by Theorem
 \ref{thm:p-graph-uniqueness}, $\succ_1\ \equiv\ \succ_2$, which is a
 contradiction.  Therefore, $\edgesof{\Gamma_{\succ_1}} \neq
 \edgesof{\Gamma_{\succ_2}}$, and for some $A$ we have $\W^2_A -
 \W^1_A \neq \emptyset$.  Lemma \ref{lemma:phi-subset} implies
 $\succ_1 \ \not \subset \ \succ_2$, which is a contradiction. \qed

\medskip

\noindent {\bf Theorem \ref{thm:pref-equiv}. }{\it
  Let $o, o' \in \univ$ s.t.   $o \neq o'$ and $\succ\  \in\  \formset_\prefset$.  Then the following conditions are equivalent:
 \begin{enumerate}
  \item $o \succ o'$;
  \item $\betterin(o, o') \supseteq Top_\succ(o, o')$;
  \item $\children{\Gamma_\succ}(\betterin(o, o')) \supseteq \betterin(o', o)$.
 \end{enumerate}
}

\smallskip
\noindent{\bf Proof.}
 Let the structure $\wha$ generate a relation equal to
 $\succ$, i.e.
\[\succ \ \equiv \ \succ_{\wha}\ \equiv \ TC \left(\bigcup_{A \in
   \attrset} \pskylinephi_{A}\right)\] 
where
  \[\pskylinephi_{A} \equiv \{(o_1, o_2)\ |\ o_1.A \attr_A o_2.A \}
  \ \cap \ \equivset{\attrset - (\W_A \cup \{A\})}.\]

\noindent
\equivproof{1}{3} Let $\children{\Gamma_\succ}(\betterin(o, o'))
\supseteq \betterin(o', o)$. W.l.o.g., take $\betterin(o,o') = \{A_1, \ldots,
A_k\}$. It is easy to check that the sequence
$(\tupleseq{o}{o'}, \attrseq{o}{o'})$ constructed as follows is a
\derivingseq for $o \succ_{\wha} o'$. Let $\attrseq{o}{o'} = 
$ $\betterin(o,o') = $ $\{A_1, $ $\ldots, A_k\}$.
Let the values of all the attributes $\attrset - (\betterin(o,o') \cup
\betterin(o',o))$ in $\tupleseq{o}{o'}$ be equal to those in
$o$ which are also equal to those in $o'$. Set $o_1$ to $o$. Now pick
$i$ from $2$ to $k$ consecutively and set the values of $\{A_i\} \cup
(W_{A_i} \cap \betterin(o',o))$ in $o_i$ to those in $o'$.  Since
$W_{A_i} = \children{\Gamma_\succ)}(A_i)$ (Theorem
\ref{thm:up-skyline}), the value of every attribute in $o_k$ will be
equal to the corresponding value in $o'$.

Now assume $\children{\Gamma_\succ}(\betterin(o, o')) \not \supseteq
\betterin(o', o)$. Thus, the set $\betterin(o', o) -
\children{\Gamma_\succ}(\betterin(o, o'))$ is nonempty. Similarly to
the proof of Lemma \ref{lemma:phi-subset}, it can be shown that no
\derivingseq exists for $o \succ_{\wha} o'$.

\noindent\equivproof{2}{3} 2 implies 3 by definition of $Top_\succ(o,o')$. Prove that
3 implies 2. Assume that 3 holds but $\exists A \in Top_\succ(o,o') -
\betterin(o,o')$.  Since $\attr_A$ is a total order, $A \in
\betterin(o',o)$. Then 3 implies that $A \not \in Top_\succ(o,o')$, which is
a contradiction.  \qed

%\subsection*{Proof of Theorem \ref{thm:general-envelope}}

\noindent {\bf Theorem \ref{thm:general-envelope}. }{\it
  Let $\succ$ be a p-skyline relation with the p-graph
  $\Gamma_\succ$, and $\set{A}, \set{B}, \set{C}, and \ \set{D}$,
  disjoint node sets of $\Gamma_\succ$.  Let the subgraphs of
  $\Gamma_\succ$ induced by those node sets be singletons or unions of
  at least two disjoint subgraphs. Then
	\begin{align*}
	 \edgeof{\set{A}}{\set{B}}{\Gamma_{\succ}} \ \wedge &
         \edgeof{\set{C}}{\set{D}}{\Gamma_{\succ}} \wedge
         \edgeof{\set{C}}{\set{B}}{\Gamma_{\succ}} \Rightarrow \\ &
         \edgeof{\set{C}}{\set{A}}{\Gamma_{\succ}} \vee
         \edgeof{\set{A}}{\set{D}}{\Gamma_{\succ}} \vee
         \edgeof{\set{D}}{\set{B}}{\Gamma_{\succ}}.
	\end{align*}
}

\smallskip
\noindent{\bf Proof.}
We prove the theorem by contradiction. Let
\begin{align*}
  \edgeof{\set{A}}{\set{B}}{\Gamma_{\succ}} \ \wedge
  \ \edgeof{\set{C}}{\set{D}}{\Gamma_{\succ}} \wedge
  \edgeof{\set{C}}{\set{B}}{\Gamma_{\succ}} \wedge \\ 
  \notedgeof{\set{C}}{\set{A}}{\Gamma_{\succ}} \wedge
  \notedgeof{\set{A}}{\set{D}}{\Gamma_{\succ}} \wedge
  \notedgeof{\set{D}}{\set{B}}{\Gamma_{\succ}}.
\end{align*}

The second part is equivalent to the following:
\begin{align*}
\exists C \in {\bf C}, & A_1, A_2 \in {\bf A}, D_1, D_2 \in {\bf D}, B \in {\bf B}\\
	& (\notedgeof{C}{A_2}{\Gamma_\succ} \wedge \tag{C-A2} \\
	& \notedgeof{A_1}{D_1}{\Gamma_\succ} \wedge \tag{A1-D1}\\
	& \notedgeof{D_2}{B}{\Gamma_\succ}) \tag{D2-B}
\end{align*}
and from the first part
\begin{align*}
 \edgeof{A_1}{B}{\Gamma_\succ} \tag{A1-B}\\
 \edgeof{A_2}{B}{\Gamma_\succ} \tag{A2-B}\\
 \edgeof{C}{D_1}{\Gamma_\succ} \tag{C-D1}\\
 \edgeof{C}{D_2 }{\Gamma_\succ}\tag{C-D2}
\end{align*}

Note that the fact that the subgraphs of $\Gamma_\succ$ induced by
{\bf A}, {\bf B}, {\bf C}, {\bf D} are singletons or unions of at
least two disjoint subgraphs implies the following four cases for
$A_1$ and $A_2$:
\begin{align*}
 \notedgesymof{A_1}{A_2}{\Gamma_\succ} \tag{Case
   A1}\\ 
\edgeof{A_1}{A_2}{\Gamma_\succ} \wedge \exists A_3 \in {\bf
   A}\ .\ \notedgesymof{A_1}{A_3}{\Gamma_\succ} \wedge
 \notedgesymof{A_2}{A_3}{\Gamma_\succ} \tag{Case
   A2}\\ 
\edgeof{A_2}{A_1}{\Gamma_\succ} \wedge \exists A_3 \in {\bf
   A}\ .\ \notedgesymof{A_1}{A_3}{\Gamma_\succ} \wedge
 \notedgesymof{A_2}{A_3}{\Gamma_\succ} \tag{Case A3}\\ A_1 \equiv A_2
 \tag{Case A4}\\
\end{align*}
Similarly, we have four cases for $D_1, D_2$:
\begin{align*}
 \notedgesymof{D_1}{D_2}{\Gamma_\succ} \tag{Case
   D1}\\ \edgeof{D_1}{D_2}{\Gamma_\succ} \wedge \exists D_3 \in {\bf
   D}\ .\ \notedgesymof{D_1}{D_3}{\Gamma_\succ} \wedge
 \notedgesymof{D_2}{D_3}{\Gamma_\succ} \tag{Case
   D2}\\ \edgeof{D_2}{D_1}{\Gamma_\succ} \wedge \exists D_3 \in {\bf
   D}\ .\ \notedgesymof{D_1}{D_3}{\Gamma_\succ} \wedge
 \notedgesymof{D_2}{D_3}{\Gamma_\succ} \tag{Case D3}\\ D_1 \equiv D_2
 \tag{Case D4}\\
\end{align*}

Notice that by our initial assumption, there exist two attributes
$A_1,A_2 \in {\bf A}$ and two attributes $D_1, D_2 \in {\bf D}$. 
Case $A4$ and $D4$ are due to the fact that $A_1, A_2$ and $D_1, D_2$ may corresponding to the same attributes in ${\bf A}$  and ${\bf D}$, respectively.

Totally we have sixteen different cases, and we need to show that all
of them lead to contradictions.  One can show that all of them
contradict the {\tt Envelope} property. We demonstrate it for the case
(A3-D2), while the other cases are handled similarly. In Figure
\ref{pic:case-a3-d2}, we show instances of the {\tt Envelope}
property.  Recall that the {\tt Envelope} property says that if a
graph has certain three edges, it must have at least one of the other
three edges. The instances we show below lead to only one possible
edge while the other two violate some conditions above. The violated
condition is shown below each corresponding edge. Finally, we show
that there is an unsatisfiable instance of the {\tt Envelope}
property.

We have exhaustively  tested the other fifteen cases and showed that
similar contradictions can be derived for them, too. \qed

\begin{figure}[ht]
\begin{tabular}{|l||c|c|c|}
      \hline
      {\bf {\tt Envelope}} & {\bf first edge} & {\bf second edge} & {\bf third edge}\\ 
      {\bf  condition}     &             &               & \\
      \hline
       	$(A_2, B),$ $(C, D_2),$ & $(D_2, B)$ & 
	${(A_2, D_2)}$ & $(C, A_2)$\\

	 $(C, B)$ &  (D2-B)  &     &  (C-A2) \\
\hline
       ${(A_2, D_2),}$ $(C, D_3),$  & $(D_3, D_2)$ & $(C, A_2)$ & $(A_2, D_3)$\\
       $(C, D_2)$ &  (D3 $\noedges$ D2)  &  (C-A2)  &    \\
\hline
       $(A_3, B),$ $(A_2, D_2),$ & $(D_2, B)$ & $(A_2, A_3)$ & $(A_3, D_2)$\\
       $(A_2, B)$ &  (D2-B)  &  (A2 $\noedges$ A3)  &    \\
\hline
       ${(A_3, D_2),}$ $(A_2, D_3),$ & ${(A_3, D_3) }$ & $(D_3, D_2)$ & $(A_2, A_3)$\\
       $(A_2, D_2)$ &    &  (D3 $\noedges$ D2)  &  (A2-A3) \\
\hline
       ${(A_2, D_3),}$ $(C, D_1),$ & ${(A_2, D_1)}$ & $(C, A_2)$ & $(D_1, D_3)$\\
       $(C, D_3)$ &     &  (C-A2)  &  (D1 $\noedges$ D3) \\
\hline
       ${(D_1, D_2),}$ $(A_3, D_3),$ & $(D_3, D_2)$ & ${(A_3, D_1)}$ & $(D_1, D_3)$\\
       $(A_3, D_2)$ &  (D3 $\noedges$ D2)  &     &  (D1 $\noedges$ D3) \\
\hline
       ${(A_3 ,D_1),}$ $(A_2, A_1),$ & $(A_2, A_3)$ & $(A_1, D_1)$ & $(A_3, A_1)$\\
       $(A_2, D_1)$ &  (A2 $\noedges$ A3)  &  (A1-D1)  &  (A3 $\noedges$ A1) \\
\hline
\end{tabular}
\caption{Case $A3$-$D2$}
\label{pic:case-a3-d2}
\end{figure}

%\subsection*{Proof of Theorem \ref{thm:min-ext-rules}}

\noindent {\bf Theorem \ref{thm:min-ext-rules}. }{\it
 Let $\succ\ \in \ \formset_\prefset$, and $T_{\succ}$ be a normalized
 syntax tree of $\succ$.  Then $\succ_{ext}$ is a \emph{minimal 
   p-extension} of $\succ$ iff the syntax tree
 $T_{\succ_{ext}}$ of $\succ_{ext}$ is obtained from $T_{\succ}$ by a
 single application of a rule from $Rule_1, \ldots, $ $Rule_4$, followed
by a single-child node elimination if necessary.
}

\smallskip

To prove Theorem \ref{thm:min-ext-rules} we introduce the
notions of \emph{frontier nodes}, and \emph{top} and \emph{bottom}
components in a syntax tree.

\begin{definition}
The \emph{top} and \emph{bottom} components of a p-skyline relation
$\succ$ are defined as follows:
\begin{enumerate}
 \item if $\succ$ is the atomic preference relation induced by an
 attribute preference relation, then top = bottom = $\succ$; 
 \item if $\succ \ = \ \succ_1 \pricomp \ldots \pricomp
 \succ_m,$ then top = $\succ_1$ and bottom = $\succ_m$.
\end{enumerate}
\end{definition}

Note that the notions of top and bottom components are undefined for
p-skyline relations defined as Pareto accumulations of p-skyline
relations.

\begin{definition}
Let $T_{\succ}$ be a normalized syntax tree of a p-skyline relation
$\succ$. Let also $C_1$ and $C_2$ be two different children nodes of a
$\parcomp$-node $C$ in $T_{\succ}$. Let $\succ_{ext}$ be a p-extension
of $\succ$. Moreover, let the subgraphs of $\Gamma_\succ$ and
$\Gamma_{\succ_{ext}}$ induced by $\formvars(C_1)$ be equal, as well
as those induced by $\formvars(C_2)$.  Let $X \in \formvars(C_1)$, $Y
\in
\formvars(C_2)$ be such that
\[\edgeof{X}{Y}{\Gamma_{\succ_{ext}}}.\] 
Then $(C_1, C_2)$ is a
\emph{frontier pair of $T_{\succ}$ w.r.t. $T_{\succ_{ext}}$}.
\end{definition}

Given a frontier pair $(C_1, C_2)$ of $T_{\succ}$
w.r.t. $T_{\succ_{ext}}$, note that
$\notedgesymof{\formvars(X)}{\formvars(Y)}{\Gamma_\succ}$ by
Proposition \ref{prop:common-anc}. By definition, a p-skyline relation
is constructed in a recursive way: a higher-level relation is defined
in terms of lower-level relations. Hence, the intuition behind the
frontier pair is as follows. When $\succ$ and $\succ_{ext}$ are
constructed, the lower-level relations $\succ_{C_1}$ and $\succ_{C_2}$
are present in both $\succ$ and $\succ_{ext}$. However, the next-level
relations defined using $\succ_{C_1}$ and $\succ_{C_2}$ in $\succ$ and
$\succ_{ext}$ are different since $\Gamma_{\succ_{ext}}$ has an edge
from a member of \mbox{$\formvars(\succ_{C_1})$} to a member of
$\formvars(\succ_{C_2})$, which is not present in
$\Gamma_{\succ}$. The next lemma shows some properties of frontier
pairs.

\begin{lemma}\label{lemma:triangle}
Let $\succ_{ext}$ be a p-extension of $\succ \ \in
\ \formset_\prefset$, and $T_{\succ}$ be a normalized syntax tree of
$\succ$. Let also $(C_1, C_2)$ (or ($C_2,C_1)$) be a frontier pair of
$T_{\succ}$ w.r.t. $T_{\succ_{ext}}$.  Denote the top and the bottom
components of $C_1$ as $A_1, B_1$, and the top and the bottom
components of $C_2$ as $A_2, B_2$.  Then
\[\edgeof{\formvars(A_1)}{\formvars(B_2)}\Gamma_{\succ_{ext}} \vee
\edgeof{\formvars(A_2)}{\formvars(B_1)}\Gamma_{\succ_{ext}}\]
\end{lemma}

\noindent{\bf Proof. } We consider the case of $(C_1,C_2)$ being a frontier
pair of $T_\succ$ w.r.t. $T_{\succ_{ext}}$. The case of $(C_2,C_1)$ is
symmetric. Since $(C_1, C_2)$ is a frontier pair of
$T_\succ$ w.r.t. $T_{\succ_{ext}}$, there are $X \in \formvars(C_1)$
and $Y \in
\formvars(C_2)$ such that
\[\edgeof{X}{Y}{\Gamma_{\succ_{ext}}}\]
Note that we have the following cases for
$X \in \formvars(C_1)$ 

\begin{center}
\begin{tabular}{|c|l|}
\hline 
$\phi_1$ & $\formvars(C_1) = \{X\}$, i.e. ($C_1 = A_1 = B_1$) \\
\hline
 $\phi_2$ & $C_1 = (A_1 \pricomp \ldots \pricomp B_1)$, $X \not \in \formvars(A_1)$ \\
\hline
 $\phi_3$ & $C_1 = (A_1 \pricomp \ldots \pricomp B_1)$, $\formvars(A_1) = \{X\}$ \\
\hline
 $\phi_4$ & $C_1 = (A_1 \pricomp \ldots \pricomp B_1)$, \\
       & $A_1 = A_1^1 \parcomp A_1^2 \ldots$, $X \in \formvars(A_1^1)$ \\
\hline
\end{tabular}
\end{center}

\smallskip
and for $Y \in \formvars(C_2)$

\smallskip
\begin{center}
\begin{tabular}{|c|l|}
\hline 
$\lambda_1$ & $\formvars(C_2) = \{Y\}$, i.e. ($C_2 = A_2 = B_2$) \\
\hline
 $\lambda_2$ & $C_2 = (A_2 \pricomp \ldots \pricomp B_2)$, $Y \not \in \formvars(B_2)$ \\
\hline
 $\lambda_3$ & $C_2 = (A_2 \pricomp \ldots \pricomp B_2)$, $\formvars(B_2) = \{Y\}$ \\
\hline
 $\lambda_4$ & $C_2 = (A_2 \pricomp \ldots \pricomp B_2)$ \\
       & $B_2 = B_2^1 \parcomp B_2^2 \ldots$, $Y \in \formvars(B_2^1)$. \\
\hline
\end{tabular}
\end{center}
\medskip

The cases $\phi_1,\phi_2$, and $\phi_3$ imply either $\edgeof{\formvars(A_1)}{X}{\Gamma_{\succ_{ext}}}$ or $\formvars(A_1) = \{X\}$ and as a result
$\edgeof{\formvars(A_1)}{Y}{\Gamma_{\succ_{ext}}}$ by transitivity of
$\Gamma_{\succ_{ext}}$. Similarly, the cases $\lambda_1, \lambda_2$, and
$\lambda_3$ imply either $\formvars(B_2) = \{Y\}$ or
$\edgeof{Y}{\formvars(B_2)}{\Gamma_{\succ_{ext}}}$. Thus every
combination of these cases implies
\edgeofx{$\formvars(A_1)$}{$\formvars(B_2)$}{$\Gamma_{\succ_{ext}}$}.  Now
consider the other combinations of the cases. All of them are handled
similar to the case ($\phi_4$, $\lambda_4$), so we consider it in
detail.
\medskip

Take the case $\lambda_4$. Take $Y' \in \formvars(B_2) -
\formvars(B_2^1)$ and apply \genenvelope to $\Gamma_{\succ_{ext}}$:
\[\edgeof{\formvars(A_2)}{Y'}{\Gamma_{\succ_{ext}}} \wedge
\edgeof{\formvars(A_2)}{Y}{\Gamma_{\succ_{ext}}} \wedge
\edgeof{X}{Y}{\Gamma_{\succ_{ext}}}\] 
which implies
\[\edgeof{\formvars(A_2)}{X}{\Gamma_{\succ_{ext}}} \vee 
\edgeof{X}{Y'}{\Gamma_{\succ_{ext}}} \vee \edgeof{Y'}{Y}{\Gamma_{\succ_{ext}}}.\]

$\notedgeof{Y'}{Y}{\Gamma_{\succ_{ext}}}$ follows from
Proposition \ref{prop:common-anc} and the fact that the subgraphs of
$\Gamma_{\succ_{ext}}$ and $\Gamma_{\succ}$ that are induced by
$\formvars(C_2)$ are the same.  
\edgeofx{$\formvars(A_2$)}{$X$}{$\Gamma_{\succ_{ext}}$} and
\edgeofx{$X$}{$\formvars(B_1)$}{$\Gamma_{\succ_{ext}}$} (following from
$\phi_4$) imply
\edgeofx{$\formvars(A_2)$}{$\formvars(B_1)$}{$\Gamma_{\succ_{ext}}$}, which
is what we need.  Hence,
\edgeofx{$\formvars(A_2)$}{$\formvars(B_1)$}{$\Gamma_{\succ_{ext}}$} or
\edgeofx{$X$}{$Y'$}{$\Gamma_{\succ_{ext}}$} for all $Y' \in \formvars(B_2)
- \formvars(B_2^1)$.  Consider $\edgeof{X}{Y'}{\Gamma_{\succ_{ext}}}$
and pick $Y'' \in \formvars(B_2^1)$. For such $Y''$ we have
$\notedgeof{Y'}{Y''}{\Gamma_{\succ_{ext}}}$ by Proposition
\ref{prop:common-anc}.  Therefore, we get a condition for
\genenvelope similar to the one above:
\[\edgeof{\formvars(A_2)}{Y''}{\Gamma_{\succ_{ext}}} \wedge
\edgeof{\formvars(A_2)}{Y'}{\Gamma_{\succ_{ext}}} \wedge
\edgeof{X}{Y'}{\Gamma_{\succ_{ext}}}\] 
implying 
\[\edgeof{\formvars(A_2)}{X}{\Gamma_{\succ_{ext}}} \vee 
\edgeof{X}{Y''}{\Gamma_{\succ_{ext}}} \vee \edgeof{Y''}{Y'}{\Gamma_{\succ_{ext}}}.\]

$\notedgeof{Y''}{Y'}{\Gamma_{\succ_{ext}}}$ by the same argument
as above. Similarly to the above, 
$\edgeof{\formvars(A_2)}{X}{\Gamma_{\succ_{ext}}}$ and
\edgeofx{$X$}{$\formvars(B_1)$}{$\Gamma_{\succ_{ext}}$} imply
\edgeofx{$\formvars(A_2)$}{$\formvars(B_1)$}{$\Gamma_{\succ_{ext}}$}, which
is what we need.  As a result, we have
\edgeofx{$\formvars(A_2)$}{$\formvars(B_1)$}{$\Gamma_{\succ_{ext}}$} or
\edgeofx{$X$}{$Y'$}{$\Gamma_{\succ_{ext}}$} $\wedge$
\edgeofx{$X$}{$Y''$}{$\Gamma_{\succ_{ext}}$}
for all $Y' \in \formvars(B_2) - \formvars(B_2^1), Y'' 
\in \formvars(B_2^1)$, that is equivalent to
\[\edgeof{\formvars(A_2)}{\formvars(B_1)}{\Gamma_{\succ_{ext}}} \vee
\edgeof{X}{\formvars(B_2)}{\Gamma_{\succ_{ext}}}.\]

Elaborating the case $\phi_4$ as above gives that 
\[\edgeof{\formvars(A_2)}{\formvars(B_1)}{\Gamma_{\succ_{ext}}} \ \vee\ 
\edgeof{\formvars(A_1)}{Y}{\Gamma_{\succ_{ext}}}.\] 
After combining these two results and applying \genenvelope to
members of $A_1$ and $B_2$, we get
\[\edgeof{\formvars(A_1)}{\formvars(B_2)}{\Gamma_{\succ_{ext}}} \ \vee\ 
\edgeof{\formvars(A_2)}{\formvars(B_1)}{\Gamma_{\succ_{ext}}}.\] 
\qed

Now we go back to the proof of Theorem \ref{thm:min-ext-rules}.

\smallskip

\noindent{\bf Proof of Theorem \ref{thm:min-ext-rules}}

\noindent\rightsideproof Let $\succ_{ext}$ be a minimal p-extension of
$\succ$. We show here that there is $\succ' \in \formset_\prefset$
obtained using a transformation rule $Rule_1, \ldots, Rule_4$ such
that
\begin{equation}\label{eq:lemma-triangle-1}
\succ \ \subset \ \succ' \ \subseteq \ \succ_{ext}.
\end{equation}
By the minimal p-extension property of $\succ_{ext}$ that implies
$\succ'\ =\ \succ_{ext}$.

Theorem \ref{thm:cnf-subset} implies that there are $X$ and $Y$ such that
$(X,$ $Y) \in \edgesof{\Gamma_{\succ_{ext}}} -
\edgesof{\Gamma_{\succ}}$.  Let $(C_1, C_2)$ be a frontier pair of
$T_{\succ}$ w.r.t. $T_{\succ_{ext}}$ such that $X \in \formvars(C_1)$
and $Y \in \formvars(C_2)$. Lemma \ref{lemma:triangle} implies that
\begin{equation}\label{eq:lemma-triangle-2}
\edgeof{\formvars(A_1)}{\formvars(B_2)}{\Gamma_{\succ_{ext}}} \vee 
\edgeof{\formvars(A_2)}{\formvars(B_1)}{\Gamma_{\succ_{ext}}}
\end{equation}
for the top $A_1, A_2$ and the bottom $B_1, B_2$ components of $C_1$
and $C_2$, correspondingly.  Consider all possible types of $C_1$ and
$C_2$. (i) Let $C_1, C_2$ be leaf nodes. Then $\succ'$ for which
\eqref{eq:lemma-triangle-1} holds may be obtained by applying
$Rule_3(T_{\succ}, C_1, C_2)$ (if the first disjunct of
\eqref{eq:lemma-triangle-2} holds) or $Rule_3(T_{\succ}, C_2, C_1)$
(if the second disjunct of \eqref{eq:lemma-triangle-2} holds). (ii)
Let $C_1$ be a $\pricomp$-node and $C_2$ be a leaf node. Then $\succ'$
may be obtained by applying $Rule_1(T_\succ, C_1, C_2)$ (if the first
disjunct of \eqref{eq:lemma-triangle-2} holds) or $Rule_2(T_\succ,
C_1, C_2)$ (if the second disjunct of \eqref{eq:lemma-triangle-2}
holds). Case (iii) when $C_1$ is a leaf node and $C_2$ is a
$\pricomp$-node is similar to the previous case. Consider case (iv)
when $C_1$ and $C_2$ are $\pricomp$-nodes. Let the first disjunct of
\eqref{eq:lemma-triangle-2} hold. The case of the second disjunct is
analogous. We note that
$\edgeof{\formvars(A_1)}{\formvars(B_1)}{\Gamma_{\succ_{ext}} }$ and
$\edgeof{\formvars(A_2)}{\formvars(B_2)}{\Gamma_{\succ_{ext}}}$.  This
with \eqref{eq:lemma-triangle-2} is a condition for
\genenvelope:
\begin{align}\label{eq:lemma-triangle-4}
\edgeof{\formvars(A_1)}{\formvars(A_2)}{\Gamma_{\succ_{ext}} } \vee 
\edgeof{\formvars(A_2)}{\formvars(B_1)}{\Gamma_{\succ_{ext}}} \ \vee \notag \\
\edgeof{\formvars(B_1)}{\formvars(B_2)}{\Gamma_{\succ_{ext}}}
\end{align}
If the first disjunct of \eqref{eq:lemma-triangle-4} holds, then
$\succ'$ can be obtained by applying $Rule_1(T_\succ, C_1, C_2)$. If
the last disjunct of \eqref{eq:lemma-triangle-4} holds, then $\succ'$
can be obtained by applying $Rule_2(T_\succ, C_2, C_1)$. Let the
second disjunct of \eqref{eq:lemma-triangle-4} hold,
i.e. $\edgeof{\formvars(A_2)}{\formvars(B_1)}{\Gamma_{\succ_{ext}}}$.
Let the child nodes of $C_1$ and $C_2$ be the sequences ($A_1 = N_1,
\ldots, N_m = B_1$) and ($A_2 = M_1, \ldots, M_n = B_2$)
correspondingly. The fact that $C_1$ and $C_2$ are $\pricomp$-nodes
implies $\edgeof{\formvars(N_i)}{\formvars(N_j)}{\Gamma_{\succ}}$ and
$\edgeof{\formvars(M_i)}{\formvars(M_j)}{\Gamma_{\succ}}$ for all $i <
j$. Since $\succ \ \subseteq \ \succ_{ext}$, the same edges are
present in $\Gamma_{\succ_{ext}}$. Note that
$\edgeof{M_1}{N_m}{\Gamma_{\succ_{ext}}}$. Pick every child of $C_2$
in its list of children from right to left and find the first index
$t$ such that
$\notedgeof{\formvars(N_1)}{\formvars(M_t)}{\Gamma_{\succ_{ext}}}$ but
$\edgeof{\formvars(N_1)}{\formvars(M_{t+1})}\Gamma_{\succ_{ext}}$. If
no such $t$ exists, then
$\edgeof{\formvars(N_1)}{\formvars(M_1)}\Gamma_{\succ_{ext}}$ and
$\succ'$ may be obtained by applying $Rule_1(T_{\succ}, C_1,
C_2)$. Assume $t \in [1, n]$.  Similarly, let $s$ be the first index
such that
$\notedgeof{\formvars(M_1)}{\formvars(N_s)}{\Gamma_{\succ_{ext}}}$ but
$\edgeof{\formvars(M_1)}{\formvars(N_{s+1})}{\Gamma_{\succ_{ext}}}$. If
$s$ does not exist, then $\succ'$ may be obtained by applying
$Rule_2(T_{\succ}, C_2, C_1)$. So assume $s \in [1, m]$.  If both $s$
and $t$ are equal to $1$, then $\succ'$ may be obtained using
$Rule_4(T_{\succ}, C_1, C_2, s, $ $t)$. In all other cases,
\genenvelope can be used to show that for all $i \in [1, s], j \in
[t+1, n]$
$\edgeof{\formvars(N_i)}{\formvars(M_j)}{\Gamma_{\succ_{ext}}}$ and
for all $i \in [1, t], j \in [s+1, m]$
$\edgeof{\formvars(M_i)}{\formvars(N_j)}{\Gamma_{\succ_{ext}}}$.
Hence $Rule_4(T_\succ, C_1, C_2, s, t)$ may be used to construct
$\succ'_{ext}$.

\noindent \leftsideproof Show that every valid application of
$Rule_1, \ldots, Rule_4$ results in a minimal extension. We do it by
case analysis. Take $Rule_3$, which results in adding the edge from
$C_i$ to $C_{i+1}$ to the p-graph. This is clearly a minimal extension
of the p-graph and hence the resulting p-skyline relation is a minimal
extension of $\succ$. The analysis pattern for the remaining rules is
as follows. We assume that some p-extension $\succ_{ext}$ obtained by
an application of $Rule_1$, $Rule_2$, or $Rule_4$ to $\succ$ is not
minimal, i.e., there is $\succ'$ s.t. $\succ
\subset \succ' \subset \succ_{ext}$. After that, we derive a contradiction
that $\Gamma_{\succ'} = \Gamma_{\succ_{ext}}$.
Take $Rule_1$. Since $\succ'$ is an extension of $\succ$ contained in
$\succ_{ext}$, there must be an edge from some $A \in \formvars(N_1)$
to some $B$ in the bottom component of $C_{i+1}$. Clearly, if 
$\formvars(N_1) = \{A\}$ and $\formvars(C_{i+1}) = \{B\}$, then 
$\Gamma_{\succ'} = \Gamma_{\succ_{ext}}$ and we get the contradiction
we want. So assume $\formvars(C_{i+1}) \neq \{B\}$. Then applying
\genenvelope to 
\begin{align*}
  \edgeof{A}{\formvars(N_2)}{\Gamma_{\succ'}} \wedge
  \edgeof{A}{\formvars(B)}{\Gamma_{\succ'}} \wedge \\
  \edgeof{\formvars(T_{i+1})}{B}{\Gamma_{\succ'}}
\end{align*}
  (where $T_{i+1}$
  is the top component of $C_{i+1}$) results in
  \edgeofx{$A$}{$\formvars(T_{i+1})$}{$\Gamma_{\succ'}$} (and hence
  $\edgeof{A}{\formvars(C_{i+1})}{\Gamma_{\succ'}}$ by transitivity of
  $\Gamma_{\succ'}$). The other alternatives are impossible:
  the corresponding edges are missing in $\Gamma_{\succ_{ext}}$ (and
  hence in $\Gamma_{\succ'}$, too). Clearly, if $\formvars(N_1) = \{A\}$,
  then we get the contradiction we need: $\Gamma_{\succ'} =
  \Gamma_{\succ_{ext}}$.  So assume $\formvars(N_1) \neq
  \{A\}$. Denote $S = \formvars(N_1) - \{A\}$.  Then applying
  \genenvelope to
\begin{align*}
  \edgeof{S}{\formvars(N_2)}{\Gamma_{\succ'}} \wedge
  \edgeof{A}{\formvars(N_2)}{\Gamma_{\succ'}} \wedge \\
  \edgeof{A}{\formvars(C_{i+1})}{\Gamma_{\succ'}}
\end{align*}
results in $\edgeof{S}{\formvars(C_{i+1})}{\Gamma_{\succ'}}$. The
other alternatives are prohibited because the corresponding p-graph
edges are not in $\Gamma_{\succ_{ext}}$ (and hence not in
$\Gamma_{\succ'}$). That results in
\edgeofx{$\formvars(N_1)$}{$\formvars(C_{i+1})$}
{$\Gamma_{\succ'}$} and the contradiction that $\Gamma_{\succ_{ext}} =
\Gamma_{\succ'}$. The case analysis for $Rule_2$ is similar. 

Now let $\succ_{ext}$ be obtained from $\succ$ by applying $Rule_4$, and
consider a p-extension $\succ'$ of $\succ$ s.t. $\succ' \subset
\succ_{ext}$. Because of this assumption, $\Gamma_{\succ'}$ has
an edge from some $A \in \formvars(N_1)$ to some $B \in \formvars(M_n)$
or from some $C \in \formvars(M_1)$ to some $D \in \formvars(N_m)$. Since
these cases are completely symmetric, take $\edgeof{A}{B}{\Gamma_{\succ'}}$.
Applying \genenvelope to 
\begin{align*}
  \edgeof{A}{\formvars(N_{s+1})}{\Gamma_{\succ'}} \wedge
  \edgeof{A}{B}{\Gamma_{\succ'}} \wedge \\
  \edgeof{\formvars(M_t)}{\formvars(M_n)}{\Gamma_{\succ'}}
\end{align*}
results in 
\begin{equation}\label{eq:super-equation1}
\edgeof{\formvars(M_t)}{\formvars(N_{s+1})}{\Gamma_{\succ'}}
\end{equation}
since all the other alternatives are impossible -- the corresponding
p-graph edges are not in $\Gamma_{\succ_{ext}}$ -- and hence not
in $\Gamma_{\succ'}$. Now apply \genenvelope to
\begin{align*}
  \edgeof{\formvars(M_t)}{\formvars(M_{t+1})}{\Gamma_{\succ'}} \wedge
  \edgeof{\formvars(M_t)}{\formvars(N_{s+1})}{\Gamma_{\succ'}} \wedge \\
  \edgeof{\formvars(N_s)}{\formvars(N_{s+1})}{\Gamma_{\succ'}},
\end{align*}
which results in 
\begin{equation}\label{eq:super-equation2}
\edgeof{\formvars(N_s)}{\formvars(M_{t+1})}{\Gamma_{\succ'}}
\end{equation}
since all the other alternatives are impossible -- the corresponding
p-graph edges are not in $\Gamma_{\succ_{ext}}$ and hence not
in $\Gamma_{\succ'}$. \eqref{eq:super-equation1}, \eqref{eq:super-equation2},
and the transitivity of $\Gamma_{\succ'}$ implies that 
$\Gamma_{\succ'} = \Gamma_{\succ_{ext}}$, which is a contradiction. 
%We showed above that every minimal 
%p-extension $\succ_{ext}$ of $\succ$ may be computed using a single
%application of one of the transformation rules. 
%Showing that every
%valid application of a transformation rule leads to a minimal
%p-extension of $\succ$ may be done by using Lemma
%\ref{lemma:triangle}, \genenvelope, and a case analysis similar to the one above.
%In particular, checking that every application of rules 1, 2, and 3
%generates a minimal p-extension is similar to the analysis above.  The
%analysis for $Rule4$ involves a number of sub-cases. First, we apply
%Lemma \ref{lemma:triangle} to show that there must be an edge-set from
%the top components of $R1'$ and $R2'$ to the bottom components of
%$R1'$ and $R2'$. After that, \genenvelope needs to be applied several
%times to show that all the remaining p-graph edges generated by
%$Rule4$ belong to the p-graph of a valid p-extension.
\qed

\medskip

\noindent {\bf Theorem \ref{thm:np-compl}. }{\it
 \favdisprob is NP-complete.
}

\smallskip
\noindent{\bf Proof.}
The favoring/disfavoring p-skyline existence problem is in NP since
checking if a p-skyline relation $\succ$ favors $G$ and disfavors $W$
in $\oset$ can be done in polynomial time by evaluating
$\winnow_{\succ}(O)$, checking $G \subseteq \winnow_{\succ}(\oset)$,
and checking if for every member of $W$ there is a member of $W$
dominating it.

To show the hardness result, we do a polynomial-time reduction from
\sat.  This is a two-step reduction. First, we show that for every
instance $\phi$ of \sat there are corresponding instances of positive
$\possystem$ and negative $\negsystem$ constraints, and $\phi$ has a
solution iff $\possystem$ and $\negsystem$ are satisfiable. Second, we
show that for every such $\possystem$ and $\negsystem$ there are
corresponding instances of $G$, $W$, and $\oset$.

Consider instances of \sat in the following form
	\[\phi(x_1,\ldots,x_n) = \psi_1(x_1, \ldots, x_n) \wedge
\ldots \wedge \psi_m(x_1, \ldots, x_n)\] 
where
\[\psi_t(x_1, \ldots, x_n) = \widehat{x_{i_t}} \vee \ldots \vee
\widehat{x_{j_t}}\]

For every instance of $\phi$, construct $\attrset = \{c, y_1,
\overline{y_1}, y_1', \ldots, $ $y_n, \overline{y_n}, y_n'\}$. 
The sets of positive and negative constraints are constructed as
follows. Let $\Gamma$ be a graph. For every variable $x_i$,
\begin{enumerate}
 \item Create positive constraints
	\begin{align*}
	\chi_{i}: & \edgeof{y_i}{c}{\Gamma} \vee
        \edgeof{\overline{y_i}}{c}{\Gamma}\\ \pi_i: &
        \edgeof{\overline{y_i}}{y_i'}{\Gamma}
	\end{align*}

 \item Create negative constraints
  \begin{align*}
   \lambda_i^1: & \notedgeof{\overline{y_i}}{y_i}{\Gamma} \\
   \lambda_i^2: & \notedgeof{y_i}{y_i'}{\Gamma} \\
   \lambda_i^3: & \notedgeof{y_i'}{c}{\Gamma}
  \end{align*}
\end{enumerate}
Now, for every $\psi_t(x_1,\ldots, x_n) = \widehat{x_{i_t}} \vee \ldots \vee \widehat{x_{j_t}}$ of $\phi$
construct the following positive constraint
\[\mu_t: \edgeof{\widehat{y_{i_t}}}{c}{\Gamma} \vee \ldots \vee \edgeof{\widehat{y_{i_t}}}{c}{\Gamma}\]
where
$
\widehat{y_i} = \left\{
\begin{array}{ll}
 y_i & \mbox{if } \widehat{x_i} = x_i \\
 \overline{y_i} & \mbox{if }  \widehat{x_i} = \overline{x_i}
\end{array} 
.
\right.$

We claim that there is a satisfying assignment $(v_1, \ldots, v_n)$
for $\phi$ iff there is a p-graph satisfying all the constraints
above.  First, assume there is a p-graph $\Gamma$ satisfying all the
constraints above.  Construct the assignment $v = (v_1, \ldots, v_n)$
as follows:

\[
v_i = \left\{
\begin{array}{ll}
 0 & \mbox{if } \edgeof{\overline{y_i}}{c}{\Gamma} \\
 1 & \mbox{if } \edgeof{y_i}{c}{\Gamma}
\end{array} 
.
\right.\] 

Since $\Gamma$ satisfies all $\chi_i$, for every $i$ we have
$\edgeof{y_i}{c}{\Gamma}$ or
$\edgeof{\overline{y_i}}{c}{\Gamma}$. Thus, every $v_i$ will be
assigned to some value according to the rule above.  Now prove that
$v_i$ is assigned to only one value, i.e., we cannot have both
$\edgeof{y_i}{c}{\Gamma}$ and $\edgeof{\overline{y_i}}{c}{\Gamma}$.
Since $\Gamma$ satisfies $\pi_i$, we have
$\edgeof{\overline{y_i}}{y_i'}{\Gamma}$.  Thus having both
$\edgeof{y_i}{c}{\Gamma}$ and $\edgeof{\overline{y_i}}{c}{\Gamma}$ and
\envelope implies
\[\edgeof{\overline{y_i}}{y_i}{\Gamma} \vee \edgeof{y_i}{y_i'}{\Gamma}
\vee \edgeof{y_i'}{c}{\Gamma}.\] 
However, the expression above violates
the constraints $\lambda_i^1, $ $\lambda_i^2,$  $\lambda_i^3$.  Therefore,
exactly one of $\edgeof{y_i}{c}{\Gamma}$,
$\edgeof{\overline{y_i}}{c}{\Gamma}$ holds.

Take every $\mu_t$. Since it is satisfied by $\Gamma$, the
corresponding $\psi_i$ must be also satisfied by the construction of
$\mu_t$. Therefore, $\phi$ is also satisfied.

Now assume that there is an assignment $(v_1, \ldots, v_n)$ satisfying
$\phi$. Show that there is a p-graph $\Gamma_{\succ}$ satisfying all
the constraints above. Here we construct such a graph.

For every $i \in [1, n]$, draw the edge 
\begin{align}
\edgeof{y_i}{c}{\Gamma_{\succ}}\ \ \ \ & \mbox{if $v_i = 1$,
  and}\tag{P1}\\ \edgeof{\overline{y_i}}{c}{\Gamma_{\succ}}, \ \ \ \ &
\mbox{otherwise} \tag{P2}
\end{align} 
This satisfies the constraint $\chi_i$. Moreover, all the constraints
$\mu_t$ are satisfied by the construction. Now, for every $i \in [1, n]$,
draw the edge
\begin{align}\label{e:p3}
\edgeof{\overline{y_i}}{y_i'}{\Gamma_{\succ}}\tag{P3}
\end{align}
which satisfies the constraint $\pi_i$.  As a result, all positive
constraints are satisfied. Moreover, none of the edges above violates
any negative constraints. Thus, all the constraints above are
satisfied.

In addition to the edges above, let us draw the following edges

\begin{enumerate}
 \item for every $i, j$ $(i \neq j)$ such that $v_i = 0, v_j = 0$,
 draw the edge
\begin{align}\label{e:p4}
 \edgeof{\overline{y_i}}{y_j'}{\Gamma_{\succ}}\tag{P4}
\end{align}

It is clear that these edges do not violate any negative constraints
above.
 \item for every $i, j$ such that $v_i = 0, v_j = 1$, draw the edge
\begin{align}\label{e:p5}
 \edgeof{\overline{y_i}}{y_j}{\Gamma_{\succ}} \tag{P5}
\end{align}
Since $i \neq j$, this edge does not violate any negative constraints
above.
\end{enumerate}

It is easy to verify that the constructed graph $\Gamma_{\succ}$
satisfies \spoenvelope and all the negative and positive constraints
above.

Now let us show that there exist sets of objects $\oset$, $G$ and $W$
which can be used to obtain the constraints $\chi_i, $ $\pi_i,$ 
$\lambda_i^1, $ $\lambda_i^2, $ $\lambda_i^3, $ $\mu_t$.  Assume that for every
attribute in $A\in \attrset$, its domain contains at least three numbers
$\{-1, 0, 1\}$, and greater values are to be preferred in the attribute preference
relation $\attr_A$. Here we construct
the sets $G$, $W$, $M$, and $\mathcal{O} = G \cup W \cup M$ that
generate the positive and negative constraints above.

\begin{enumerate} 
 \item Let $G$ consist of a single object $g$ with all attributes
   values equal to $0$.
 \item Let $W = \{b_1, \ldots, b_n, u_1, \ldots, u_n, w_1, \ldots, w_m\}$
	be const\-ruc\-ted as follows:

	\begin{itemize}
	 \item for every $i \in [1, \ldots, n]$, let all the
           attributes of $b_i$ be equal to $0$, except for the value of
           $\overline{y_i}$, which is $-1$, and the value of $y_i'$, which is $1$.

	\item for every $i \in [1, \ldots, n]$, let all the attributes
          of $u_i$ be equal to $0$, except for the value of $y_i,
          \overline{y_i}$, which is  $-1$, and the value of $c$, which is
          $1$.

	\item for every $t \in [1, \ldots, m]$, let $\mu_t:
          \edgeof{\widehat{y_{i_t}}}{c}{\Gamma} \vee \ldots \vee
          \edgeof{\widehat{y_{j_t}}}{c}{\Gamma}$, where $\widehat{y_i}
          \in \{y_i, \overline{y_i}\}$. Let all attributes of $w_t$ be
          equal to $0$, except for the value of $c$, which is  $1$, and the
          values of $\widehat{y_{i_t}}, \ldots, \widehat{y_{j_t}}$
          (whatever they are), which are $-1$.
	\end{itemize}

  \item Let $M = \{m_1^1, m_1^2, m_1^3, \ldots, m_n^1, m_n^2, m_n^3\}$
    be constructed as follows. For all $i \in [1,\ldots, n]$,
	\begin{itemize}
	 \item Let all attributes of $m_i^1$ be $0$, except for the value $y_i$, which is $-1$,
           and the value of $\overline{y_i}$ which is $1$.

	 \item Let all attributes of $m_i^2$ be $0$,  except for the value of $y_i$, which is $1$,
           and the value of $y_i'$, which is $-1$.

	 \item Let all attributes of $m_i^3$ are $0$, except for the value of $y_i'$, which is $1$,
            and the value of $c$, which is  $-1$.
	\end{itemize}	
\end{enumerate}

It can be easily shown that these sets of objects induce the set of
constructed constraints (see Example \ref{ex:induce-constraints}).\qed

\begin{figure}
	\begin{tikzpicture}[xscale=1, yscale=1.5]
		\tikzstyle{cir} = [draw=black,circle,inner sep=1pt]
                \node[cir] (y1)  at (1,2) {{\small $y_1$}};
                \node[cir] (y1n) at (1,1) {{\small $\overline{y_1}$}};
                \node[cir] (y1p) at (1,0) {{\small $y_1'$}};

                \node[cir] (y2)  at (3,2) {{\small $y_2$}};
                \node[cir] (y2n) at (3,1) {{\small $\overline{y_2}$}};
                \node[cir] (y2p) at (3,0) {{\small $y_2'$}};

                \node[cir] (y3)  at (5,2) {{\small $y_3$}};
                \node[cir] (y3n) at (5,1) {{\small $\overline{y_3}$}};
                \node[cir] (y3p) at (5,0) {{\small $y_3'$}};

                \node[cir] (c)   at (-1,1.5) {{\small $c$}};

		\draw[->, thick] (y1n) -- (y1p);
		\draw[->, thick] (y2n) -- (y2p);
		\draw[->, thick] (y3n) -- (y3p);

		\draw[->, thick] (y2n) -- (y1);
		\draw[->, thick] (y2n) -- (y3);
		\draw[->, thick] (y2n) -- (c);

		\draw[->, thick] (y1) -- (c);
		\draw[->, thick] (y3) -- (c);
	\end{tikzpicture}
        \caption{Example \ref{ex:induce-constraints}}
        \label{pic:ex:induce-constraints}
\end{figure}

\begin{example}\label{ex:induce-constraints}
        Take $n = 3$ and 
\[\phi(x_1, x_2, x_3) = (x_1 \vee x_2 \vee \overline{x_3}) 
\wedge (\overline{x_1} \vee x_2 \vee x_3).\]
Then $\attrset = \{c, y_1, \overline{y_1}, y_1', y_2, \overline{y_2},
y_2', y_3, \overline{y_3}, y_3'\}$. The constraints $\mu_1, \mu_2$ are
\begin{align*}
        \mu_1 : \edgeof{y_1}{c}{\Gamma} \vee \edgeof{y_2}{c}{\Gamma} \vee
\edgeof{\overline{y_3}}{c}{\Gamma} \\
        \mu_2 : \edgeof{\overline{y_1}}{c}{\Gamma} \vee \edgeof{y_2}{c}{\Gamma} \vee
\edgeof{y_3}{c}{\Gamma}
\end{align*}
Take the assignment $v = (1,0,1)$ satisfying $\phi$. By construction
above, we get the graph $\Gamma$ as in Figure \ref{pic:ex:induce-constraints}.Now let
us construct the sets $G$, $W$ and $M$ as above.

\noindent\begin{tabular}{c|c|c|c|c|c|c|c|c|c|c}
        & $y_1$ & $\overline{y_1}$ & $y_1'$ & 
          $y_2$ & $\overline{y_2}$ & $y_2'$ & 
          $y_3$ & $\overline{y_3}$ & $y_3'$ & $c$ \\
\hline
$g$         & $ 0$ & $ 0$ & $ 0$ & $ 0$ & $ 0$ & $ 0$ & $ 0$ & $ 0$ & $ 0$ & $ 0$ \\
$b_1$       & $ 0$ & -$1$ & $ 1$ & $ 0$ & $ 0$ & $ 0$ & $ 0$ & $ 0$ & $ 0$ & $ 0$ \\
$b_2$       & $ 0$ & $ 0$ & $ 0$ & $ 0$ & -$1$ & $ 1$ & $ 0$ & $ 0$ & $ 0$ & $ 0$ \\
$b_3$       & $ 0$ & $ 0$ & $ 0$ & $ 0$ & $ 0$ & $ 0$ & $ 0$ & -$1$ & $ 1$ & $ 0$ \\
$u_1$       & -$1$ & -$1$ & $ 0$ & $ 0$ & $ 0$ & $ 0$ & $ 0$ & $ 0$ & $ 0$ & $ 1$ \\
$u_2$       & $ 0$ & $ 0$ & $ 0$ & -$1$ & -$1$ & $ 0$ & $ 0$ & $ 0$ & $ 0$ & $ 1$ \\
$u_3$       & $ 0$ & $ 0$ & $ 0$ & $ 0$ & $ 0$ & $ 0$ & -$1$ & $ 1$ & $ 0$ & $ 1$ \\
$w_1$       & -$1$ & $ 0$ & $ 0$ & -$1$ & $ 0$ & $ 0$ & $ 0$ & -$1$ & $ 0$ & $ 1$ \\
$w_2$       & $ 0$ & -$1$ & $ 0$ & -$1$ & $ 0$ & $ 0$ & -$1$ & $ 0$ & $ 0$ & $ 1$ \\
$m_1^1$     & -$1$ & $ 1$ & $ 0$ & $ 0$ & $ 0$ & $ 0$ & $ 0$ & $ 0$ & $ 0$ & $ 0$ \\
$m_1^2$     & $ 1$ & $ 0$ & -$1$ & $ 0$ & $ 0$ & $ 0$ & $ 0$ & $ 0$ & $ 0$ & $ 0$ \\
$m_1^3$     & $ 0$ & $ 0$ & $ 1$ & $ 0$ & $ 0$ & $ 0$ & $ 0$ & $ 0$ & $ 0$ & $ 1$ \\
$m_2^1$     & $ 0$ & $ 0$ & $ 0$ & -$1$ & $ 1$ & $ 0$ & $ 0$ & $ 0$ & $ 0$ & $ 0$ \\
$m_2^2$     & $ 0$ & $ 0$ & $ 0$ & $ 1$ & $ 0$ & -$1$ & $ 0$ & $ 0$ & $ 0$ & $ 0$ \\
$m_2^3$     & $ 0$ & $ 0$ & $ 0$ & $ 0$ & $ 0$ & $ 1$ & $ 0$ & $ 0$ & $ 0$ & $ 1$ \\
$m_3^1$     & $ 0$ & $ 0$ & $ 0$ & $ 0$ & $ 0$ & $ 0$ & -$1$ & $ 1$ & $ 0$ & $ 0$ \\
$m_3^2$     & $ 0$ & $ 0$ & $ 0$ & $ 0$ & $ 0$ & $ 0$ & $ 1$ & $ 0$ & -$1$ & $ 0$ \\
$m_3^3$     & $ 0$ & $ 0$ & $ 0$ & $ 0$ & $ 0$ & $ 0$ & $ 0$ & $ 0$ & $ 1$ & $ 1$ \\
\end{tabular}

\medskip

Then $G = \{g\}$, $W = \{b_1,b_2,b_3,u_1,u_2,u_3,w_1,w_2\}$, $M = 
\{m_1^1, \ldots, m_3^3\}$. For $W$ to be a set of inferior examples,
$g$ must be preferred to each member of $W$. Take for instance, $g
\succ b_1$. By Theorem \ref{thm:pref-equiv}, that is equivalent to
$\edgeof{\overline{y_1}}{y_1'}{\Gamma_{\succ}}$, which corresponds to
$\pi_1$. Similarly, $g \succ u_1$ results in
$\edgeof{y_1}{c}{\Gamma_{\succ}} \vee
\edgeof{\overline{y_1}}{c}{\Gamma_{\succ}}$, which corresponds to
$\chi_1$. $g \succ w_1$ results in $\edgeof{y_1}{c}{\Gamma_\succ} \vee
\edgeof{y_2}{c}{\Gamma_\succ} \vee
\edgeof{\overline{y_3}}{c}{\Gamma_\succ}$, which corresponds to
$\mu_1$. The other members of $W$ are handled similarly (resulting
in the remaining positive constraints).

For $G$ to be superior, no member of $M \cup W$ must be preferred to
$g$ according to $\succ$. Clearly, for a p-skyline relation $\succ$
(which is an SPO), this is equivalent to saying that no member of only
$M$ must be preferred to $g$: above we already have constraints that
$g$ is preferred to every member of $W$, and $\succ$ is
irreflexive. $m_1^1 \not
\succ g$ results in $\notedgeof{\overline{y_i}}{y_1}{\Gamma_\succ}$,
which corresponds to $\lambda_1^1$. The other members of $M$ are
handled similarly, resulting in the remaining negative constraints.
\end{example}

%\subsection*{Proof of Proposition \ref{prop:polytime-comput}}  

\noindent {\bf Proposition \ref{prop:polytime-comput}. }{\it
 Let $\succ$ be a p-skyline relation, $\oset$ a finite set of tuples,
 and $G$ and $W$, disjoint subsets of $\oset$. Then the next two operations
 can be done in polynomial time:
 \begin{enumerate}
  \item verifying if $\succ$ is maximal favoring $G$ and disfavoring $W$ in $\oset$;
  \item constructing a maximal p-skyline relation $\succ_{ext}$ that favors
    $G$, disfavors $W$ in $\oset$ and is a p-extension of $\succ$ (under the assumption that $\succ$ favors
  $G$ and disfavors $W$ in $\oset$).
 \end{enumerate}
}

\smallskip
\noindent{\bf Proof.}
To check if $\succ$ favors $G$ and disfavors $W$ in $\oset$, we need
to compute $\winnow_{\succ}(\oset)$, check $G \subseteq
\winnow_{\succ}(\oset)$, and verify that for every $o \in W$, there is
$o' \in G$ such that $o'
\succ o$. All those tasks can clearly be performed in polynomial time. If some of these
conditions fails, $\succ$ is obviously not maximal. Otherwise, we need
to check if each of its minimal p-extensions favors $G$ and disfavors
$W$. Note that since $\succ$ disfavors $W$ in $\oset$, each of its
p-extensions also disfavors $W$ in $\oset$.  Hence, $\succ$ is not
maximal if at least one minimal p-extension favors $G$ in $\oset$, and
it is maximal otherwise.  Corollaries \ref{cor:polynomial-time-to-ext}
and \ref{cor:number-of-min-ext} imply that all minimal p-extensions of
$\succ$ can be constructed in polynomial time.
 
To construct a maximal p-extension $\succ'$ of $\succ$, we take
$\succ$, construct all of its minimal p-extensions and verify if
at least one of them favors $G$ in $\oset$.  If some of them does, we
select it and repeat for it the same procedure. We do it until for
some $\succ'$ none of its minimal p-extensions favors $G$ in
$\oset$. This implies that $\succ'$ is a maximal p-skyline relation
favoring $G$ and disfavoring $W$ in $\oset$. Moreover, $\succ'$ is a
superset of $\succ$ by construction. Corollaries
\ref{cor:polynomial-time-to-ext}, \ref{cor:number-of-min-ext},
and \ref{cor:min-ext-length}
imply that such a computation can be done in polynomial time. \qed

\medskip

\noindent {\bf Theorem \ref{thm:func-exist-fnp}. }{\it
 \funcfavdisprob is \fnpcomplete
}

\smallskip
\noindent{\bf Proof.}
  Given two disjoint subsets $G$ and $W$ of $\oset$ and $\succ \in
  \formset_\prefset$, checking if $\succ$ favors $G$ and disfavors $W$
  in $\oset$ can be done in polynomial time (Lemma
  \ref{prop:polytime-comput}). Hence, \funcfavdisprob is in \fnp.

  Now show that \funcfavdisprob is \fnphard. To do that, we use a
  reduction from \fsat.  In particular, we find functions $R$ and $S$,
  both computable in logarithmic space, such that 1) for each instance
  $x$ of \fsat, $R(x)$ is an instance of \funcfavdisprob, and 2) for
  each correct output $z$ of $R(x)$, $S(z)$ is a correct output of $x$.
  For such a reduction, we use the construction from the proof of
  Theorem \ref{thm:np-compl}.  There we showed how a relation (denote
  it as $\succ$) satisfying all the constraints (and thus favoring/disfavoring the
  constructed $G$ and $W$) may be obtained. In the current reduction,
  if there is a p-skyline relation favoring $G$ and disfavoring $W$ in
  $\oset$, then the relation $\succ$ itself is returned.  Otherwise,
  ``no'' is returned.

 The function $R$ mentioned above has to convert an instance of \fsat
 to an instance of \funcfavdisprob (i.e., $G$, $W$, and $\oset$).  In
 the reduction shown in the proof of Theorem \ref{thm:np-compl}, such
 a transformation is done via a set of constraints. However, it is
 easy to observe that such a construction can be performed using the
 corresponding instance of \fsat.  By the construction, the sets $G$,
 $M$, and the subset $\{b_1, \ldots, b_n, u_1, $ $\ldots, u_n\}$ of $W$
 are common for every instance of \fsat with $n$ variables.  To
 construct the subset $\{w_1, \ldots, w_m\}$ of $W$, one can use the
 expression $\psi_t$ instead of the corresponding constraint
 $\mu_t$. It is clear that the function $R$ performing such a
 transformation can be evaluated in logarithmic space.

 We construct the function $S$ as follows. If the instance of
 \funcfavdisprob returns ``no'', $S$ returns ``no''. Otherwise, it
 constructs the satisfying assignment $(v_1, \ldots, v_n)$ in the
 following way: for every $i$, $v_i$ is set to $1$ if the p-graph
 contains the edge $\edgeof{y_i}{c}{\Gamma_\succ}$, and $0$ otherwise.
 It is clear that such a computation may be done in logarithmic
 space. \qed

\medskip

\noindent {\bf Theorem \ref{thm:computation-hard}. }{\it
 \funcoptfavdisprob is \fnpcomplete
}

\smallskip
\noindent{\bf Proof.}
  Given $\succ \in \formset_\prefset$, checking if it is maximal
  favoring $G$ and disfavoring $W$ can be done in polynomial time
  (Proposition \ref{prop:polytime-comput}). Hence, \funcoptfavdisprob is in
  \fnp.

 We reduce from \funcfavdisprob to show that it is \fnphard.  Here we
 construct the function $F$ that takes a p-skyline relation or ``no''
 and returns a p-skyline relation or ``no''.  $F$ returns ``no'' if
 its input is ``no''. If its input is a p-skyline relation $\succ$, it
 returns a maximal p-extension of $\succ$ as shown in Proposition
 \ref{prop:polytime-comput}. As a result, $F$ returns a maximal
 favoring/dis\-fa\-vor\-ing p-skyline relation iff the corresponding
 favoring/dis\-fa\-vor\-ing p-skyline relation exists. The functions
 $R$ and $S$ transforming inputs of \funcfavdisprob to inputs of
 \funcoptfavdisprobx and outputs of \funcoptfavdisprob to outputs of
 \funcfavdisprob correspondingly are trivial and hence are computable
 in log\-space. Therefore, the problem \funcoptfavdisprob is
 \fnpcomplete. \qed

\medskip

\noindent {\bf Proposition \ref{prop:edge-diff}. }{\it
 Let a relation $\succ\ \in \formset_\prefset$ be a maximal
 $M$-favoring relation, and a p-extension $\succ_{ext}$ of
 $\succ$ be $(M \cup \{A\})$-favoring. Then every edge in
 $\edgesof{\Gamma_{\succ_{ext}}} - \edgesof{\Gamma_{\succ}}$ starts or
 ends in $A$.
}

\smallskip
\noindent{\bf Proof.}
 Take $\Gamma_{\succ_{ext}}$ and construct $\Gamma'$ from it by
 removing all edges going from or to $A$. Clearly, $\Gamma'$ is an
 SPO. Now consider the \envelope property. Pick four nodes of
 $\Gamma_{\succ}$ different from $A$. Since $\Gamma_{\succ_{ext}}$ is
 a p-graph, the \envelope property holds for the graph induced by
 these four nodes in $\Gamma_{\succ_{ext}}$. \envelope also holds for
 the corresponding subgraph of $\Gamma'$. Thus, $\Gamma'$ satisfies
 the {\tt Envelope} property as well, i.e., it's a p-graph of a
 p-skyline relation $\succ'$. Moreover, $\edgesof{\Gamma_\succ}
 \subseteq \edgesof{\Gamma_{\succ'}}$ since $\Gamma_{\succ}$ has no
 edges from/to $A$ and $\edgesof{\Gamma_\succ} \subseteq
 \edgesof{\Gamma_{\succ_{ext}}}$. Since $\succ$ is maximal
 $M$-favoring, $\edgesof{\Gamma_{\succ}} =
 \edgesof{\Gamma'}$. Therefore, all edges in
 $\edgesof{\Gamma_{\succ_{ext}}} - \edgesof{\Gamma_{\succ}}$ go from
 or to $A$. \qed

%\subsection*{Proof of Proposition \ref{prop:cons-modif}}

\noindent {\bf Proposition \ref{prop:cons-modif}. }{\it
 Let a relation $\succ\ \in \formset_\prefset$ satisfy a system of
 negative constraints $\negsystem$.  Construct the system of negative
 constraints $\negsystem'$ from $\negsystem$ in which every constraint $\tau' \in
 \negsystem'$ is created from a constraint $\tau$ of $\negsystem$ in the
 following way:

\begin{itemize}
 \item $\clhs{\tau'} = \clhs{\tau}$
 \item $\crhs{\tau'} = \crhs{\tau} - \{B \in \crhs{\tau}\ |\ \exists A
   \in \clhs{\tau} \ .\ \edgeof{A}{B}{\Gamma_\succ}\}.$
\end{itemize}

\noindent Then every p-extension $\succ'$ of $\succ$
satisfies $\negsystem$ iff $\succ'$ satisfies
$\negsystem'$.
}

\smallskip
\noindent{\bf Proof.}

\noindent \leftsideproof Take $\tau'$ from $\negsystem'$ with the
corresponding $\tau \in \negsystem$. By construction, $\clhs{\tau} =
\clhs{\tau}, \crhs{\tau'} \subseteq \crhs{\tau}$.  Now assume $\succ'$
satisfies $\tau'$. This means that
\begin{equation}\label{eq:cons-modif-1}
\exists B \in \crhs{\tau'} \ \forall A \in \clhs{\tau'}
\ :\ \notedgeof{A}{B}{\Gamma_{\succ'}}
\end{equation}
Now recall that $\crhs{\tau'} \subseteq \crhs{\tau}$. Thus $B \in
\crhs{\tau}$. This together with $\clhs{\tau} = \clhs{\tau'}$ and
\eqref{eq:cons-modif-1} gives
\[\exists B \in \crhs{\tau} \ \forall A \in \clhs{\tau}
\ .\ \notedgeof{A}{B}{\Gamma_{\succ'}},\]
i.e., $\Gamma_{\succ'}$ satisfies $\tau$.

\noindent \rightsideproof Now let $\succ'$ satisfy $\tau$. This means
\begin{equation}\label{eq:cons-modif-2}
\exists B \in \crhs{\tau} \ \forall A \in \clhs{\tau} \ .\ \notedgeof{A}{B}{\Gamma_{\succ'}}
\end{equation}
Since $\succ \ \subseteq \ \succ'$, $\edgesof{\Gamma_{\succ}}
\subseteq \edgesof{\Gamma_{\succ'}}$. Thus, if there is no edge from
$\clhs{\tau}$ to $B$ in $\Gamma_{\succ'}$, then there is no such edge
in its subset $\Gamma_{\succ}$. Recall that $\tau'$ is a
\emph{minimized} version of $\tau$ w.r.t. $\succ$.  Thus, the lack of
edge from $\clhs{\tau}$ to $B$ in $\Gamma_{\succ}$ implies $B \in
\crhs{\tau'}$.  This together with $\clhs{\tau} = \clhs{\tau'}$ and
\eqref{eq:cons-modif-2} gives
\[\exists B \in \crhs{\tau'} \ \forall A \in \clhs{\tau'}
\ .\ \notedgeof{A}{B}{\Gamma_{\succ'}},\] i.e., $\Gamma_{\succ'}$
satisfies $\tau'$. \qed

\medskip

\noindent {\bf Proposition \ref{prop:cons-checking}. }{\it
 Let a relation $\succ\ \in \formset_\prefset$ satisfy
 a system of negative constraints $\negsystem$, and $\negsystem$ be
 minimal w.r.t. $\succ$. Let $\succ'$ be a
 p-extension of $\succ$ such that every edge in $\edgesof{\Gamma_{\succ'}} -
 \edgesof{\Gamma_{\succ}}$ starts or ends in $A$.  Denote the
 \emph{new} parents and children of $A$ in $\Gamma_{\succ'}$ as $P_A$
 and $C_A$ correspondingly. Then $\succ'$ violates $\negsystem$ iff
 there is a constraint $\tau \in \negsystem$ such that
 
 \begin{enumerate}
   \item $\crhs{\tau} = \{A\} \wedge P_A \cap \clhs{\tau} \neq \emptyset, \mbox{ or}$
   \item $A \in \clhs{\tau} \wedge \crhs{\tau} \subseteq C_A$
 \end{enumerate}
}

\smallskip

\noindent{\bf Proof.}

\noindent \leftsideproof Trivial since the two conditions above imply
violation of $\negsystem'$ by $\succ$.

\noindent \rightsideproof Assume that there is no constraint $\tau$
for which the two conditions hold, but some $\tau' \in \negsystem$ is
violated, i.e.,
\[\children{\Gamma_\succ}(\clhs{\tau'}) \supseteq \crhs{\tau'}.\] By
Theorem \ref{thm:cnf-subset}, $\edgesof{\Gamma_{\succ}} \subset
\edgesof{\Gamma_{\succ'}}$. We also know that all the new edges in
$\Gamma_{\succ'}$ start or end in $A$. Since $\Gamma_{\succ}$
satisfies $\tau'$ but $\Gamma_{\succ'}$ does not, we get that either
$A \in \clhs{\tau'}$ or $A \in \crhs{\tau'}$. If $A$ is in
$\crhs{\tau'}$ then the fact that $\tau'$ is violated by
$\Gamma_{\succ'}$ implies that $\crhs{\tau'} = \{A\}$.  Moreover, the
fact that $\tau'$ is minimal w.r.t. $\succ$ implies $P_A \cap
\clhs{\tau'} \neq \emptyset$.  If $A \in \clhs{\tau'}$, then the
minimality of $\tau'$ implies that $\tau'$ is violated because of
$\crhs{\tau'} \subseteq C_A$. \qed

\medskip

\noindent {\bf Theorem \ref{thm:main2-runtime}.} {\it
 The function {\tt elicit} returns a syntax tree of a maximal
 p-skyline relation favoring $G$ in $\oset$. Its running time is
 $O(|\negsystem| \cdot |\attrset|^3)$.
}

\smallskip

\noindent{\bf Proof.}
First, we prove that {\tt elicit} always returns a maximal p-skyline
relation satisfying $\negsystem$.  By construction, the p-skyline
relation returned by {\tt elicit} satisfies the constructed system
of negative constraints $\negsystem$.  Now prove that $\succ$ returned
by {\tt elicit} is a maximal p-skyline relation satisfying
$\negsystem$.  A simple case analysis shows that {\tt push} picks
every p-skyline relation
\begin{enumerate}
\item which is a minimal p-extension of $\succ$ represented by the parameter $T$,
  and
\item whose p-graph has only edges going between the nodes $M \cup
  \{A\}$,
\end{enumerate}
until it finds one not violating $\negsystem$ (of course, given the
fact that the p-skyline relation, whose p-graph is obtained from $\Gamma_\succ$ by
removing edges going to/from $A$, is maximal $M$-fa\-vo\-ring).  Recall
that $T$ constructed in line 2 of {\tt elicit} represents a maximal
$M$-fa\-vor\-ing p-skyline relation satisfying $\negsystem$, for a
singleton $M$. Now assume that $T_{\succ}$ at the end of some iteration of the
{\bf for}-loop of {\tt elicit} represents a non-maximal $M_1$-favoring
p-skyline relation $\succ$. Take the first such an iteration of the {\bf
for}-loop.  It implies that there is an $M_1$-favoring p-skyline relation $\succ^*$ which strictly contains
$\succ$ and satisfies $\negsystem$. By Theorem
\ref{thm:cnf-subset}, $\edgesof{\Gamma_{\succ^*}}$ also strictly
contains $\edgesof{\Gamma_{\succ}}$. Take an edge
$\edgeof{X}{Y}{\Gamma_{\succ^*}}$ which is not in
$\edgesof{\Gamma_\succ}$. Let $\succ'$ be the relation constructed in the
{\bf for}-loop in {\tt elicit} when $A$ was equal to $X$ or $Y$,
whatever was the last one.  Take the corresponding set of attributes $M_2$. According
to the argument above, $\succ'$ is maximal $M_2$-fa\-vo\-ring.  Since
$\succ' \ \subseteq \ \succ$, $\Gamma_{\succ'}$ does not contain the edge $(X,Y)$.
At the same time, if we take $\Gamma_{\succ^*}$
and leave in it only the edges going to and from the elements of $M_2$,
it will strictly contain $\Gamma_{\succ'}$ and not violate
$\negsystem$. Hence, $\succ'$ is not maximal $M_2$-favoring, which is a
contradiction. That implies that {\tt elicit} returns a maximal
$\attrset$-favoring (or simply favoring) p-skyline relation satisfying
$\negsystem$.

Now let us show that the running time of the algorithm is $O(|\negsystem|
\cdot |\attrset|^3)$.  First, let us consider the running time of the
sub-procedures. The running time of {\tt minimize} and {\tt
check\-Constr} is $O( |\negsystem| \cdot |\attrset|)$. The time needed
to modify the syntax tree using a transformation rule is
$O(|\attrset|)$: every rule creates, deletes, and modifies a constant
number of nodes of a syntax tree, but updating their
$\formvars$-variables is done in $O(|\attrset|)$. Similarly, syntax
tree normalization runs in time $T_{normalizeTree} = O(|\attrset|)$
for such modified syntax trees. As a result, the time needed to
execute the bodies of the loops (lines 5-8, 11-14, 18-36) of {\tt
push} is $T_{rule} = O(|\negsystem| \cdot |\attrset|)$.
 
\begin{figure}
	\centering
 \begin{tikzpicture}
  \node[initial,state] 		(S1) 	at (0,1)	{$S_{\parcompsymbol}$};
  \node[state,accepting]	(S3) 	at (1.5,0)	{$S_3$};
  \node					at (1.5,1.2)	{$Rule_3$};
  \node[state] 			(S2) 	at (3,1)	{$S_{\pricompsymbol}$};

  \path[->] 	(S1) edge [loop above] node {$Rule_1, Rule_2$} ()
		     edge              (S2);
  \path[->] 	(S2) edge [loop above] node {$Rule_1, Rule_2$} ();

  \path[->]	(S1) edge (S3);
  \path[->]	(S2) edge (S3);
 \end{tikzpicture}
 \caption{Using {\tt push} for computation of a maximal $(M \cup \{A\})$-favoring
p-skyline relation}
 \label{pic:state-diag}
\end{figure}
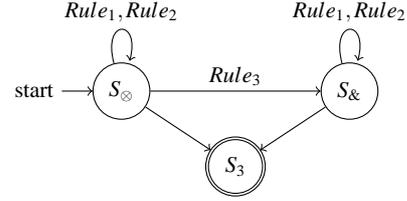

Now let $T$ be a syntax tree of a maximal $M$-favoring p-skyline
relation.  Consider the way {\tt push} is used in {\tt elicit} to
construct a maximal $(M \cup \{A\})$-favoring p-skyline relation. The
state diagram of this process is shown in Figure
\ref{pic:state-diag}.  It has three states: $S_{\parcompsymbol}$ and
$S_{\pricompsymbol}$ which correspond to $T$ in which $A$ is a child
of a $\parcompsymbol$- and $\pricompsymbol$-node, respectively; and
$S_3$ which corresponds to the case when no transformation rule can be
applied to $T$, or every rule application violates $\negsystem$.

The starting state is $S_{\parcompsymbol}$, because in the starting
$T$, $A$ is a child of the topmost $\parcompsymbol$-node. After
applying the transformation rules $Rule_1$ and $Rule_2$ in lines 21 and 25
respectively, $A$ becomes a child node of another
$\parcompsymbol$-node of the modified $T$.  After applying $Rule_3$
(lines 30 and 34), $A$ becomes a child of a $\pricompsymbol$-node in
the modified $T$, and we go to the state $S_{\pricompsymbol}$.  When
in $S_{\pricompsymbol}$, we can only apply $Rule_1$ or $Rule_2$ from
lines 6 and 12 respectively. Note that after applying these rules, $A$
is still a child of the same $\pricompsymbol$-node in the modified
$T$. When no rule can be applied to $T$ at some state, we go to the
accepting state $S_3$ and return $false$.

Consider the total number of nodes of $T$ enumerated in the loops
(lines 4-8, 10-14, and 17-36) of {\tt push} to construct a maximal $(M
\cup \{A\})$-favoring p-skyline relation. Note that when we go from
$S_{\parcompsymbol}$ to $S_{\parcompsymbol}$ by applying $Rule_1$ or
$Rule_2$, $A$ becomes a descendent of the $\parcompsymbol$-node whose
child it was originally. Hence, when in $S_{\parcompsymbol}$ we
enumerate the nodes $C_i$ to apply $Rule_1$ or $Rule_2$ to, we never
pick any $C_i$ which we picked in the previous calls of {\tt push}. In
the process of going from $S_{\pricompsymbol}$ to itself via an
application of $Rule_1$ or $Rule_2$, we \emph{may} enumerate the same
node $C_{i+1}$ more than once because $A$ does not change its parent
$\pricompsymbol$-node as a result of these applications.  To avoid
checking these rules against the same nodes $C_{i+1}$ more than once,
one can keep track of the nodes which have already been picked and
tested.

The total number of nodes in a syntax tree is $O(|\attrset|)$, hence
the tests $\formvars(C_{i+1}) \subseteq M$ (lines 4, 10) and
$\formvars(C_i) \subseteq M$ (line 17) are performed $O(|\attrset|)$
times and the rules are applied to the tree $O(|\attrset|)$
times. Each of the containment tests above requires time
$O(|\attrset|)$, given the bitmap representation of sets. Hence, to
compute the syntax tree of a maximal $(M \cup \{A\})$-favoring from the syntax tree
of a maximal $M$-favoring p-skyline relation, we need time
$O(|\negsystem| \cdot |\attrset|^2)$. Finally, the running time of
{\tt elicit} is $O(|\negsystem| \cdot |\attrset|^3)$. \qed

\medskip

\noindent {\bf Theorem \ref{thm:negsystimpl}. }{\it 
  \negsystimpl is co-NP complete
}

\smallskip
\noindent{\bf Proof.}
We show that checking the existence of $\succ \in \formset_\prefset$ satisfying
$\negsystem_1$ but not satisfying $\negsystem_2$ is
NP-complete. Clearly, this problem is in NP: we can guess $\succ \in
\formset_\prefset$ and in polynomial time check if it satisfies every $\tau \in
\negsystem_1$ (i.e., if there is a member of $\crhs{\tau}$ which has
no parent in $\clhs{\tau}$) but violates some $\tau' \in
\negsystem_2$. Now prove that checking if there's $\succ$ satisfying
$\negsystem_1$ but violating $\negsystem_2$ is NP-hard.

Here we show the reduction from SAT. Consider ins\-tan\-ces of SAT in the
following form
\[\varphi(x_1, \ldots, x_n) = \phi_1(x_1, \ldots,x_n) \wedge \ldots
\wedge \phi_m(x_1, \ldots,x_n)\]
where 
\[\phi_t(x_1, \ldots, x_n) = \widetilde{x_{i_t}} \vee \ldots \vee
\widetilde{x_{j_t}}\] 
and $\widetilde{x_i} \in \{ x_i, \overline{x_i}
\}$. For every instance $\varphi$, we construct
\[\attrset = \{x_1, \overline{x_1}, \ldots, x_n, \overline{x_n}, T, F\}.\]

Construct $\negsystem_1$ as follows:
\begin{enumerate}
 \item for every $\phi_t(x_1, \ldots, x_n) = \widetilde{x_{i_t}}
   \vee \ldots \vee \widetilde{x_{j_t}}$, create a constraint
   $\tau^1_t$ as follows:
\begin{align*}
 \clhs{\tau^1_t} & = \{F\}\\
 \crhs{\tau^1_t} & = \{\widetilde{x_{i_t}}, \ldots, \widetilde{x_{j_t}}\}
\end{align*}
%% \begin{example}
%%  Let $\phi_t = x_1 \vee \overline{x_2} \vee x_3$. Then $\clhs{\tau_t}
%%  = \{F\}$ and $\crhs{\tau_t} = \{x_1, \overline{x_2}, x_3\}$.
%% \end{example}

 \item for every variable $x_i$ of $\varphi$, create two constraints
   $\tau^2_i$ and $\tau^3_i$:
\begin{align*}
 \clhs{\tau_i^2} & = \{T\}\\
 \crhs{\tau_i^2} & = \{x_i, \overline{x_i}\}
\end{align*}
and
\begin{align*}
 \clhs{\tau_i^3} & = \{F\}\\
 \crhs{\tau_i^3} & = \{x_i, \overline{x_i}\}
\end{align*}
\end{enumerate}

Now we construct $\negsystem_2$ consisting of a single constraint
$\kappa$ as follows.
\begin{align*}
 \clhs{\kappa} & = \{T, F\}\\
 \crhs{\kappa} & = \{x_i, \overline{x_i}, \ldots, x_n, \overline{x_n}\}
\end{align*}

We prove that there is a satisfying assignment to $\varphi$ iff there
is a p-graph $\Gamma$ satisfying $\negsystem_1$ and not
satisfying $\negsystem_2$. First, assume that there is a
satisfying assignment $y = ({y_1}, $ $\ldots,$ $y_n)$ to $\varphi$. We
construct the graph $\Gamma$ as follows.  For every $i \in [1, n]$,
\begin{enumerate}
 \item if ${y_i} = 1$, then 
   $\edgeof{T}{x_i}{\Gamma}$ and $\edgeof{F}{\overline{x_i}}{\Gamma}$;
 \item if ${y_i} = 0$, then 
   $\edgeof{F}{x_i}{\Gamma}$ and $\edgeof{T}{\overline{x_i}}{\Gamma}$;
 \item $\Gamma$ has no other edges.
\end{enumerate}

Clearly, $\Gamma$ satisfies {\tt SPO} (every node has either an
incoming or outgoing edge, but not both) and \envelope (every node has
at most one incoming edge) and hence is a p-graph. We show that
$\Gamma$ satisfies $\negsystem_1$.
\begin{enumerate}
 \item Consider every constraint $\tau^1_t$ for every $\phi_t(x_1,
   \ldots, x_n) = \widetilde{x_{i_t}} \vee \ldots \vee
   \widetilde{x_{j_t}}$.  Since $y$ satisfies $\phi_t$, at least one
   of the conjuncts of $\phi_t$ (say, $\widetilde{x_{i_t}}$) is
   $1$. If $\widetilde{x_{i_t}} = x_{i_t}$, then $y_{i_t} = 1$, and
   $\notedgeof{F}{x_{i_t}}{\Gamma}$ by construction. If
   $\widetilde{x_{i_t}} = \overline{x_{i_t}}$, then $y_{i_t} = 0$ and
      $\notedgeof{F}{\overline{x_{i_t}}}{\Gamma}$. Hence, $\tau^1_t$ is
   satisfied.
 \item Consider $\tau^2_i$ and $\tau^3_i$ for every $x_i$. By
   construction of $\Gamma$, they are satisfied because
    it cannot be the case that  \edgeofx{$T$}{$x_i$}{$\Gamma$} and
   $\edgeof{T}{\overline{x_i}}{\Gamma}$ or 
   $\edgeof{F}{x_i}{\Gamma}$ and
   $\edgeof{F}{\overline{x_i}}{\Gamma}$. Hence, $\tau^2_i$ and
   $\tau^3_i$ are satisfied.
\end{enumerate}
Now consider $\negsystem_2$ and the constraint
$\kappa$. By construction, for every $i \in [1,n]$, the component
$y_i$ of $y$ is set to 0 or 1.  Hence,  
$\edgeof{T}{x_i}{\Gamma}$ and $\edgeof{F}{\overline{x_i}}{\Gamma}$ or 
$\edgeof{T}{\overline{x_i}}{\Gamma}$ and
$\edgeof{F}{x_i}{\Gamma}$. Therefore, $\kappa$ is violated by $\Gamma$.
\medskip

Now we show that if $\negsystem_1$ is satisfied by a p-graph $\Gamma$
and $\negsystem_2$ is not, then there is a satisfying assignment $y$
to $\varphi$. Take such a p-graph $\Gamma$.  We construct $y$ as
follows: $$y_i = \left\{ \begin{array}{ll} 1 & \mbox{if }
\edgeof{T}{x_i}{\Gamma}\\ 0 & \mbox{if } \edgeof{F}{x_i}{\Gamma},
\end{array}
\right .$$

First, we show that $y_i$ is well defined, i.e., exactly one of the following holds for every $i \in [1,n]$:
$\edgeof{T}{x_i}{\Gamma}$ and $\edgeof{F}{x_i}{\Gamma}$.
Since $\kappa \in \negsystem_2$ is violated
by $\Gamma$, for every $i \in [1,n]$
\begin{align}\label{eq:conp-proof-1}
\forall i \in [1, n]\ .\ (\edgeof{T}{x_i}{\Gamma} \vee
\edgeof{F}{x_i}{\Gamma}) \wedge \notag\\
(\edgeof{T}{\overline{x_i}}{\Gamma}
\vee \edgeof{F}{\overline{x_i}}{\Gamma})
\end{align}
Since $\negsystem_1$ is satisfied, 
\begin{equation}\label{eq:conp-proof-2}
 \forall i \in [1,n]\ .\ \notedgeof{T}{x_i}{\Gamma} \vee
 \notedgeof{T}{\overline{x_i}}{\Gamma},
\end{equation}
which follows from the satisfaction of $\tau^2_i$, and
\begin{equation}\label{eq:conp-proof-3}
 \forall i \in [1,n]\ .\ \notedgeof{F}{x_i}{\Gamma} \vee
 \notedgeof{F}{\overline{x_i}}{\Gamma},
\end{equation}
which follows from the satisfaction of $\tau^3_i$. Therefore, 
\eqref{eq:conp-proof-1}, \eqref{eq:conp-proof-2}, and \eqref{eq:conp-proof-3}
imply 
\begin{align}\label{eq:conp-proof-4}
 \forall i \in [1,n]\ .\ &
 \edgeof{T}{x_i}{\Gamma}\wedge\notedgeof{F}{x_i}{\Gamma} \wedge
 \edgeof{F}{\overline{x_i}}{\Gamma} \wedge \notag \\ 
& \notedgeof{T}{\overline{x_i}}{\Gamma} \vee 
 \edgeof{F}{x_i}{\Gamma} \wedge \notedgeof{T}{x_i}{\Gamma}\wedge \notag \\
& \edgeof{T}{\overline{x_i}}{\Gamma} \wedge
 \notedgeof{F}{\overline{x_i}}{\Gamma}
\end{align}

Now we show that $y$ satisfies $\varphi$. Since every $\tau^1_t$ is
satisfied, at least one of conjuncts of $\phi_t$ (say,
$\widetilde{x_{i_t}}$) does not have an incoming edge from $F$. If
$\widetilde{x_{i_t}} = x_{i_t}$ (i.e.,
$\notedgeof{F}{x_{i_t}}{\Gamma}$) then by \eqref{eq:conp-proof-4}
$\edgeof{T}{x_{i_t}}{\Gamma}$ and hence $y_{i_t} = 1$. Thus $\phi_t$
is satisfied.  Similarly, if $\widetilde{x_{i_t}} =
\overline{x_{i_t}}$ then $\edgeof{F}{x_i}{\Gamma}$ and hence $y_{i_t}
= 0$. Thus $\phi_t$ is satisfied.  Finally, $\varphi$ is
satisfied. Hence, we proved coNP-completeness of $\negsystimpl$.\qed

\medskip

\noindent {\bf Theorem \ref{thm:negsystimplsubset}. }{\it
  \negsystimplsubset is co-NP complete
}

\smallskip

\noindent{\bf Proof.}
The co-NP-completeness of \negsystimplsubset follows from the
co-NP-completeness of \negsystimpl. Namely, the membership test is the
same as in \negsystimplx.  To show co-NP-hardness of
\negsystimplsubset, we reduce from \negsystimpl.
We use the observation that $\negsystem_1$ implies $\negsystem_2$ iff
$\negsystem_1 \cup \negsystem_2$ is equivalent to $\negsystem_1$. \qed

\end{document}